\pdfoutput=1

\documentclass[11pt,a4paper]{article} 
\usepackage{mystyle}

\usepackage{amsmath,amssymb,epsfig,bm,amsfonts,yfonts}
\usepackage{tensor}
\usepackage{mathtools}
\usepackage{commath}
\usepackage{placeins}
\usepackage{mathrsfs}
\usepackage{listings}
\usepackage{lstautogobble}
\usepackage{todonotes}
\usepackage{graphicx}
\usepackage{subcaption,tikz}
\usepackage[utf8]{inputenc}

\relax
\title{Fluid dynamics of heavy ion collisions with Mode expansion (Fluid{\it u}M)}

\author{Stefan Floerchinger,}
\author{Eduardo Grossi}
\author{and Jorrit Lion}

\affiliation{Institut f\"{u}r Theoretische Physik, Universit\"{a}t Heidelberg, Philosophenweg 16, 69120 Heidelberg, Germany}
\emailAdd{stefan.floerchinger@thphys.uni-heidelberg.de}
\emailAdd{e.grossi@thphys.uni-heidelberg.de}
\emailAdd{j.lion@thphys.uni-heidelberg.de}

\abstract{
The fluid dynamics of a relativistic fireball with longitudinal and transverse expansion is described using a background-fluctuation splitting. Symmetry representations of azimuthal rotations and longitudinal boosts are used for a classification of initial state configurations and their fluid dynamic propagation in terms of a mode expansion. We develop an accurate and efficient numerical scheme based on the pseudo-spectral method to solve the resulting hyperbolic partial differential equations. Comparison to the analytically known Gubser solution underlines the high accuracy of this technique. We also present first applications of Fluid{\it u}M to central heavy ion collisions at the LHC energies featuring a realistic thermodynamic equations of state as well as shear and bulk viscous dissipation.}
\begin{document}
\maketitle

\section{Introduction}

High energy nuclear collisions at the LHC, at RHIC and elsewhere provide an interesting opportunity to study a fundamental quantum field theory -namely QCD- in an evolving state out-of-equilibrium.
It has been learned in recent years, that much of the bulk dynamics of a heavy ion collision can be described in terms of {\it relativistic fluid dynamics} \cite{ALICE:2011ab, Chatrchyan:2012ta, ATLAS:2012at, Adare:2011tg, Adamczyk:2013gw}.
This should be understood as a low-energy effective description of QCD dynamics in a situation of high energy density and for non-equilibrium dynamics, see \cite{Gyulassy:2004zy,Schafer:2009dj,Huovinen:2013wma,Heinz:2013th,Gale:2013da,shen2014standard,Jaiswal:2016hex,Romatschke:2017ejr,Busza:2018rrf,Dubla:2018czx} for recent reviews.

It is highly interesting to study the detailed relation between a microscopic realization as a relativistic quantum field theory and the macroscopic description  as a relativistic quantum fluid. This includes theoretical as well as experimental investigations. In practice, this enterprise is not always easy for various reasons. Experimentally, one can only access the final state after complicated dynamics involving the collision itself, different stages of  non-equilibrium evolution, hadronization,  kinetic freeze-out and finally hadron resonance decays, before  long-lived hadrons can be observed in particle detectors. Thus, the final observable in the detector provides only integrated information of all these stages.

Theoretically, complications arise mainly due to the non-perturbative character of QCD and because of the non-equilibrium character of the evolving state. Oftentimes, quite involved and computationally expensive numerical codes need to be used \cite{Song:2007ux,Bozek:2011ua,Schenke:2010nt,Karpenko:2013wva,Shen:2014vra,del2013relativistic,Habich:2014jna}. This makes advancements of understandings difficult. For example, for state-of-the art comparisons between experimental data and theoretical models based on Bayesian analysis \cite{Bernhard:2016tnd,Auvinen:2017nrm,Moreland:2018gsh}, the numerical evolution is a severe limiting factor. 

In the present work we aim at improving the situation by developing a theoretical approach and accurate numerical techniques based on two statistical symmetries. Azimuthal rotation symmetry and Bjorken boost symmetry are arguably available in realistic experiments. Both symmetries are not exact for a single event, but they can be used advantageously to classify the initial state configurations at the point where a fluid dynamics description becomes valid and they can be used to reduce the numerical effort of the fluid dynamical simulation, as we will explain. In addition, the formulation allows to gain valuable analytic insight into the dynamics of heavy ion collisions. 

The basic underlying idea of a mode expansion approach to the fluid dynamics of heavy ion collisions has been formulated in ref.\ \cite{Floerchinger2014b}. Initial states have been classified in refs.\ \cite{Floerchinger:2013vua,Petersen:2013cta} and the kinetic freeze-out in this formalism has been discussed in ref.\ \cite{floerchinger2014kinetic}. Statistical properties of initial state models in this framework have been analyzed in ref.\ \cite{Floerchinger2014a}. One can in fact formulate a systematic perturbative expansion for deviations from a symmetric situation, with favorable convergence properties \cite{Floerchinger2014, Brouzakis:2014gka}. What is missing until today is an efficient numerical algorithm to solve fluid evolution equations for the symmetric background as well azimuthal and rapidity dependent perturbations in an accurate way. This gap will be closed with the present paper. 

Recently, we have already discussed some properties of the radial expansion dynamics, in particular the mathematical structure of the corresponding partial differential equation with an emphasis on well-poseness and causality \cite{Floerchinger:2017cii}. 
In the present work we continue this discussion and provide in particular a detailed documentation of the numerical algorithm that can solve the hyperbolic partial differential equations for both the symmetric background and for various perturbations around it. 

One advantage of concentrating on symmetry properties as well as the mathematical structure of the evolution equations (hyperbolicity) is that the various extensions can be implemented in a straight-forward way. While we concentrate here on high energy collisions described by relativistic fluid dynamics without any conserved quantum number except energy and momentum, one may in the future wish to include non-vanishing baryon number, electric charge, electromagnetic fields or QCD fields such as the chiral order parameter, coherent pion fields etc. Because of the reduced numerical effort in the mode expansion description it may be also feasible to follow directly the two point function or higher-order correlation functions.
Because of the generic causality structure of relativistic theories and due to general symmetry arguments, the numerical scheme needed for such investigations would be relatively similar to what we develop here.

\section{Symmetries, coordinates and hyperbolic evolution equations}
\label{eq:Symmetries}
In this section we discuss a useful coordinate system and symmetry transformations for a description of high-energy collisions in terms of a set of hyperbolic partial differential equations. Here, we do not assume a specific form of these differential equations, except for causality in the relativistic sense, so that characteristic velocities are bound from above by the velocity of light. In this case one can provide initial conditions on a Cauchy surface with time-like (or, as a limit, light-like) normal vectors. The equations are then evolved from one Cauchy surface to the next.

\subsection{Coordinate system and symmetries}

For the description of high energy nuclear collisions it is convenient  to choose a coordinate system with the origin at the collision point, in the center of the fireball. The laboratory time $t$ (after the collision) and the longitudinal coordinate $z$ (parallel to the beam axis) are conveniently parametrized by the new Bjorken time coordinate or proper time $\tau=\sqrt{t^2-z^2}$ and the rapidity $\eta=\text{arctanh}(z/t)$, such that $t=\tau \cosh(\eta)$ and $z=\tau \sinh(\eta)$. Technically, the coordinate origin in rapidity and the transverse plane can be posed such that $\int_\Sigma d\Sigma_\mu x_\nu T^{\mu\nu}=0$, where the integral runs over a hypersurface of constant Bjorken time $\tau$ for $\tau\to 0$. In the transverse plane it is convenient to use cylindrical coordinates such that $r=\sqrt{x^2+y^2}$, $\phi=\arctan(y/x)$ and $x=r \cos(\phi)$, $y=r \sin(\phi)$. 

The coordinate system $\tau$, $r$, $\phi$ and $\eta$ is particularly suited to discuss two important symmetry transformations. The first one corresponds to azimuthal rotations around the beam axis, $\phi\to\phi+\Delta\phi$. Any field configuration at fixed Bjorken time $\tau$ can be decomposed into a linear superposition of irreducible representations with respect to this $U(1)$ symmetry. These are plane waves $e^{im\phi}$, transforming as $e^{im\phi} \to e^{im\Delta\phi}e^{im\phi}$. Because $\phi=\phi+2\pi$, the wave number $m$ is quantized and must be integer $m\in \mathbb{Z}$. 

The second useful class of symmetry transformations are longitudinal boosts $\eta\to \eta+\Delta\eta$. In a similar way as for azimuthal rotations, field configurations at fixed Bjorken time $\tau$ can also be decomposed into irreducible representations with respect to this translational or $\mathbb{R}^1$ symmetry. These are  plane waves $e^{ik\eta}$, transforming as $e^{ik\eta}\to e^{ik\Delta\eta} e^{ik\eta}$, where $k\in \mathbb{R}$ can now be any real number. This decomposition into irreducible representations is particularly convenient for a mode expansion framework to solve the equations, we will describe next.


\subsection{Quasi-linear and hyperbolic partial differential equations}
\label{sec_quasilinear}
For the following discussion we assume that we can describe our system in terms of a number of fields collected into the ``Nambu vector'' or ``Nambu spinor'' $\mathbf{\Phi}$ with $N$ components. In practice, $\mathbf{\Phi}$ might contain the independent components of temperature, fluid velocity, shear stress, bulk viscous pressure and any other field necessary for a local description.

Moreover, we assume that the evolution is determined by a set of hyperbolic, quasi-linear partial differential equations. It can be written in the symbolic form 
\begin{align}
\begin{split}
\mathbf{A}(\mathbf{\Phi},\tau,r)\cdot\partial_\tau \mathbf{\Phi}+\mathbf{B}(\mathbf{\Phi},\tau,r)\cdot \partial_r \mathbf{\Phi}
+\mathbf{C}(\mathbf{\Phi},\tau,r)\cdot \partial_\phi \mathbf{\Phi}
+\mathbf{D}(\mathbf{\Phi},\tau,r)\cdot \partial_\eta \mathbf{\Phi}
-\mathbf{S}(\mathbf{\Phi},\tau,r)=0.
\end{split}
\label{eq_eompertnotyettruncated}
\end{align}
We use here the $N\times N$ coefficient matrices $\mathbf{A}$, $\mathbf{B}$, $\mathbf{C}$ and $\mathbf{D}$. The ``source term'' $\mathbf{S}$ is an $N$-component vector. The explicit dependence on $\tau$ and $r$ originates from the choice of coordinates.

\subsection{Background-fluctuation splitting and mode expansion}

We will be interested in situations where the transformation behaviour with respect to  azimuthal rotation and longitudinal Bjorken boost are useful guiding principles. This does not imply that  field configurations are invariant under these symmetry transformations. We rather take these to be \emph{statistical symmetries}  and we will use a background - fluctuation splitting assuming that the fluctuation part (that breaks the symmetries) is not too large with respect to the symmetric background. This allows us to set up an expansion scheme. Under such circumstances, the symmetries are still very helpful to classify field configurations and to evolve them in time as we will discuss below. Technically, we write at some Bjorken time $\tau$
\begin{align}
\mathbf{\Phi}(\tau, r, \phi, \eta) = \mathbf{\Phi}_0(\tau,r)+\epsilon \, \mathbf{\Phi}_1(\tau,r,\phi,\eta), 
\label{eq_perturbation}
\end{align}
where the background field $\mathbf{\Phi}_0(\tau,r)$ is invariant under azimuthal rotations and Bjorken boosts, and the deviations from this symmetric situation are parametrized by the perturbation or fluctuation fields $\mathbf{\Phi}_1(\tau,r,\phi,\eta)$. We take $\epsilon$ as a formal expansion parameter, but will set $\epsilon\to1$ at the end. (The real expansion principle is the deviation from a symmetric situation.) Obviously, if \eqref{eq_perturbation} is used in \eqref{eq_eompertnotyettruncated}, one obtains
\begin{align}
\begin{split}
\mathbf{A}(\mathbf{\Phi}_0+\epsilon \mathbf{\Phi}_1,\tau,r)\cdot\partial_\tau (\mathbf{\Phi}_0+\epsilon \mathbf{\Phi}_1)+\mathbf{B}(\mathbf{\Phi}_0+\epsilon \mathbf{\Phi}_1,\tau,r)\cdot \partial_r(\mathbf{\Phi}_0+\epsilon \mathbf{\Phi}_1)\\
+\mathbf{C}(\mathbf{\Phi}_0+\epsilon \mathbf{\Phi}_1,\tau,r)\cdot \partial_\phi(\mathbf{\Phi}_0+\epsilon \mathbf{\Phi}_1)
+\mathbf{D}(\mathbf{\Phi}_0+\epsilon \mathbf{\Phi}_1,\tau,r)\cdot \partial_\eta(\mathbf{\Phi}_0+\epsilon \mathbf{\Phi}_1)\\
-\mathbf{S}(\mathbf{\Phi}_0+\epsilon \mathbf{\Phi}_1,\tau,r)=0.
\end{split}
\label{eq_eompertnotyettruncated2}
\end{align}
From \eqref{eq_eompertnotyettruncated2} one can then obtain the equations of motion for the background fields $\mathbf{\Phi}_0$ by considering only terms of zeroth order in $\epsilon$. The linearized equations for the perturbations are then obtained by taking only terms of first order and including higher order terms gives rise to quadratic and higher mode interactions.

The equations of motion for the background fields are now partial differential equations reduced to $1+1$ dimensions given by
\begin{align}
\mathbf{A}_0(\mathbf{\Phi}_0,\tau,r)\cdot\partial_\tau \mathbf{\Phi}_0(\tau,r)+\mathbf{B}_0(\mathbf{\Phi_0},\tau,r)\cdot \partial_r\mathbf{\Phi}_0(\tau,r)-\mathbf{S}_0(\mathbf{\Phi}_0,\tau,r)=0.
\label{eq_eombg}
\end{align}
Because of symmetry constraints, the background fields $\mathbf{\Phi}_0$ in general have fewer independent components than $\mathbf{\Phi}$. For example, in Israel-Stewart type fluid dynamics, as used below, one may take $\mathbf{\Phi}_0=(T,v, \pi^\phi{}_\phi, \pi^\eta{}_\eta, \pi_\text{bulk})$. The matrices $\mathbf{A}_0$ and $\mathbf{B}_0$ essentially  correspond to the projection of the matrices $\mathbf{A}$ and $\mathbf{B}$ to the reduced space of independent components, evaluated on the background configuration $\mathbf{\Phi}_0$. Although \eqref{eq_eombg} are still non-linear partial differential equations, solving them is easier than solving the set \eqref{eq_eompertnotyettruncated} in $3+1$ dimensions.

For the perturbations, we find at linear order in $\epsilon$,
\begin{align}
\begin{split}
\mathbf{A}_1(\mathbf{\Phi}_0,\tau,r)\cdot\partial_\tau \mathbf{\Phi}_1+\mathbf{B}_1(\mathbf{\Phi}_0,\tau,r)\cdot \partial_r\mathbf{\Phi}_1+\mathbf{C}_1(\mathbf{\Phi}_0,\tau,r)\cdot \partial_\phi\mathbf{\Phi}_1\\
+\mathbf{D}_1(\mathbf{\Phi}_0,\tau,r)\cdot \partial_\eta\mathbf{\Phi}_1-\mathbf{S}_1(\mathbf{\Phi}_0,\tau,r)\cdot\mathbf{\Phi}_1=0.
\end{split}
\label{eq_eomPerturbation}
\end{align}
The matrices $\mathbf{A}_1$, $\mathbf{B}_1$, $\mathbf{C}_1$ and $\mathbf{D}_1$ simply correspond to $\mathbf{A}$, $\mathbf{B}$, $\mathbf{C}$ and $\mathbf{D}$ evaluated on the background configuration $\mathbf{\Phi}_0$. In contrast, the source term matrix $\mathbf{S}_1$ contains also contributions from the linearization of $\mathbf{A}$ and $\mathbf{B}$ around the background field,
\begin{equation}
\mathbf{S}_1(\mathbf{\Phi}_0,\tau,r) = \frac{\partial}{\partial \mathbf{\Phi}} \left[ \mathbf{S}(\mathbf{\Phi}, \tau, r) - \mathbf{A}(\mathbf{\Phi}, \tau, r) \cdot \partial_\tau \mathbf{\Phi}_0 - \mathbf{B}(\mathbf{\Phi}, \tau, r) \cdot \partial_r \mathbf{\Phi}_0 \right]_{\mathbf{\Phi}=\mathbf{\Phi}_0}.
\end{equation}
Note here that also the dependence of thermodynamic and transport properties on the (fluid-) fields needs to be taken into account. 

In practice, the matrix expressions can be algebraically rather complex. These equations not only depend on the background fields, but also on their derivatives, as well as derivatives of thermodynamic and transport coefficients. This fact makes it necessary to have differentiable transport coefficients and smooth background fields. Shock formations in the background will cause problems in the linearized equations. Shocks are suppressed though (but not necessarily fully prevented) by viscosity and smooth initial conditions.

It is now useful to expand the fluctuation or perturbation fields into Fourier modes,
\begin{equation}
\mathbf{\Phi}_1(\tau, r, \phi, \eta) = \sum_{m=-\infty}^\infty \int \frac{dk}{2\pi} e^{im\phi+ik\eta} \; \tilde{\mathbf{\Phi}}_1(\tau, r, m, k).
\label{eq:FourierRepresentationPerturbationFields}
\end{equation}
The evolution equation for the perturbations can then be written as
\begin{align}
\begin{split}
\mathbf{A}_1(\mathbf{\Phi}_0,\tau,r)\cdot\partial_\tau \tilde{\mathbf{\Phi}}_1+\mathbf{B}_1(\mathbf{\Phi}_0,\tau,r)\cdot \partial_r \tilde{\mathbf{\Phi}}_1+i m \, \mathbf{C}_1(\mathbf{\Phi}_0,\tau,r)\cdot \tilde{\mathbf{\Phi}}_1\\
+i k \, \mathbf{D}_1(\mathbf{\Phi}_0,\tau,r)\cdot \tilde{\mathbf{\Phi}}_1-\mathbf{S}_1(\mathbf{\Phi}_0,\tau,r)\cdot\tilde{\mathbf{\Phi}}_1=0.
\end{split}
\label{eq_eomPerturbationFourier}
\end{align}
%
Note that these are now again partial differential equations reduced to $1+1$ dimensions. In the following, we will sometimes also write $\mathbf{\Phi}_1$ instead of $\tilde{\mathbf{\Phi}}_1$ for the Fourier transformed field but this should not lead to confusion.

\subsection{Scalar, vector and tensor modes and their parity}
\label{sec:ScalarVectorTensorModesParity}
In a relativistic setting, fields are usually classified into scalars, vectors and tensors with respect to Lorentz symmetry. (At some point one may wish to evolve also spinor fields, but this is beyond our current setup.) For our purpose, mainly the subgroups of azimuthal rotations and Bjorken boosts are of relevance. In a coordinate system with Bjorken time $\tau$, rapidity $\eta$, radius $r$ and rapidity $\eta$, these transformations have become translations, and components of Lorentz vectors and tensors, such as e.\ g.\ the fluid velocity components $u^\tau$, $u^r$, $u^\phi$ and $u^\eta$, transform formally simply like scalars. 

It is sometimes useful to extend the radial coordinate $r$ to negative values, in particular to circumvent the boundary at $r=0$ (or actually to treat it properly). This leads to a double coverage of coordinate space because the coordinate point $(\tau, r, \phi, \eta)$ agrees with $(\tau, -r, \phi+\pi, \eta)$. Fields can be classified with respect to their behaviour under the parity transformation $\mathscr{R}$ corresponding to $(r,\phi)\to (- r, \phi+\pi)$. In real space, Lorentz scalars like temperature $T$ are even with respect to $\mathscr{R}$ and so are Lorentz vector components in time, rapidity and azimuthal directions such as $u^\tau$, $u^\phi$ and $u^\eta$. In contrast, radial components such as $u^r$ are odd under $\mathscr{R}$ in the sense that $u^r\to - u^r$. More general, components of tensors are odd (even) under this parity if they contain $r$ as an index an odd (even) number of times. For example $\pi^{r\phi}$ is odd, while $\pi^{rr}$ is even. 

In the Fourier representation of fields defined in equation \eqref{eq:FourierRepresentationPerturbationFields}, the $\mathscr{R}$ parity of each field receives an additional factor $(-1)^m$ due to $e^{im\phi}\to e^{im(\phi+\pi)}=(-1)^m e^{im\phi}$. In summary, Fourier modes with azimuthal wave number $m$ of tensor field components where the index $r$ appears $n$ times, have $\mathscr{R}$ parity $(-1)^{m+n}$.

We need to discuss also the boundary conditions of various fields for $r\to 0$. Scalar fields such as temperature are expected to be regular and smooth for $r\to 0$. In cartesian transverse coordinates $x$ and $y$, the field should be analytic, i.\ e.\ expandable into a Taylor series, at the origin $x=y=0$. From this condition one can infer that the Fourier modes $\tilde{\mathbf{\Phi}}(\tau,r,m,k)$ corresponding to scalar fields go like $r^{|m|}$ for $r\to 0$. A faster growth with $r$ would correspond to a kind of conical singularity at $r=0$, which is not expected. 

For transverse vector fields such as $(u^x,u^y)$ the argument is similar. Their Fourier components vanish like $r^{|m|}$ for $r\to 0$. In terms of transverse coordinates $r$ and $\phi$ this changes somewhat, however, because of transformation Jacobians such as in
\begin{equation}
\begin{split}
u^r = & \cos(\phi) u^x + \sin(x) u^y=e^{i\phi} \frac{u^x-i u^y}{2} + e^{-i\phi} \frac{u^x+iu^y}{2}, \\
u^\phi = & - \frac{\sin(\phi)}{r} u^x + \frac{\cos(\phi)}{r} u^y = \frac{e^{i\phi}}{r} \frac{i u^x +u^y}{2} + \frac{e^{-i\phi}}{r} \frac{-iu^x+u^y}{2}.
\end{split}
\end{equation}
This shows that the Fourier components of the {\it positive circular velocity}
\begin{equation}
u^+ = \frac{u^r+i r u^\phi}{\sqrt{2}} = e^{-i\phi}\frac{u^x+iu^y}{\sqrt{2}}
\end{equation}
go for $r\to 0$ like $r^{|m+1|}$ and the Fourier components of the {\it negative circular velocity}
\begin{equation}
u^- = \frac{u^r-i r u^\phi}{\sqrt{2}} = e^{i\phi} \frac{u^x-iu^y}{\sqrt{2}}
\end{equation}
go for $r\to 0$ like $r^{|m-1|}$. In a similar way one can analyze linear combinations of transverse tensor components.

\section{Characterization of initial conditions}
\label{sec_expansionofinitialconditions}
If the evolution equations are hyperbolic differential equations, initial conditions must be provided on an appropriate Cauchy surface, for example corresponding to constant Bjorken time $\tau=\tau_0$. This concerns the background field as well as the fluctuations or perturbations around it.

\subsection{Background configuration}
For the background field $\mathbf{\Phi}_0$ it is particularly convenient to choose it symmetric under azimuthal rotations and Bjorken boosts in the longitudinal direction. Moreover, this background is typically fixed for a given class of events, i.\ e.\ it does not have statistical fluctuations. One might take the background to correspond to an expectation value or event average for an appropriately defined ensemble of events, but this is not strictly necessary. At a given initialization time $\tau_0$ one needs to specify then initial conditions  as functions of radius $r$, only. Schematically, we write
\begin{equation}
\mathbf{\Phi}_0(\tau_0, r) = \mathbf{b}_0(r).
\end{equation}
In general, the initialization includes scalar modes such as energy density, but also the radial components of vector fields, such as velocity $u^r$, or tensor components allowed by azimuthal rotation and Bjorken boost symmetry, could be initialized as part of the background. 

\subsection{Fluctuating modes}
\label{sec:FluctuatingModes}

For the fluctuating part of the fields $\mathbf{\Phi}_1$, two additional complications arise. First, they should not be taken as symmetric with respect to azimuthal rotations and longitudinal boost, and second, these fields fluctuate from even to event. They can also be subject to quantum and statistical (e.\ g.\ thermal) fluctuations. 

On a Cauchy hypersurface of constant Bjorken time $\tau_0$ where initial conditions are specified, one can expand the perturbation fields into Fourier modes according to equation \eqref{eq:FourierRepresentationPerturbationFields}. Note that the field in Fourier space $\tilde{\mathbf{\Phi}}_1(\tau_0, r, m, k)$ is characterized by the azimuthal wave number $m$ and longitudinal wave number $k$ but it is still a function of radius $r$. It is now very convenient to use also an expansion in terms of a set of basis functions $q_{m,l}(r)$ with radial wave number $l$. (The basis functions $q_{m,l}(r)$ need to depend also on $m$ because the boundary conditions at $r\to 0$ are $m$-dependent.) This amounts to writing $\tilde{\mathbf{\Phi}}_1(\tau_0, r, m, k) = \sum_l q_{m,l}(r) \, \mathbf{h}_{m,l}(k) $ or, using \eqref{eq:FourierRepresentationPerturbationFields},
\begin{equation}
\mathbf{\Phi}_1(\tau_0, r, \phi, \eta) = \sum_{m=-\infty}^\infty \sum_{l} \int \frac{dk}{2\pi} e^{im\phi+ik\eta} \, q_{m,l}(r) \,{\mathbf{h}}_{m,l}(k).
\end{equation}
One can then characterize the initial conditions for a single fluctuating field as an amplitude $\mathbf{h}_{m,l}(k)$ that depends on three wave numbers: $m$ for the azimuthal dependence, $k$ for the longitudinal and $l$ for the radial dependence. An ensemble of events can be characterized by a probability  distribution $p[\mathbf{h}]$ or, equivalently, by corresponding moments, cumulants or correlation functions, such as e.\ g.\ $\langle \mathbf{h}_{m,l}(k) \mathbf{h}_{m^\prime,l^\prime}(k^\prime) \rangle$.

\begin{figure}[t!]
\centering
\includegraphics[width=0.33\textwidth]{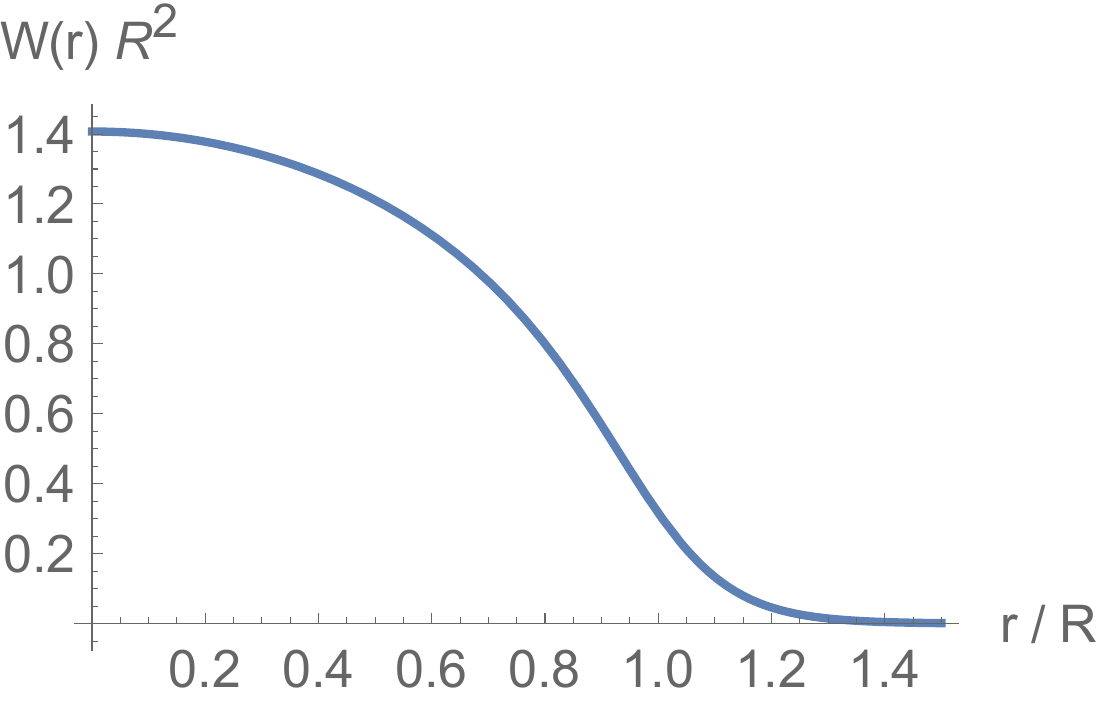}\hspace{0.4cm}
\includegraphics[width=0.33\textwidth]{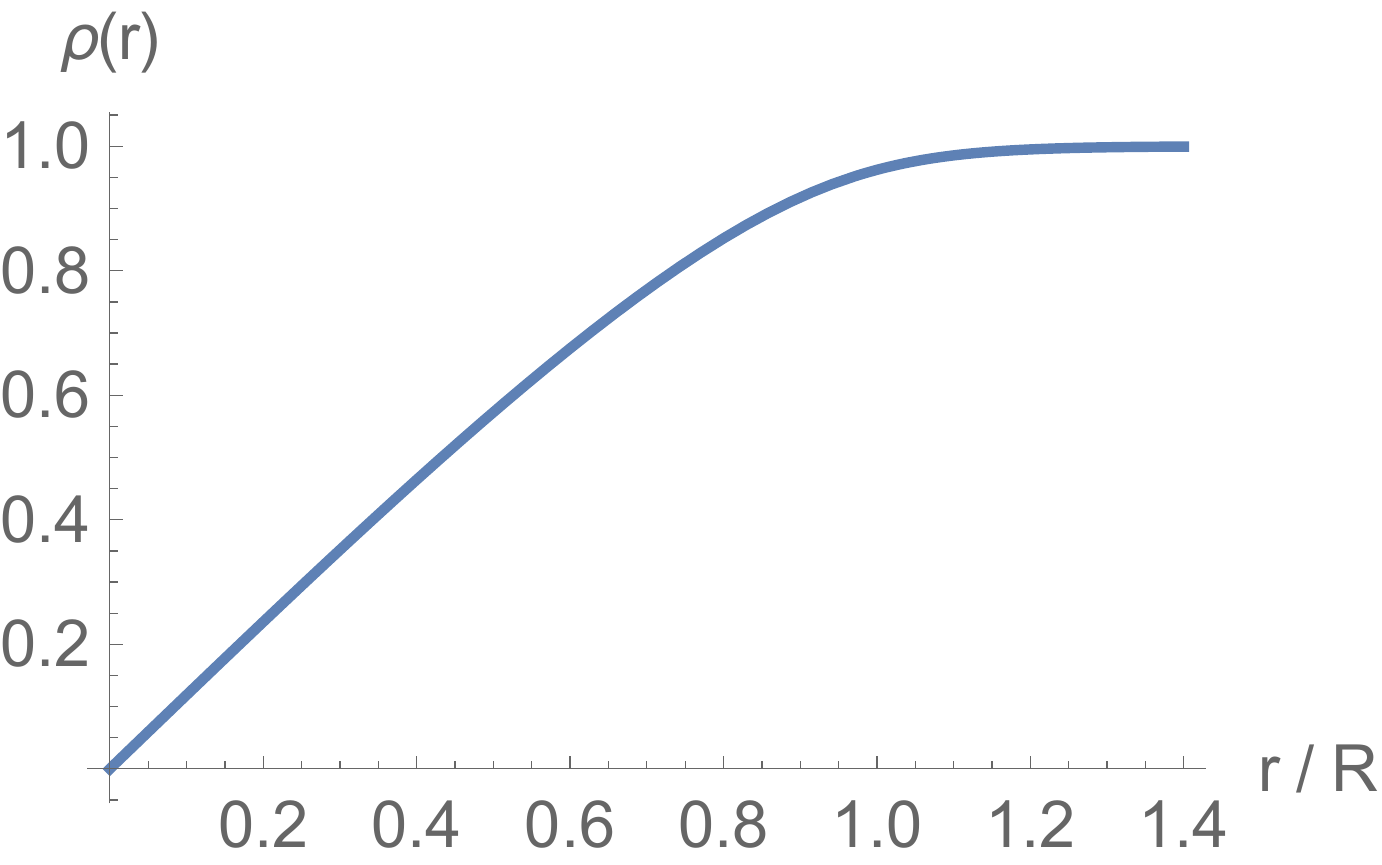}
\caption{Left panel: Example for the function $W(r)$ as constructed by integrating a Woods-Saxon profile along the longitudinal direction $z$ and normalizing the resulting function according to equation \eqref{eq:Wnormalization}. The units are set by the Woods-Saxon radius $R$. We also choose the surface thickness parameter such that $a/R=0.082$. Right panel: the function $\rho(r)$ as defined in equation \eqref{eq:definitionrhor}.}
\label{functionWfunctionrho}
\end{figure}

To construct a set of basis functions $q_{m,l}(r)$, it is convenient to 
employ a function $W(r)$ which is positive at small radius $r$ and falls off for large $r$, see fig.\ \ref{functionWfunctionrho} for an example. It could give the shape of the expected energy density in the transverse plane, averaged over events and azimuthal orientations. It will be convenient to normalize $W(r)$ such that
\begin{equation}
2\int_0^\infty dr \, r \, W(r) = 1.
\label{eq:Wnormalization}
\end{equation}
Note that with this normalization, $W(r)$ has dimensions of inverse squared length. One may consider $W(r) = dQ/(r dr)$ as a transverse {\it density} of some quantity $Q$ which happens to integrate to the total value $\int dQ = \int r dr W(r)=1/2$. Typically, $W(r)$ will be non-zero inside some radius $R$ and decay quickly (e.\ g.\ exponentially) outside of it.
The transverse density $W(r)$ is useful as a normalization factor for other transverse densities. For example, one may write a transverse particle density as
\begin{equation}
\frac{dN}{r dr d\phi} = W(r) f(r,\phi) = \frac{dQ}{r dr} f(r,\phi).
\end{equation}
Under coordinate transformations $r\to \tilde r(r)$, the function $W(r)$ transforms as a density, $\tilde W= \frac{dQ}{\tilde r d\tilde r} = \frac{r dr}{\tilde r d\tilde r} W$ (it picks up a Jacobian factor) while $f(r,\phi)$ transforms as a simple function, $\tilde f= f$. 

Moreover, based on $W(r)$ one can also construct a map from the unbounded domain of possible radii $r\in (0,\infty)$ to a radial variable on a finite interval $\rho\in(0,1)$ by setting
\begin{equation}
\rho(r) = \sqrt{ 2 \int_{0}^{r}dr'\,r'\, W(r') }, \quad\quad\quad \frac{d \rho(r)}{dr} = \frac{r W(r)}{\rho(r)}.
\label{eq:definitionrhor}
\end{equation}
By construction, $\rho$ is linear in $r$ for small radii while $\rho\to 1$ for $r\to \infty$. If one again considers $W(r) = dQ/(r dr)$ as a transverse {\it density}, this density transforms to a uniform density with respect to the new coordinate, $dQ/(\rho d\rho) =1$. In other words, the coordinate $\rho(r)$ is constructed such that $\frac{1}{2}\rho^2(r)$ counts the integrated quantity $Q$ inside the radius $r$. In fig.\ \ref{functionWfunctionrho} we show an example for both the normalized function $W(r)$ and the corresponding map function $\rho(r)$.

Based on $W(r)$, we also define the following scalar product for dimensionless functions of radius, $f(r)$ and $g(r)$,
\begin{equation}
(f,g) = \int_0^\infty dr \, r \, W(r) \, f^*(r) \, g(r).
\label{eq:scalarproductr}
\end{equation}
The appearance of $W(r)$ makes sure that the region of small radii (where the density is non-vanishing) typically dominates the integral. Note that under the variable transform $r\to \rho(r)$ the scalar product \eqref{eq:scalarproductr} becomes
\begin{equation}
(f,g) = \int_0^1 d\rho \, \rho \, f^*(r(\rho)) \, g(r(\rho)).
\label{eq:scalarproductrho}
\end{equation}
We assumed here implicitly that $f(r)$ and $g(r)$ are ordinary functions and not {\it densities}. The latter would pick up an additional Jacobian weight factor from the transformation $r\to \rho$. As illustrated above, such densities can be constructed from ordinary functions $f(r)$ by multiplying with (an appropriate power) of $W(r)$.



An expansion scheme based on a complete and orthonormal set of Bessel functions has been developed in \cite{Petersen:2013cta,Floerchinger2014b,Floerchinger:2013vua} and refined in appendix A of ref.\ \cite{Floerchinger2014a}. In the following we recall the main elements of this scheme. We concentrate on scalar fields and take the transverse energy density $\epsilon(r,\phi)$ as an example, but vector or tensor fields can be treated similarly \cite{Floerchinger:2013vua}.

We first introduce azimuthal Fourier modes as usual,
\begin{equation}
\epsilon(r,\phi) = \sum_{m=-\infty}^\infty e^{im\phi} \epsilon_{m}(r), \quad\quad\quad \epsilon_{m}(r) = \frac{1}{2\pi} \int_0^{2\pi} d\phi \, e^{-im\phi} \, \epsilon(r, \phi).
\end{equation}
As argued in section \ref{sec:ScalarVectorTensorModesParity}, scalar modes  in Fourier space such as $\epsilon_m(r)$ behave like $r^{|m|}$ for small radii $r\to 0$. The Bessel functions of the first kind $J_m(z)$ have this property, and they can also be chosen such as to fulfill an orthogonality relation with respect to the scalar product \eqref{eq:scalarproductrho}. Together with the discussion above, this motivates to expand 
\begin{equation}
\epsilon_m(r) = \sum_{l=1}^\infty \epsilon_{m,l} \, W(r) \, J_{m}{\big (}z^{(m)}_l \rho(r) {\big )}.
\label{eq:expansionepsilonmBessel}
\end{equation}
Here, $\epsilon_{m,l}$ are expansion coefficients and $z^{(m)}_l$ are real numbers, corresponding to the $l$'th zero crossing of $zJ^\prime_m(z)+ c J_m(z)$, where $c$ is some arbitrary constant. The functions $J_{m}{\big (}z^{(m)}_l \rho {\big )}$ satisfy by construction the Robin boundary condition $\rho f^\prime(\rho) + c f(\rho)=0$ at the outer boundary $\rho=1$. In ref.\ \cite{Floerchinger2014a} Dirichlet boundary conditions $f(\rho)=0$ where employed, corresponding to the limit $c\to \infty$ of the more general Robin boundary employed here. In the following, we will mainly employ the opposite limit $c\to 0$, where the $z^{(m)}_l$ correspond to the zero crossings of $J^\prime_m(z)$ and the boundary conditions are of von Neumann type. In principle, the Bessel functions form a complete basis on $\rho\in(0,1)$ for any value of $c$, but the convergence properties might depend on this choice. Von Neumann boundary conditions have the advantage that the value $f(\rho)$ remains unconstrained for $\rho\to 1$.

\begin{figure}[t!]
\centering
\includegraphics[width=0.33\textwidth]{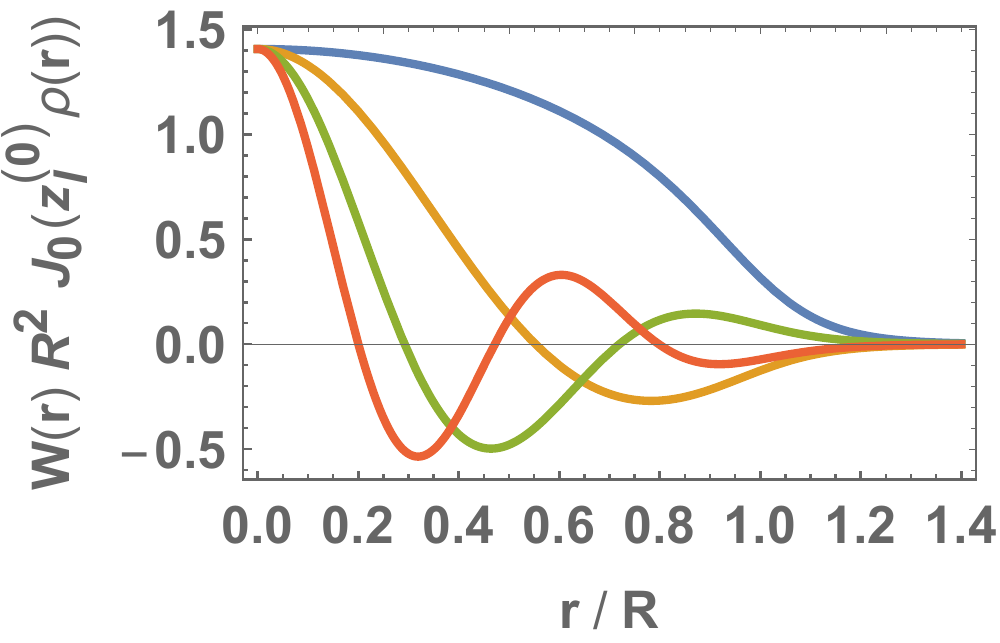}\hfill
\includegraphics[width=0.33\textwidth]{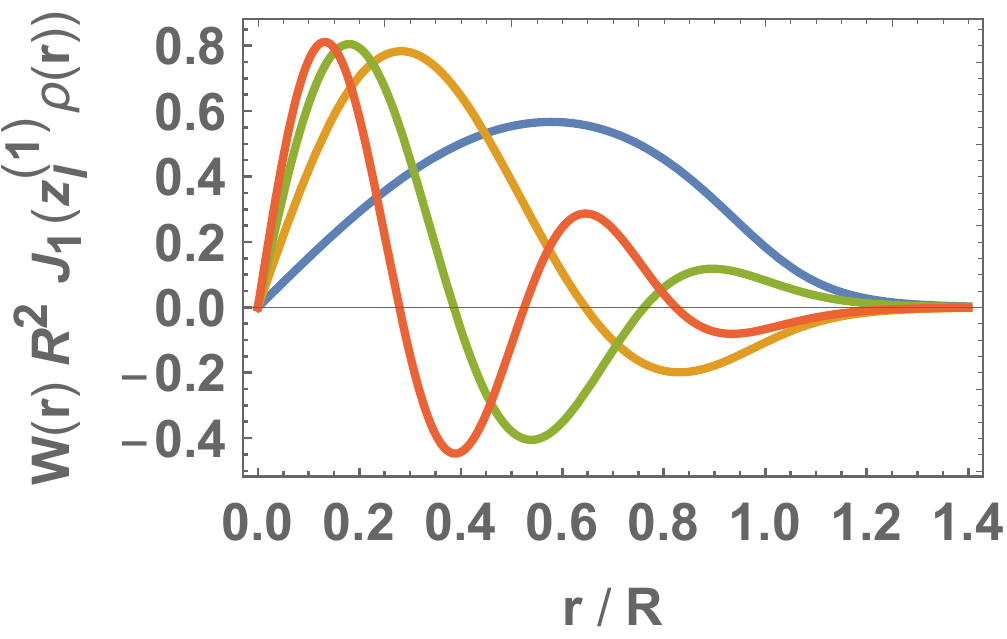}\hfill
\includegraphics[width=0.33\textwidth]{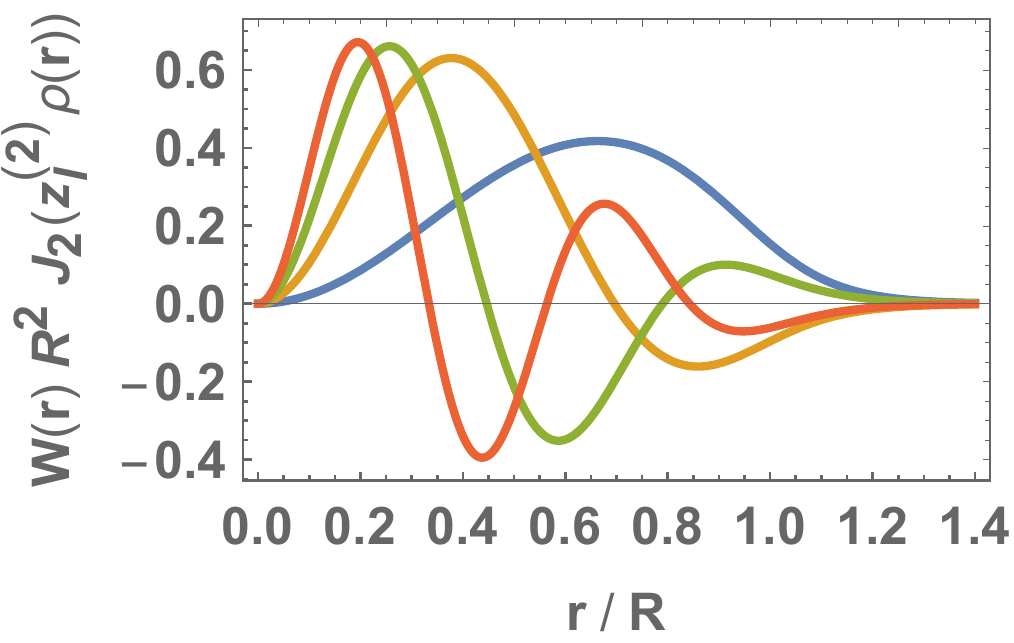}
\caption{Basis functions used to represent the radial dependence of transverse densities with Bessel functions according to \eqref{eq:basisfunctionsBessel}. We compare $m=0$ (left panel) to $m=1$ (center panel) and $m=2$ (right panel). In each case we show the functions with $l=1,2,3,4$ as can be seen from the number of zero crossings. The numbers $z^{(m)}_l$ have been chosen as the $l$'th zero crossing of $zJ^\prime_m(z)$ corresponding to von Neumann boundary conditions at $\rho=1$. Units are set by the Woods-Saxon radius $R$.}
\label{basisfunctionsBessel}
\end{figure}

The Bessel functions have the orthogonality relation
\begin{equation}
\int_0^\infty dr \, r \, W(r) \, J_{m}{\big (}z^{(m)}_l \rho(r) {\big )} \, J_{m}{\big (}z^{(m)}_{l^\prime} \rho(r) {\big )} = \int_0^1 d\rho \, \rho \, J_{m}{\big (}z^{(m)}_l \rho {\big )} \, J_{m}{\big (}z^{(m)}_{l^\prime} \rho {\big )} = c_{m,l} \, \delta_{l l^\prime},
\end{equation}
with coefficients $c_{m,l}$ given by
\begin{equation}
c_{m,l} = \frac{\left[ {\big (}z_l^{(m)}{\big )}^2 - m^2 \right] J_m^2{\big (}z^{(m)}_l{\big )} + {\big (}z^{(m)}_l{\big )}^2 J_m^{\prime 2}{\big (}z^{(m)}_l{\big )}}{2{\big (} z^{(m)}_l{\big )}^2}.
\end{equation}
We have now constructed a complete and orthogonal basis of mode functions with the radial wave number $l$. In summary, transverse densities can be expanded in terms of Bessel functions by choosing
\begin{equation}
q_{m,l}(r) = W(r) J_{m}{\big (}z^{(m)}_l \rho(r) {\big )}.
\label{eq:basisfunctionsBessel}
\end{equation}
We show the first few ($l=1,2,3,4$) of these basis functions in Fig.\ \ref{basisfunctionsBessel} for $m=0,1,2$. It is interesting to note that for $m=0$ and $l=1$ one recovers the background density function, $q_{0,1}(r) = W(r)$. In general, the basis functions have $l-1$ zero crossings or nodes between $r=0$ and $r\to \infty$. By construction, they are all concentrated in the region where the background density $W(r)$ is non-vanishing.

We also note here that the expansion coefficients $\epsilon_{m,l}$ in \eqref{eq:expansionepsilonmBessel} are different from the commonly used event eccentricities. There is, however, a connection which is worth to be discussed. In our notation, event eccentricities $\varepsilon_m$ and the corresponding event angles $\psi_m$ are defined by,
\begin{equation}
{\cal E}_{m} = \varepsilon_{m} e^{-im\psi_{m}} = \frac{\int_{r,\phi} \epsilon(r,\phi) r^m e^{-im\phi}}{\int_{r,\phi} \epsilon(r,\phi) r^m} = \frac{\int_0^\infty dr \, r^{m+1} \, \epsilon_m(r)}{\int_0^\infty dr \, r^{m+1} \, \epsilon_0(r)}.
\end{equation}
Using the expansion \eqref{eq:expansionepsilonmBessel} and the scalar product \eqref{eq:scalarproductr}, one can write this as
\begin{equation}
{\cal E}_{m} = \frac{\sum_l \epsilon_{m,l} \left(r^m, J_m{\big (}z^{(m)}_l \rho(r){\big )}\right)}{\sum_l \epsilon_{0,l} \left(r^m, J_0{\big (}z^{(0)}_l \rho(r){\big )} \right)}.
\end{equation}
The denominator is dominated by the $l=1$ term which corresponds to the background density and one can assume $\epsilon_{0,1}=1$. To linear order in deviations from the background one obtains thus
\begin{equation}
{\cal E}_{m} = \sum_l \epsilon_{m,l} \frac{\left(r^m, J_m{\big (}z^{(m)}_l \rho(r){\big )}\right)}{\left(r^m, 1 \right)}.
\label{eq:relationEccentricitiesExpansionCoefficients}
\end{equation}
The complex event eccentricity ${\cal E}_{m}$ corresponds to a linear superposition of the expansion coefficients $\epsilon_{m,l}$. Numerically, one finds that the coefficient of the term with $l=1$ is largest (the corresponding basis function is the only one without nodes) and to good approximation one has in fact ${\cal E}_{m} \approx \epsilon_{m,1}\,d_m$ where the coefficient $d_m$ is independent of the event and corresponds to the ratio of scalar products appearing for the $l=1$ term in \eqref{eq:relationEccentricitiesExpansionCoefficients}. In principle, one could devise an alternative expansion scheme, where the event eccentricity appears as the lowest order coefficient and all higher order terms correspond to orthogonal functions without any contribution to eccentricity. This would lead to a basis of polynomials instead of Bessel functions, but we do not explore this further here.

\section{Relativistic fluid dynamics}
\label{sec:RelativisticFluidDynamics}

\subsection{Equations of motion}

The equation of motion of relativistic fluids dynamics follow from energy and momentum conservation and other conservation laws (like e.\ g.\ net baryon number conservation related to a global U(1) symmetry of QCD), supplemented with additional constitutive relations. We discuss here a relativistic fluid without (or negligible) net baryon number or any other conserved charge. The relevant conservation law is then the one for energy and momentum,
\begin{equation}
\nabla_\mu T^{\mu}_{\;\;\nu} =0.
\end{equation}
It is useful to decompose the energy-momentum tensor as
\begin{equation}
T^{\mu}_{\;\;\nu} = \epsilon u^\mu u_\nu + (p+\pi_\text{bulk}) \Delta^{\mu}_{\;\;\nu} + \pi^{\mu}_{\;\;\nu}.
\end{equation}
The fluid velocity $u^\mu$ is defined in the so-called Landau frame as the time-like eigenvector of $T^\mu_{\;\;\nu}$ and the energy density $\epsilon$ is the corresponding eigenvalue. The fluid velocity is normalized to  $u^\mu u_\mu=-1$. We also use the projector $\Delta^\mu_{\;\;\nu} =u^\mu u_\nu +  \delta^\mu_\nu$ orthogonal to the fluid velocity $u^\mu$. The pressure $p$ and the energy density $\epsilon$ are related through the thermodynamic equation of state as in equilibrium, $p=p(\epsilon)$, while the bulk viscous pressure $\pi_{\text{bulk}}$ measures the deviation of the isotropic pressure from this. The  symmetric shear stress tensor $\pi^{\mu\nu}$ is traceless, $\pi^{\mu}_{\;\;\mu}=0$, and orthogonal to the fluid velocity $u_\mu \pi^{\mu}_{\;\;\nu}=0$.

From energy-momentum conservation one obtains evolution equations for energy density and fluid velocity,
\begin{equation}
\begin{split}
\label{eq:energy_momentum_conservation}
u^\mu \partial_\mu \epsilon + \left( \epsilon + p +\pi_\text{bulk} \right) \nabla_\mu u^\mu + \pi^{\mu}_{\;\;\nu} \nabla_\mu u^\nu = & 0, \\
(\epsilon + p + \pi_\text{bulk}) u^\nu \nabla_\nu u^\mu + \Delta^{\mu\nu} \partial_\nu (p+\pi_\text{bulk}) + \Delta^{\mu\nu} \nabla_\rho \pi^{\rho}_{\;\;\nu} = & 0.
\end{split}
\end{equation}
In this form, the system of evolution equations is not closed, but needs to be supplemented by additional constitutive relations for the stress tensor $\pi^{\mu}_{\;\;\nu}$ and the bulk viscous pressure $\pi_\text{bulk}$. These could be provided by constraint equations or by additional evolution laws. 

The relativistic generalization of the Navier-Stokes equation \cite{nla.cat-vn725379, PhysRev.58.919} follows the former principle; $\pi^{\mu\nu}$ as well as $\pi_\text{bulk}$ can be expressed there in terms of gradients of fluid velocity. However this approximation has been shown to violate the relativistic causality principle and to be linearly unstable \cite{PhysRevD.31.725, Hiscock:1983zz}. 

Another possibility is to provide the constitutive relation as dynamical equations for the shear stress tensor $\pi^{\mu}_{\;\;\nu}$ and bulk viscous pressure $\pi_\text{bulk}$; this idea was first introduced by M\"{u}ller as well as Israel and Stewart \cite{Israel:1979wp,Muller:1967zza}. 
Oftentimes such equations are organised in terms of  Knudsen and Reynolds numbers. 
The Knudsen number Kn is the ratio between a microscopic scale like the mean free path and a macroscopic one, like the size over which  the macroscopic fields change effectively. The Reynolds number is the ratio of the macroscopic length to the scale where perturbation are damped by the viscosity. 

The equation of motion up to second second order in Knudsen number Kn and inverse Reynolds number Re$^{-1}$ have been obtained in ref.\ \cite{Denicol:2012cn}. Here we include terms of order ${\cal O}(\text{Re}^{-2})$ and ${\cal O}(\text{Kn } \text{Re}^{-1})$ but drop terms of order ${\cal O}(\text{Kn}^2)$ because they are not compatible with a hyperbolic structure and relativistic causality\cite{Muronga:2003ta,Baier:2007ix,Romatschke:2009kr,Denicol:2012cn, Floerchinger:2017cii}. The evolution equation for shear stress in then
\begin{equation}
\label{eq:shear_tensor_conservation_DNMR}
\begin{split}
 P^{\mu\;\,\rho}_{\;\;\nu\;\,\sigma} \left[
 \tau_\text{shear} \left( u^\lambda \nabla_\lambda \pi^{\sigma}_{\;\;\rho} - 2  \pi^{\sigma\lambda}\omega_{\rho\lambda} \right) 
 + 2 \eta \nabla_\rho u^\sigma  
 -\varphi_7\,  \pi^\lambda_{\;\;\rho} \pi^{\sigma}_{\;\;\lambda} 
+\tau_{\pi \pi }\, \pi^{\sigma}_{\;\;\lambda} \sigma^{ \lambda}_{\;\;\rho} -\lambda_{\pi \Pi }\, \pi_{\text{bulk}} \nabla_\rho u^\sigma 
 \right] 
\\
+ \pi^{\mu}_{\;\;\nu} \left[ 1+ \delta_{\pi\pi} \nabla_\rho u^\rho
-\varphi_6\, \pi_{\text{bulk}} \right]= 0.
\end{split}
\end{equation}
The projector to the symmetric, transverse and trace-less part of a tensor is defined here by
\begin{equation}
P^{\mu\nu}_{\;\;\;\rho\sigma}=
\frac{1}{2}\Delta^{\mu}_{\;\;\rho}\Delta^\nu_{\;\;\sigma}+\frac{1}{2}\Delta^{\mu}_{\;\;\sigma}\Delta^\nu_{\;\;\rho}-\frac13 \Delta^{\mu\nu}\Delta_{\rho\sigma}.
\end{equation}
We also use the following abbreviations for symmetric and anti-symmetric combinations of fluid velocity gradients,
\begin{equation}
\sigma_{\mu\nu} = P_{\mu\nu}^{\;\;\;\rho\sigma} \nabla_\rho u_\sigma, \quad\quad\quad \omega_{\mu\nu} = \frac{1}{2}\left( \nabla_\mu u_\nu - \nabla_\nu u_\mu \right)= \frac{1}{2}\left( \partial_\mu u_\nu - \partial_\nu u_\mu \right).
\end{equation}
Similarly, the evolution equation for $\pi_{\mathrm{bulk}}$ is given by 
\begin{equation}
\label{eq:bulk_pressure_conservation_DNMR}
\tau_\text{bulk} \, u^\mu \partial_\mu \, \pi_\text{bulk} + \pi_\text{bulk} + \zeta \nabla_\mu u^\mu  +\delta_{\Pi\Pi} \pi_\text{bulk}\nabla_\mu u^\mu-\varphi_1 \pi_\text{bulk}^2-\lambda_{\Pi \pi}\pi^{\mu\nu} \nabla_\mu u_\nu-\varphi_3 \pi^{\mu}_{\;\;\nu}\pi^{\nu}_{\;\;\mu}= 0.
\end{equation} 
Among the various transport coefficients introduced in \eqref{eq:shear_tensor_conservation_DNMR} and \eqref{eq:bulk_pressure_conservation_DNMR}, the most important ones are 
the shear viscosity $\eta$ and the bulk viscosity $\zeta$ which also appear in the Navier-Stokes approximation, and the relaxation times $\tau_\text{shear}$ and $\tau_\text{bulk}$. The latter determine how fast the shear stress tensor and the bulk viscous pressure relax towards their asymptotic values $\pi^{\mu}_{\;\;\nu}=-2\eta \sigma^{\mu}_{\;\;\nu}$ and $\pi_\text{bulk} = - \zeta \nabla_\rho u^\rho$, respectively. The additional second order transport coefficient  $\tau_{\pi\pi}$, $\delta_{\pi\pi}$, $\lambda_{\pi\Pi}$, $\delta_{\Pi\Pi}$ and $\lambda_{\Pi\pi}$ that are of order $\mathcal{O}(\text{Kn}\;\text{Re}^{-1})$ and 
$\varphi_7$, $\varphi_6$, $\varphi_1$ and $\varphi_3$ of order $\mathcal{O}(\text{Re}^{-2})$, can be understood as non-linear modifications to the relaxation type equation \cite{Molnar:2013lta}.

The equations \eqref{eq:shear_tensor_conservation_DNMR} and \eqref{eq:bulk_pressure_conservation_DNMR} together with \eqref{eq:energy_momentum_conservation} form now a closed system of first order, quasi-linear partial differential equations for the energy density (or any other independent thermodynamical field such as enthalpy density or temperature), the independent components of fluid velocity and shear stress, and for bulk viscous pressure. It was shown that these equation are actually hyperbolic \cite{Floerchinger:2017cii} and in particular they can be cast into the form \eqref{eq_eompertnotyettruncated} introduced previously. This is the set of equations we will work with in the present paper, but note here that much of our formalism can also be used when the set of equations is extended, as long as such extensions again lead to quasi-linear, hyperbolic evolution equations.

In practice it is of course convenient to choose an explicit parametrization of the fluid fields. In the following we will parametrize the thermodynamic fields either by temperature $T$ or by enthalpy density $w=\epsilon+p$. The former is a free parameter for many microscopic calculations of thermodynamic and transport properties in the grand canonical ensemble and is therefore particularly convenient for the background evolution. Enthalpy density is convenient for the parametrization of the linear perturbations because it makes explicit that the thermodynamic equation of state enters mainly in terms of the velocity of sound. The fluid velocity is conveniently parametrized in term of the spatial components $u^r$, $u^\phi$ and $u^\eta$ with the temporal component $u^\tau$ related to this by the normalization $u^\mu u_\mu = -1$.  The shear stress has five independent components which may be chosen as $\pi^{\phi}{}_{\phi}$, $\pi^{r}{}_{\phi}$, $\pi^{r}{}_{\eta}$, $\pi^{\eta}{}_{\eta}$ and $\pi^{\phi}{}_{\eta}$. Finally, we have the bulk viscous pressure $\pi_\text{bulk}$.

\section{Thermodynamic equation of state and transport properties}
\label{eq:Thermodynamics}

\subsection{Thermodynamic equation of state}
Relativistic fluid dynamics depends on the thermodynamic equation of state. At vanishing baryon and electric charge chemical potentials, all thermodynamic information can be derived in the grand canonical ensemble from the pressure as a function of temperature $p(T)$. In the regime of the quark-gluon plasma, the equation of state is now rather well known from lattice QCD calculations  \cite{Borsanyi:2016ksw, Bazavov:2014pvz}. At temperatures below the crossover transition to a fluid dominated by hadronic degrees of freedom, one can use a hadron resonance gas approximation. 

For our purpose of solving the fluid evolution equations in a background-fluctuation splitting approach, and for the numerical treatment with the pseudo-spectral method, it is particularly important to have a regular enough equation of state with a continuous sound velocity. Because the numerical solution extends also into the hadronic phase, it is important that thermal properties are physical also at low temperatures even though such regions are already outside the freeze-out surface and should therefore not affect experimental observables. For a numerical treatment it is particularly convenient to have a parametrization of the equation of state in an analytic form. For fluid dynamic perturbation theory and the mode expansion, we need also derivatives of $p(T)$, for example to calculate the velocity of sound. However, parametrizations of the equation of state available in the literature are formulated for the trace anomaly $\epsilon-3p$  (which makes a numerical integration necessary) \cite{Borsanyi:2016ksw} or become unphysical at low temperatures because of a pole in the Pad\'e approximation \cite{Bazavov:2014pvz}.

For this reason, we have developed a parametrization of the available numerical results from lattice QCD calculations, that fulfills our requirements. At temperatures above $T_\text{crit} \approx 154$ MeV, our fit reproduced the results published in ref.\ \cite{Borsanyi:2016ksw}, while at smaller temperatures it reproduces a hadron resonance gas approximation following ref.\ \cite{huovinen2010qcd} based on vacuum masses (up to $2$ GeV) and vanishing chemical potentials reasonably well. 

The parametrization of pressure as a function of temperature is taken as the following combination of exponential and rational functions,
\begin{equation}
\label{eq:eosparamentrization}
\dfrac{p(T)}{T^4}=\exp\left[-\dfrac{c^2}{(T/T_c)}-\dfrac{d^2}{(T/T_c)^2}\right] \left[\dfrac{
\dfrac{(16+\frac{21}{2}N_f)\pi^2}{90}+a_1 \left(\dfrac{T_c}{T}\right)+a_2 \left(\dfrac{T_c}{T}\right)^2+a_3 \left(\dfrac{T_c}{T}\right)^3+a_4 \left(\dfrac{T_c}{T}\right)^4
}{
1+b_1 \left(\dfrac{T_c}{T}\right)+b_2 \left(\dfrac{T_c}{T}\right)^2+b_3 \left(\dfrac{T_c}{T}\right)^3+b_4 \left(\dfrac{T_c}{T}\right)^4
}\right].
\end{equation}
 Note that for asymptotically large temperatures $p(T)$ approaches the result for free gluons and $N_f$ free quarks. Below we take $N_f=3$ and $T_c=154 $ MeV. The best fit results for the fit parameter $a_j$, $b_j$, $c$ and $d$ are reported in table \ref{tab:fitparamenter}.
\begin{table}[h]
	\centering
	\begin{tabular}{|cc|cc|cc|cc|cc|}
		\hline
		$a_1 $& -0.752335&$a_2$ &-1.8151 & $a_3$& -2.83317 &$a_4$& 4.20517 & $c$& 0.547521\\
		\hline
		$b_1$ & -1.68716 &$b_2$& 7.83336 &$b_3$& -13.3421 &$b_4$ &9.22752 & $d$& 0.0148163 \\
		\hline
	\end{tabular}
	\caption{Best fit parameter for the thermodynamic equation of state as parametrized in equation \eqref{eq:eosparamentrization}.}
	\label{tab:fitparamenter}
\end{table}
\begin{figure}[t]
	\centering
		\includegraphics[height=0.32\linewidth]{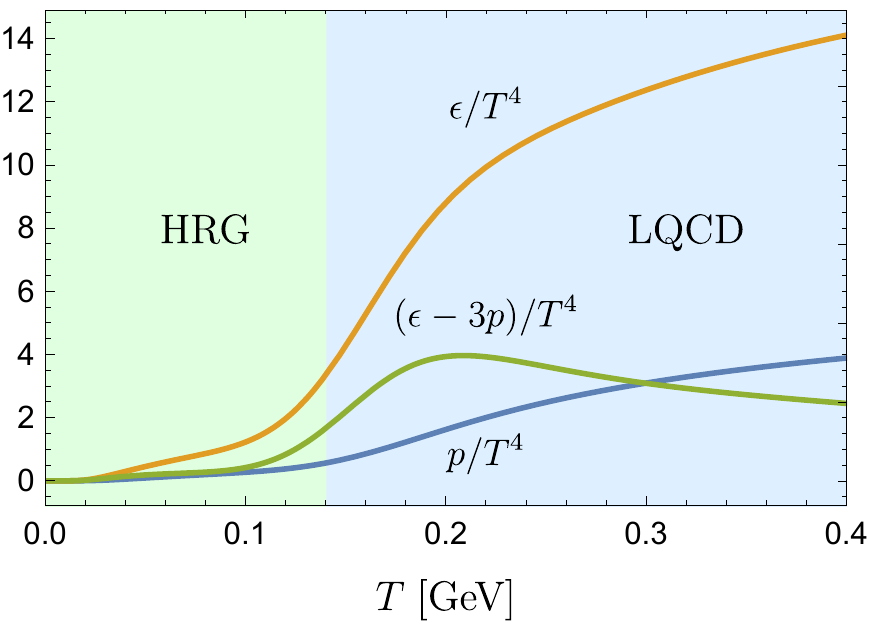} \hfill
		\includegraphics[height=0.32\linewidth]{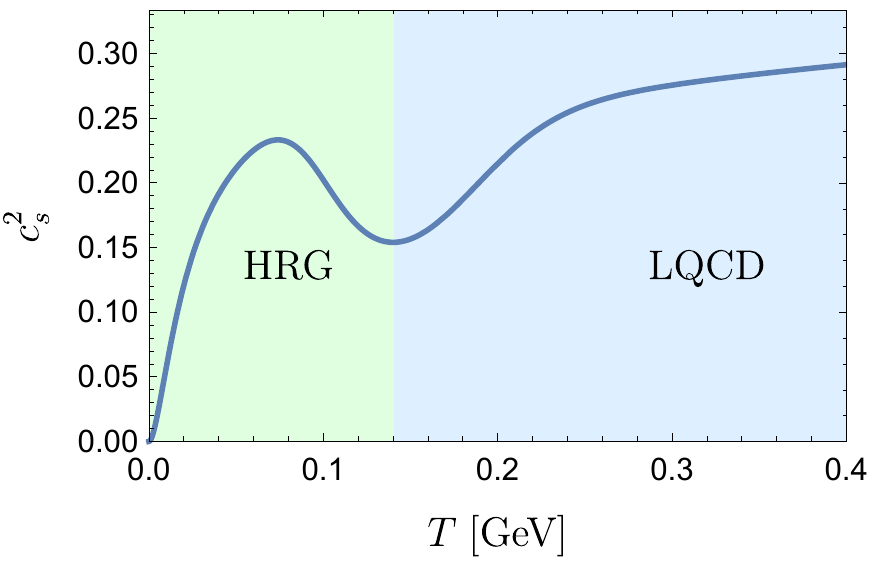}
\caption{The thermodynamic equation of state $p(T)$ as parametrized in equation \eqref{eq:eosparamentrization}. We show energy density $\epsilon$, pressure $p$ and the trace anomaly $\epsilon-3$ in units of $T^4$ in the left panel and the squared sound velocity $c_s^2(T)$ in the right panel. Lattice QCD data underlying the fit at high temperatures are taken from ref.\ \cite{Borsanyi:2016ksw} and ref.\ \cite{Bazavov:2014pvz}, the hadron resonance gas approximation used at low temperatures was calculated following ref.\ \cite{huovinen2010qcd}. In the transition region both results were smoothly connected.}
\label{epcs}
\end{figure}
The exponential terms in the prefactor in eq.\ \eqref{eq:eosparamentrization} help in particular to reproduce the hadron resonance gas regime while the rational term parametrizes the crossover to a quark-guon plasma.

In the left panel of fig.\ \ref{epcs} we show the resulting energy density $\epsilon$, pressure $p$ and trace anomaly $\epsilon-3p$ in units of $T^4$ as a function of temperature. The right panel shows the square of the thermodynamic velocity of sound $c_s^2$ as a function of temperature. The latter is particularly important for the fluid dynamic evolution and determines for example the characteristic velocities in the absence of dissipative stresses. 

\begin{figure}[t]
	\centering
		\includegraphics[height=0.32\linewidth]{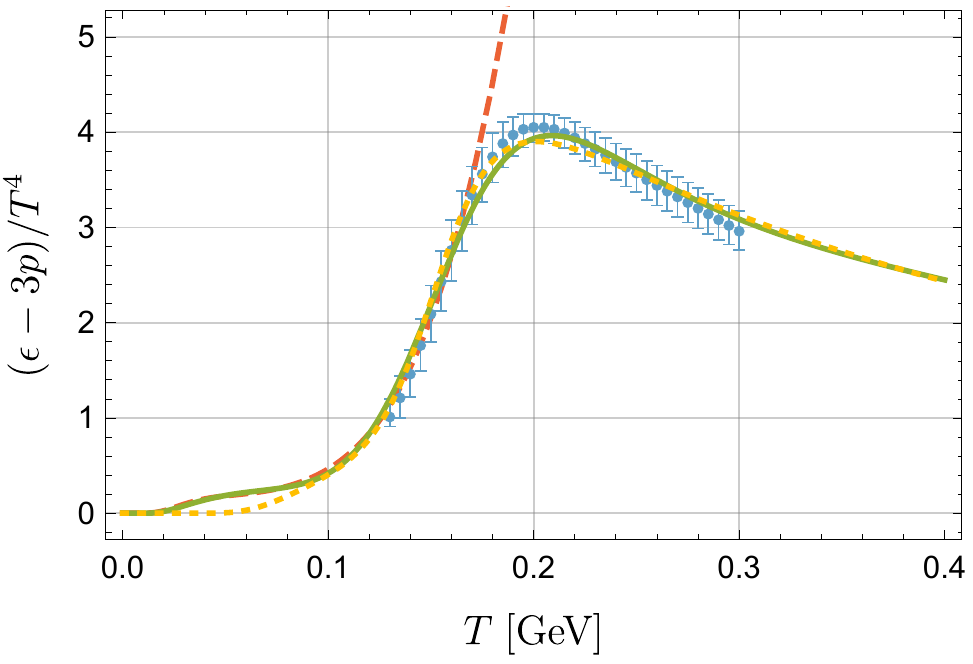} 
\caption{The trace anomaly $(\epsilon-3p)/T^4$ as a function of temperature $T$. We show our parametrization in eq.\ \eqref{eq:eosparamentrization} (solid line) together with numerical data from the HotQCD collaboration \cite{Bazavov:2014pvz} (for $2+1$ quark flavors, symbols with error bars), a parametrization of lattice QCD data from ref.\ \cite{Borsanyi:2016ksw} (for $2+1+1$ quark flavors, dotted line) and results of a hadron resonance gas approximation (dashed line). As becomes apparent, our parametrization interpolates continuously between known results in the different regimes.}
\label{comparison}
\end{figure}

To develop the fit \eqref{eq:eosparamentrization} we have considered the trace anomaly $\epsilon-3p$. In fig.\ \ref{comparison} we show our fit (solid curve), together with available numerical data from the HotQCD collaboration \cite{Bazavov:2014pvz} (for $2+1 $ quark flavors, symbols with error bars), an analytic parametrization of lattice QCD data from ref.\ \cite{Borsanyi:2016ksw} (for $2+1+1$ flavors, dotted curve) and the hadron resonance gas approximation (dashed line). As becomes apparent, our parametrization captures both the low temperature hadron resonance regime and the high temperature lattice QCD results to reasonable accuracy.

\subsection{Transport properties}
\label{sec_transportcoeff}

In addition to the thermodynamic equation of state, relativistic fluid dynamics needs also transport properties such as shear and bulk viscosity, corresponding relaxation times and other second order coefficients. Unfortunately, transport properties are not yet understood from first principles as  good as the thermodynamic equilibrium properties \cite{KARSCH2008217}. For a relatively recent computation of shear viscosity using SU(3) Yang Mills theory as an approximation to QCD as well as further references, see ref \cite{PhysRevLett.115.112002}. Recently a comparison of theoretical calculations to experimental data was done on this basis  \cite{Dubla:2018czx}.

To make progress, we make rather simplifying assumptions for the present paper. The ratio of shear viscosity to entropy density $\eta/s$ is taken (for now) as a constant value between $0.08$ and $0.3$ (if no value is specified we take $\eta/s=0.2$), relatively close to the KSS bound \cite{Kovtun:2004de,policastro2001shear}. For the shear stress relaxation time we assume $\tau_\text{shear}=5 \eta/(sT)$. We also set $\delta_{\pi\pi}=4 \tau_\text{shear}/3$.  Following ref.\ \cite{Moreland:2018gsh}, we parametrize the ratio of bulk viscosity to entropy density as a Lorentz curve,
\begin{align}
\frac{\zeta}{s}=\frac{0.032}{1+\left(\frac{T-0.175\,\mathrm{GeV}}{0.024\,\mathrm{GeV}}\right)^2},
\label{eq_bulkbass}
\end{align}
and take for the corresponding relaxation time
\begin{align}
\tau_{\mathrm{bulk}}=\frac{\zeta}{sT}\frac{1}{15\left(\frac{1}{3}-c_s^2\right)}.
\label{eq:bulkrelaxationTime}
\end{align}
All other second order transport coefficients are neglected here but can be added in future applications of our formalism without additional effort. Table \ref{tab_parameters} summarizes the transport properties and other parameters used for the numerical evolution.




\section{Numerical Methods}
\label{sec_numerics}

Equations \eqref{eq_eombg} for the time evolution of the background configuration constitutes a set of non-linear (but quasi-linear), hyperbolic partial differential equations. In a similar way, \eqref{eq_eomPerturbationFourier} for perturbations around this background constitutes a set of linear, hyperbolic differential equations. Solutions to such equations can in general not be found in closed form and numerical methods need to be employed. In the present section we discuss the numerical scheme we have developed for this purpose. The main idea is to discretize the radial coordinate $r$ and to solve the resulting ordinary differential equations by standard methods. For the spatial discretization we have developed a pseudo-spectral scheme that allows to reach rather high numerical accuracy with comparatively little computational effort \cite{Tadmor,Gottlieb1985,GOTTLIEB200183,Boyd2000,hesthaven_gottlieb_gottlieb_2007}.  Particular challenges are posed by the inner boundary $r\to 0$ and the open boundary conditions for large radii $r\to \infty$. For comparison and for benchmarking, we will also employ a more standard finite difference scheme. Further numerical schemes exist, of course, but will not be discussed here. A popular approach is for example the higher order finite volume method. An overview over different approaches can be found in ref.\ \cite{rezzolla2013relativistic}.

For the specific situation of a fluid with conformal symmetry, an analytic solution for the radial expansion has been put forward by Gubser \cite{gubser2010symmetry}. It will be convenient to use this solution to benchmark our numerical solution of the background equations. The numerical methods used to solve the linearized equations for the perturbations can in turn be benchmarked against the background solution for an azimuthally symmetric situation, as we discuss below.

In the following we will first briefly discuss how we solve the system of ordinary differential equations (ODE) resulting from a specific discretization of the radial direction. 

\subsection{The method of lines}
\label{sec_adams}
A common procedure to solve partial differential equations (PDE) is the {\it method of lines}. The spatial directions are discretized, but time (in our case Bjorken time $\tau$) remains, at least in a first step, as a continuous evolution parameter. Let us first write the system of equations \eqref{eq_eombg} in the form
\begin{align}
\partial_\tau \mathbf{\Phi}(\tau,r)+(\mathbf{A}^{-1}\mathbf{B})(\mathbf{\Phi}(\tau,r),\tau,r)\cdot \partial_r\mathbf{\Phi}(\tau,r)-(\mathbf{A}^{-1}\mathbf{S})(\mathbf{\Phi}(\tau,r),\tau,r)=0.
\label{eq_eombgAInverted}
\end{align}
We assumed here the the coefficient matrix $\mathbf{A}$ can be inverted and we have dropped the index $0$ denoting background fields. In the next step we assume {\it some} discretization of the radial direction $r\to r_j$, leading with $\mathbf{\Phi}_j(\tau) = \mathbf{\Phi}(\tau, r_j)$ to
\begin{align}
\partial_\tau \mathbf{\Phi}_j(\tau)+(\mathbf{A}^{-1}\mathbf{B})(\mathbf{\Phi}_j(\tau),\tau,r)\cdot \sum_k D_{jk} \mathbf{\Phi}_k(\tau)-(\mathbf{A}^{-1}\mathbf{S})(\mathbf{\Phi}_j(\tau),\tau,r)=0.
\label{eq_eombgDiscretized}
\end{align}
The matrix $D_{jk}$ represents the radial derivative acting as a linear operation on the variables $\Phi_k$. Its explicit form depends on the discretization scheme. As it stands, \eqref{eq_eombgDiscretized} is now a set of ordinary differential equations that can be solved by standard numerical methods, such as e.\ g.\ Adams method.

\subsection{Pseudospectral discretization method}
\label{sec:Pseudospectraldiscretization}

Partial differential equations can be solved via spectral methods. At a given instant of time, the solution is approximated as a linear superposition of certain basis functions. The latter have typically support on the entire spatial domain. Spatial derivatives are represented as linear operations on the coefficients of this expansion. In some sense, spectral methods provide a global approach to the solution which is in contrast to finite difference schemes at low order where derivatives are represented by rather local, sparse matrices. 

One advantage of spectral methods is the fast convergence with the number of basis functions. For well posed problems, they are less CPU expensive than e.\ g.\ finite difference schemes and can reach higher accuracy. A disadvantage of standard implementations using continuous basis functions is that they struggle with possible discontinuities in the solution and the reconstruction of shock waves. The most common spectral method is probably the Fourier spectral method, applicable for periodic boundary conditions.

Quite generally, a function $u(r)$ on a finite domain can be approximated by a set of basis functions $B_j(r)$ as
\begin{align}
u_N(r)=\sum_{j=0}^{N}c_j B_j(r).
\label{2real}
\end{align}
For example, for periodic boundary conditions on the domain $r \in (0,2\pi)$, the basis could be chosen as $B_j(r) = e^{i j r}$. In our case, since the equations \eqref{eq_eombgAInverted} are solved in cylindrical coordinates, we need to specify boundary conditions at the coordinate origin $r=0$ and for large radii $r\to \infty$. As was discussed in section \ref{sec:ScalarVectorTensorModesParity}, one can formally continue $r$ to negative values and use the $\mathscr{R}$ parity $(r,\phi) \to (- r,\phi+\pi)$ to find appropriate boundary conditions for the different fields at $r=0$. For large radii $r\to \infty$, dimension-full physical fields such as energy density must go to zero, which is an example for a {\it behavioral boundary condition} \cite{Boyd2000}.

\begin{figure}[t]
\centering
	\includegraphics[width=0.39\linewidth]{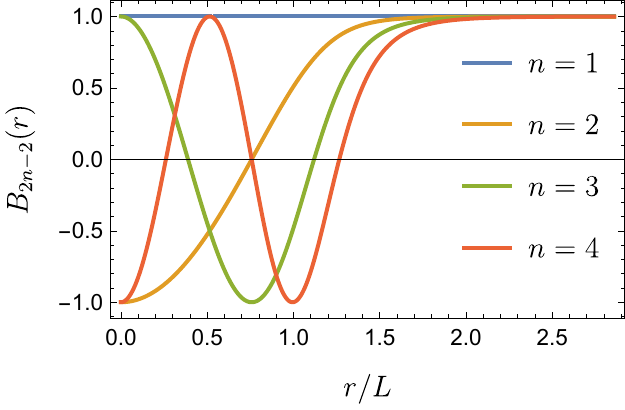} \hspace*{0.6cm}
	\includegraphics[width=0.39\linewidth]{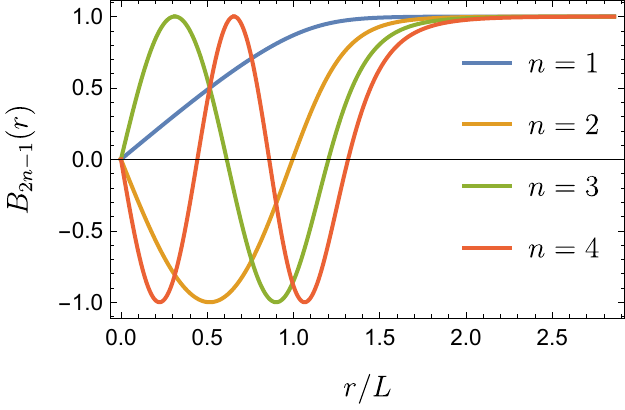}
\caption{Basis functions as defined in \eqref{eq:polynomial_basis} with even $\mathscr{R}$ parity $B_{2n-2}(r)$ (left panel) and odd $\mathscr{R}$ parity $B_{2n-1}(r)$ (right panel) as a function of $r/L$. We use here the exponent $\alpha=1/10$.}
\label{chebs}
\end{figure}

A convenient set of basis functions for our purpose are the Chebyshev polynomials of the first kind $T_n(x)$, defined via the following trigonometric representation,\footnote{Another possible choice would be the one sided Jacobi polynomials (see \cite{Boyd2000} page 385).} 
\begin{equation}
T_{n}(x) =\text{cos} (n\; \text{arccos}(x)),\quad \quad\quad x\;\in (-1,1).
\end{equation}
To map the semi-infinite domain of radii $r\in(0,\infty)$ to a finite interval $x\in(0,1)$, we shall use the map
\begin{equation}
r=\frac{L x }{(1-x^2)^\frac{1}{\alpha}}=\frac{L \cos \theta  }{\sin^\frac{2}{\alpha}(\theta)},
\label{rofth}
\end{equation}
with some characteristic length $L$ and an exponent $\alpha>0$. In the second equation we have substituted $x=\cos(\theta)$, with $\theta\in(0,\pi)$.
With the help of the relation in \eqref{rofth} we define the desired basis functions as
\begin{equation}
\label{eq:polynomial_basis}
B_{n}(r)=T_n(x(r))=\cos (n \theta(r)).
\end{equation}
In fig.\ \ref{chebs} we show a few of these basis functions. Note that the functions $B_{2n-1}$ are odd with respect to the $\mathscr{R}$ parity discussed in section \ref{sec:ScalarVectorTensorModesParity}, while the even functions $B_{2n}$ are even. In summary,
\begin{equation}
B_{n}(r)=(\pm 1)^n B_n(-r ).
\end{equation}  
For an expansion of functions with definite $\mathscr{R}$ parity it is natural to use the representations
\begin{equation}
\label{functional_expansion}
f^e(r)=\sum_{n=1}^\infty  a_n B_{2n-2}(r),\quad\quad\quad
f^o(r)=\sum_{n=1}^\infty b_n B_{2n-1}(r).
\end{equation}

This set of polynomials inherits an orthogonality relation form Chebyshev polynomials, and consequently from Fourier modes, due to the transformation chain  $r=r(x)=r(\theta)$ defined in \eqref{rofth}. It reads  
\begin{equation}
\int_{0}^{\pi} d\theta \, \cos(n\theta)\cos(m\theta) =\int_{-1}^{1} \frac{dx}{\sqrt{1-x^2}} \, T_n(x)T_m(x)=\int_{-\infty}^{\infty} d r \, w(r) \, B_n(r)B_m(r) =\frac{\pi c_n}{2}\delta_{mn},
\end{equation}
where $c_n=1$ for $n\neq 0$ and $c_0=2$. In the last relation we also used the weight function
\begin{equation}
w(r)= \frac{|x'(r)|}{\sqrt{1-x^2(r)}}=|\theta'(r)|.
\end{equation}
It is now natural to define a scalar product in the $L^2$ functional space spanned by the functions $B_n$ with a corresponding induced norm, 
\begin{equation}
\label{inner_norm}
(f,g)_{L_w^2[\mathbb{R}]}=\int_{-\infty}^{\infty} dr \, w(r) \, f(r) g(r),\quad\quad\quad\quad \norm{f}_{L_w^2[\mathbb{R}]}^2=(f,f)_{L_w^2[\mathbb{R}]}.
\end{equation}
 Within the definition \eqref{inner_norm} the coefficients of the expansion \eqref{functional_expansion} can be computed,
 \begin{equation}
 \label{coefficient_scalar_product}
 a_n=\frac{2}{\pi c_{2n-2}} (f^e,B_{2n-2})_{L_w^2[\mathbb{R}]},\quad\quad\quad \text{and}\quad\quad\quad\quad b_n=\frac{2}{\pi c_{2n-1}} (f^o,B_{2n-1})_{L_w^2[\mathbb{R}]}.
\end{equation}  

While the expansion in \eqref{functional_expansion} is not yet an approximation (for sufficiently regular functions), it becomes a useful scheme for numerical calculations if the summations are truncated at some finite order $N$. We write
\begin{equation}
\label{truncated_functional_expansion}
f^e(r)=\sum_{n=1}^N a_n B_{2n-2}(r),\quad\quad\quad
f^o(r)=\sum_{n=1}^N b_n B_{2n-1}(r).
\end{equation}
The spectral expansion coefficients should be obtainable for given functions $f^o$ or $f^e$ from an approximate version of \eqref{coefficient_scalar_product}. In practice this means that the integral for the scalar product in \eqref{inner_norm} must be approximated somehow by a finite sum. The classical solution of this last problem is to find an appropriate quadrature rule for the integrals. 

Using the parity properties and the transformation rules we can write \eqref{coefficient_scalar_product} as
\begin{equation} 
\begin{split}
a_n&=\frac{4}{\pi c_{2n-2}} \int_{0}^{\infty} dr \, w(r) \, f^e(r)B_{2n-2}(r) = \frac{4}{\pi c_{2n-2}} \int^{\frac{\pi}{2}}_{0} d\theta \, f^e(r(\theta))\cos\left[(2n-2) \theta \right],\\
 b_n&=\frac{4}{\pi c_{2n-1}} \int_{0}^{\infty} dr \, w(r) f^o(r)B_{2n-1}(r) = \frac{4}{\pi c_{2n-1}} \int^{\frac{\pi}{2}}_{0} d \theta \, f^o(r(\theta))\cos\left[(2n-1) \theta \right].
\end{split}
\end{equation}  
An approximation to these integrals can be found using the midpoint rule in the interval $[0,\pi/2]$ with $N$ nodes and constant weight in $\theta$-space, or otherwise using the corresponding nodes mapped with the transformation \eqref{rofth} and the weights modified correspondingly. More concrete, the evaluation points in $\theta$ are
\begin{equation}
	\theta_J= \frac{\left(J -\frac{1}{2} \right)\pi}{2N },\quad\quad\quad  J=1,\ldots, N.
	\label{thetaI}
\end{equation}
In terms of radius $r$ these correspond to
\begin{equation}
r_J=\frac{L \cos\left(  \frac{\left(J -\frac{1}{2} \right)\pi}{2N } \right)}{\sin^\frac{2}{\alpha}\left( \frac{\left(J -\frac{1}{2} \right)\pi}{2N } \right)} ,\quad\quad\quad J=1,\cdots N .
\label{eq:rIdiscretizationpoints}
\end{equation}
The resulting approximated integrals are
\begin{equation} 
\begin{split}
a_n&= \frac{2}{N c_{2n-2}}  \sum_{J=1}^N f^e(r_J)\cos\left[\left(n-1\right)\left(J -\frac{1}{2} \right)\frac{\pi}{N } \right],\\
b_n&= \frac{2}{N c_{2n-1}}  \sum_{J=1}^N f^o(r_J)\cos\left[\left(n-\frac{1}{2}\right)\left(J -\frac{1}{2} \right)\frac{\pi}{N } \right].
\end{split}\label{eq:coefficientsanbn}
\end{equation} 
In general, the midpoint rule or rectangular rule approximation to the integral of a generic function can be a rather crude approximation, but for periodic function it yields exponentially accurate results. Because we are using a map to a Fourier basis, this approximation allows us to reach rather high accuracy in the determination of the coefficients $a_n$ and $b_n$.    

Equation \eqref{eq:coefficientsanbn} makes an interesting relation between the basis expansion \eqref{eq:polynomial_basis} and Fourier transformations explicit. If the discretization points $r_J$ are chosen according to \eqref{eq:rIdiscretizationpoints}, and with the midpoint quadrature rule, the coefficients $a_n$ and $b_n$ are related to the functional values $f^e(r_J)$ and $f^o(r_J)$ via a discrete Fourier transform. On the other side, one has the inverse relation
\begin{equation}
\begin{split}
f^e(r_J)&=\sum_{n=1}^{N} a_n \cos\left[\left(n-1 \right)\left(J-\frac{1}{2} \right)\frac{\pi}{N}\right],\\
f^o(r_J)&=\sum_{n=1}^{N} b_n \cos\left[\left(n-\frac{1}{2} \right)\left(J-\frac{1}{2} \right)\frac{\pi}{N}\right].
\end{split}
\label{eq:backwarddiscreteFourier}
\end{equation}
It is of course very convenient to have such relations in terms of discrete Fourier transforms, also because fast algorithms exist to implement them. In appendix \ref{app:ConventionFourier} we express the relations in a standardized form that is directly suitable for an algorithmic implementation.

A major task for the formalism developed above is to yield an expression for spatial derivatives. Formally, radial derivatives of even functions in the expansion \eqref{functional_expansion} can be computed as follows,
\begin{equation}
\label{eq:derivativeeven}
\begin{split}
\frac{\partial}{\partial r } f^e (r) &=\sum_{n=1}^\infty a_n \frac{\partial}{\partial r } B_{2n-2}(r)= \sum_{n=1}^\infty a_n \frac{\partial}{\partial r } \cos \left[ (2n-2) \theta (r ) \right]\\
&=-\frac{\partial \theta}{\partial r }
\sum_{n=1}^\infty a_n (2n-2)   \sin \left[ (2n-2) \theta (r ) \right].
\end{split}
\end{equation}
Similarly, for odd functions one obtains
\begin{equation}
\label{eq:derivativeodd}
\frac{\partial}{\partial r } f^o (r) = -\frac{\partial \theta}{\partial r }
\sum_{n=1}^\infty b_n (2n-1)   \sin \left[ (2n-1) \theta (r ) \right].
\end{equation}
If one truncates the summations to $n=1,\ldots, N$ and specializes to the evaluation points \eqref{eq:rIdiscretizationpoints}, one obtains
\begin{equation}
\label{eq:derivativeevendiscrete}
\begin{split}
\frac{\partial}{\partial r } f^e (r_J) = & -\frac{\partial \theta(r_J)}{\partial r }
\sum_{n=1}^N a_n (2n-2)   \sin \left[ (n-1) \left(J-\frac{1}{2} \right) \frac{\pi}{N}\right], \\
\frac{\partial}{\partial r } f^o (r_J) = & -\frac{\partial \theta(r_J)}{\partial r }
\sum_{n=1}^N b_n (2n-1)   \sin \left[ \left(n-\frac{1}{2}\right)\left(J-\frac{1}{2}\right) \frac{\pi}{N} \right].
\end{split}
\end{equation}
Note that these expressions can be understood as discrete sine transformations. In summary, to calculate a derivative with respect to $r$, one first performs a discrete cosine transform \eqref{eq:coefficientsanbn} to the space of spectral coefficients and uses then the sine transform in \eqref{eq:derivativeevendiscrete} to obtain a position space expression for the derivative. Using this approximation for derivatives, one can easily compute the right-hand side of the semi-discrete equation \eqref{eq_eombgDiscretized} and solve the corresponding set of ordinary differential equations. In appendix \ref{app:ConventionFourier} we collect useful formulas for an algorithmic implementation.

\subsection{Numerical spectral viscosity}
\label{sec:NumericalSpectralViscosity}
A generic strategy to ensure the stability of a numerical scheme for hyperbolic problems is to modify the equation of motion by 
adding a diffusive term, typically a higher (even) derivative with a coefficient that vanishes in the formal limit $N\to \infty $. This is commonly called {\it numerical viscosity}. The diffusive term has to be big enough to avoid spurious oscillations but sufficiently small in order to not destroy the accuracy of the numerical scheme. In finite difference schemes, stability is reached by adding a small second order derivative to the equation, while in finite volume schemes this is usually done by adopting a limiter function in the reconstruction of the intercell value needed to compute the fluxes. From a pseudo-spectral point of view, the spurious oscillations originate  from the high frequency part of the spectrum, which does not decay sufficiently fast or even grows with the number of points. A simple solution to recover the decay of the spectrum in the UV with increasing number of points $N$ is to adopt a filtering technique: at each time step, the spectrum of the unknowns is filtered with a continuous function that reduces the high frequency part but leaves the (physical) low frequency part unchanged. 

To explain how this can be implemented time step by time step, let us consider first the introduction of second (or higher) order derivatives to an hyperbolic problem with periodic boundary condition $\theta=\theta+2\pi$,
\begin{equation}
\frac{\partial}{\partial t }u=\mathcal L u+Q^{2p}u.
\end{equation}
Here, $\mathcal L u$ represents the discretized version of the hyperbolic problem and 
\begin{equation}
Q^{2p}u=\epsilon_N(-1)^{p+1} \frac{\partial^{2p}}{\partial \theta^{2p}}u.
\end{equation}
The coefficient $\epsilon_N$ is supposed to vanish in the continuum limit $N\to \infty$. 
In Fourier space, this operator is diagonal, so it acts independently on each mode,
\begin{equation}
Q^{2p}\cos(n \theta)= -\epsilon_N n^{2p} \cos(n \theta).
\end{equation}
During the evolution this diffusive operator leads to an effective damping to the spectrum. Indeed, working in 
Fourier space and neglecting the term $\mathcal L u$ for a moment, we have
the solution 
\begin{equation}
\hat u_n(t+\Delta t)= \exp[-\epsilon_N n^{2p} \Delta t ] \, \hat u_n(t).
\end{equation}
Choosing properly $\epsilon_N\simeq 1/ N^{2p}$, it is possible to implement an effective damping of the high frequency modes 
and leave the low part of the spectrum nearly unchanged.

For a Chebyshev expansion, the operator $Q$ gets modified according to \cite{cheb,hesthaven_gottlieb_gottlieb_2007,GELB20003,GOTTLIEB200183}
\begin{equation}
Q^{2p}u=\epsilon_N(-1)^{p+1}\left[\sqrt{1-x^2} \frac{\partial}{\partial x}\right]^{2p}u,
\end{equation}
because one wants to preserve the property of being diagonal in the space of modes,
\begin{equation}
Q^{2p}T_n(x)= -\epsilon_N n^{2p} T_n( x ).
\end{equation}
This still implements a damping in time of the high frequency modes as becomes clear from the transformation law between Chebyshev and Fourier modes. Is clear now how to generalize this type of exponential filter operator to our basis $B_n(r)$ in \eqref{eq:polynomial_basis}. Using the chain rule, we have 
\begin{equation}
\frac{\partial}{\partial \theta }\quad \to\quad\sqrt{1-x^2} \frac{\partial}{\partial x } \quad \to\quad\frac{\partial r}{\partial \theta } \frac{\partial}{\partial r},
\end{equation} 
and 
\begin{equation}
Q^{2p}u=\epsilon_N(-1)^{p+1}\left[\frac{\partial r}{\partial \theta } \frac{\partial}{\partial r} 
\right]^{2p}u.
\end{equation}
Using the basis expansion  \eqref{truncated_functional_expansion}, the discretization points \eqref{eq:rIdiscretizationpoints} and 
the transformation \eqref{rofth}, the actual expressions for the numerical spectral viscosity operators for the odd and even expansions read
\begin{equation}
\begin{split}
Q^{2p}f^o(r_J)&=-\epsilon_N\sum_{n=1}^N b_n (2n-1)^{2p} \cos\left[\left(n-\frac12\right)\left(J-\frac12\right)\frac{\pi}{N}\right],\\
Q^{2p}f^e(r_J)&=-\epsilon_N\sum_{n=1}^N a_n (2n-2)^{2p} \cos\left[\left(n-1\right)\left(J-\frac12\right)\frac{\pi}{N}\right].
\end{split}
\label{eq:spectralviscosityoperators}
\end{equation}
This filter has been shown to perform well and to recover spectral accuracy for some hyperbolic problems like Burgers' equation  and to also mitigate the Gibbs phenomena in the presence of shocks \cite{GELB20003}.
More recently  a slight modification of this operator  was proposed to reduce the dissipation in the smooth part of the solution while still guaranteeing the stabilization of the numerical scheme in presence of 
discontinuities \cite{Tadmor2012}.

%
%

\subsection{Validation against Gubser flow}
%
%
To verify and validate our numerical scheme, it is useful to compare against a known
analytic (or semi-analytic) solution. For Israel-Stewart type theories, such a solution with azimuthal rotation symmetry, longitudinal boost symmetry and an additional conformal symmetry has been found by Gubser \cite{Gubser:2010ze, Gubser:2010ui, Marrochio:2013wla}.

Consider the minimal set of equations for the evolution of temperature, fluid velocity and shear stress in the presence of a conformal symmetry,
\begin{equation}
\begin{split}
\frac{u^\lambda \nabla_\lambda T}{T} +\frac{\nabla_\mu u^\mu}{3} +\frac{ \pi^\mu_{\nu}\sigma^{\nu}_\mu}{3s T}=0,
\\
u^\lambda \nabla_\lambda u^\mu+\frac{\Delta^\mu_{\lambda}\nabla^{\lambda} T}{T}+\frac{\Delta^{\mu}_\lambda\nabla_{\alpha} \pi^{\alpha \lambda}}{ s T }=0,
\\
\frac{\tau_\text{shear}}{s T }\left( \Delta^\mu_\alpha \Delta^\nu_\beta u^\lambda \nabla_{\lambda} \pi^{\alpha \beta} +\frac{4}{3} \nabla_{\lambda}u^\lambda \pi^{\mu \nu} \right) +\frac{\pi^{\mu\nu}}{sT} = -\frac{2 \eta }{s T} \sigma^{\mu\nu}.
\end{split}
\end{equation}
As a result of the conformal symmetry, the thermodynamic equation of state is $p=\epsilon / 3$ and $\eta/s$ as well as $\tau_\text{shear}/{sT}$ are constant. The Gubser solution to the fluid evolution equations can be most directly obtained from the following Weyl rescaling of the Minkowski metric, 
\begin{equation}
ds ^2 = -d\tau^2 +dr^2+r^2 d\phi^2+\tau^2d\eta^2= \tau^2  \left[ -d\rho^2 +\cosh^2(\rho)d\theta^2+\cosh^2(\rho)\sin^2(\theta)d\phi^2+d\eta^2 \right].
\label{eq:MinkowskiConformal}
\end{equation}
\begin{figure}[t]
	\centering
	\includegraphics[width=0.48\textwidth]{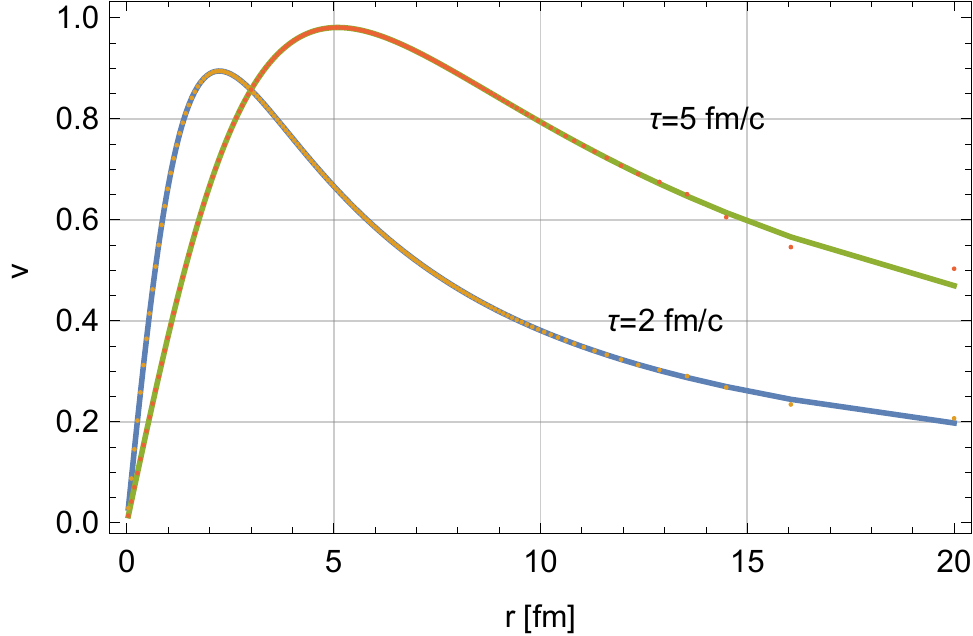} \hfill
	\includegraphics[width=0.48\textwidth]{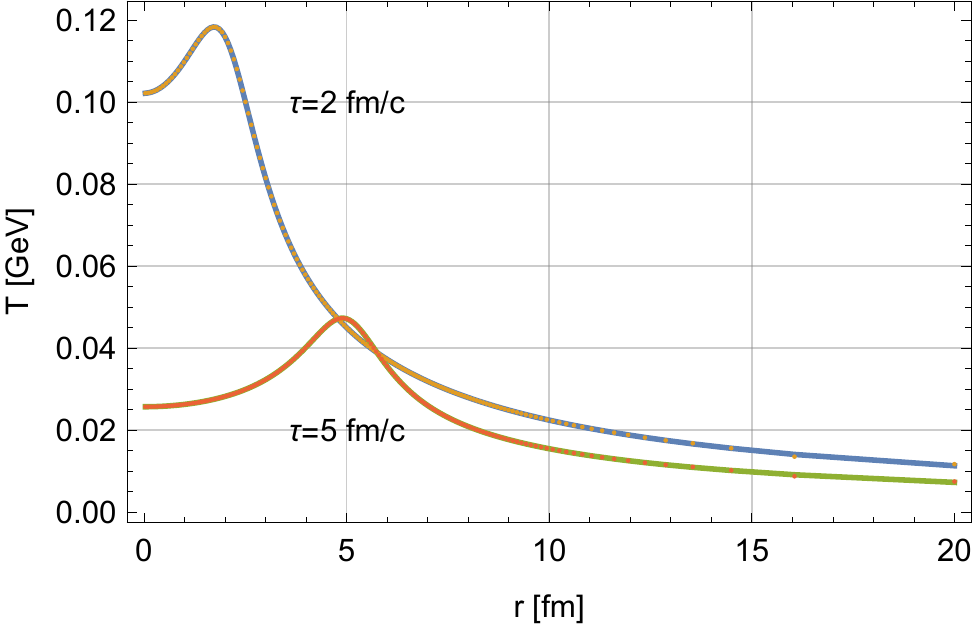} \\
	\includegraphics[width=0.48\textwidth]{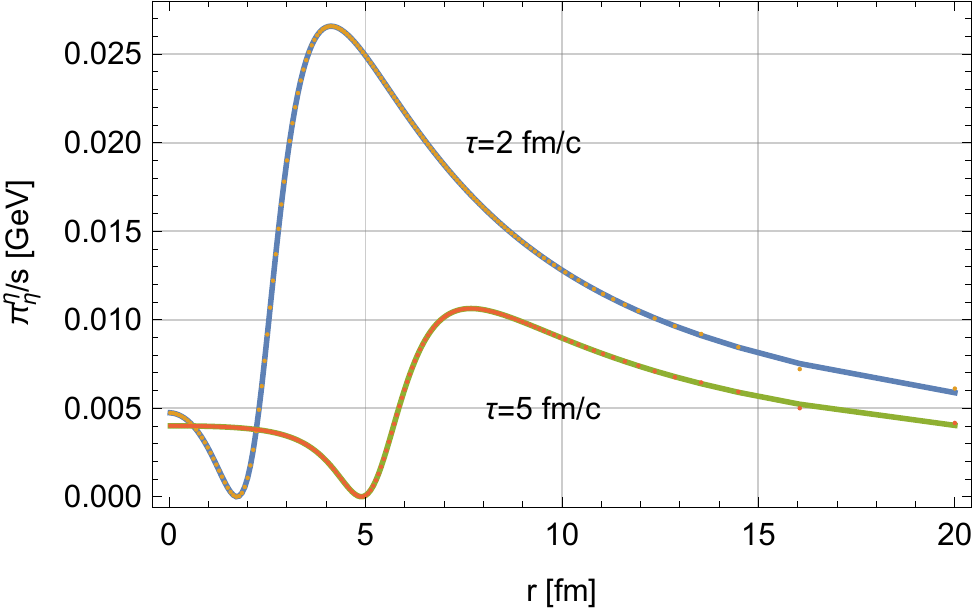} \hfill
	\includegraphics[width=0.48\textwidth]{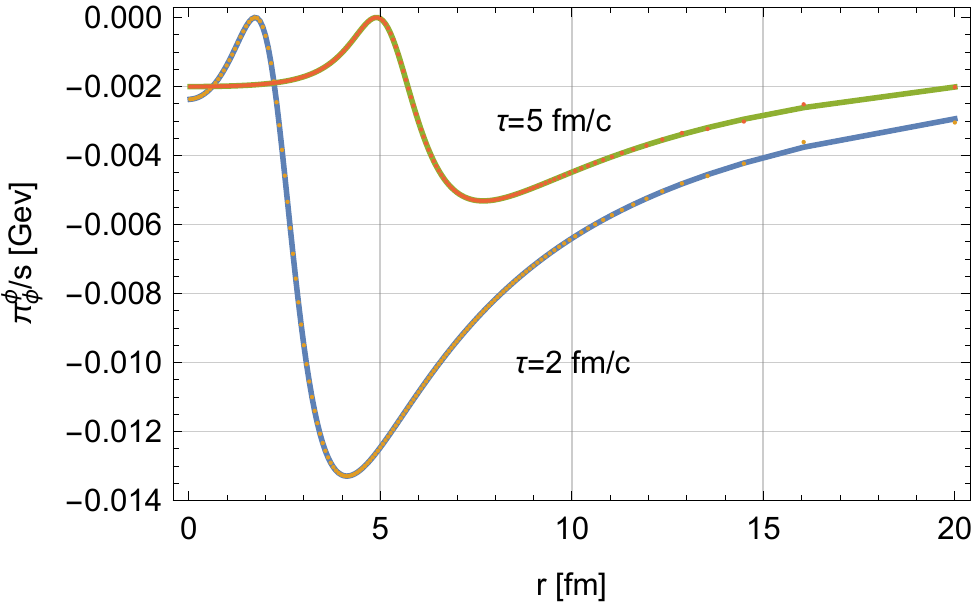}
	\caption{Radial fluid velocity $v=u^r/u^\tau$ (upper left panel), temperature $T$ (upper right panel) and the shear stress components divided by entropy density $\pi^\eta_\eta/s$  (lower left panel) and $\pi^\phi_\phi/s$ (lower right panel) as a function of radius $r$ at Bjorken times $\tau=2$ fm/c and $\tau=5$ fm/c. The lines correspond to the semi-analytic Gubser solution, while the points give our numerical results obtained with the pseudo-spectral method with $N=150$ discretization points. Within the line width there is no disagreement except for a few points at large radius where the density drops and the distance between neighboring discretization points increases. We have here chosen the maximal radius to be $20$ fm. For applications to more realistic situations we choose somewhat larger values so that the region where the discretization points become sparse is further to the right.}
\label{eq:GubserValidation}
\end{figure}
The change of variables is defined here by
\begin{equation}
\sinh(\rho )= -\frac{1-\tau^2+r^2}{2\tau},\quad\quad\quad \tan(\theta) = \frac{2 r }{1+ \tau^2 -r^2}.
\end{equation}
Apart from the conformal factor $\tau^2$, the metric on the right hand side of \eqref{eq:MinkowskiConformal} is the one of three-dimensional de Sitter space times the real line, $\text{dS}_3\otimes \mathbb{R}$. In that space, the fluid equations can be solved rather directly in the presence of rotational and translational symmetries on hypersurfaces of constant de Sitter time coordinate $\rho$. For example, the fluid velocity $\hat u^\mu$ is simply constant and points into the time direction $\rho$, temperature is a function of de Sitter time only, $\hat T=\hat T(\rho)$, and only one component of shear stress, $\hat \pi^\eta_\eta(\rho)$, is independent. The independent equations of motion are then
\begin{equation}
\begin{split}
&\frac{1}{\hat T}\frac{\mathrm{d}}{\mathrm{d} \rho} \hat T +\frac{2}{3} \tanh \rho =\frac13 \bar \pi^{\eta}_{\eta} \tanh \rho,\\
&\frac{c}{\hat T }\frac{\eta}{\hat \hat s } \left[ \frac{\mathrm{d}}{\mathrm{d} \rho} \bar \pi^{\eta}_{\eta}+\frac{4}{3}\left(\bar \pi^{\eta}_{\eta}\right)^2 \tanh \rho   \right] + \bar \pi^{\eta}_{\eta}=\frac{4}{3} \frac{\eta}{\hat s \hat T }\tanh \rho,
\end{split}
\end{equation}
where we have parametrized the shear stress in terms of the dimensionless ratio $\bar \pi^{\eta}_{\eta}= \hat  \pi^{\eta}_{\eta}/ (\hat T \hat s )$. To recover the fluid fields in our conventions one can use the relations 
\begin{equation}
u_{\mu} = \tau \frac{\partial \hat x^\nu}{\partial x^\mu} \hat u_\nu, \quad\quad\quad T = \frac{\hat T}{\tau}, \quad\quad\quad 
\pi_{\mu\nu}  =   \frac{1}{\tau ^2 } \frac{\partial \hat x^\alpha}{\partial x^\mu}  \frac{\partial \hat x^\beta}{\partial x^\nu} \hat \pi_{\alpha \beta },
\end{equation}
where $x^\mu= (\tau, r, \phi ,\eta)$ are Bjorken coordinates and $\hat x^\mu = (\rho, \theta, \phi, \eta )$ are the coordinates of $\text{dS}_3\otimes \mathbb{R}$.

In figure \ref{eq:GubserValidation} we show a comparison between our numerical solution using the pseudo-spectral method with $N=150$ discretization points and the semi-analytic Gubser solution. We have used here the initialization time $\tau_0= 1 \text{ fm/c}$, the shear viscosity to entropy ratio $\eta/s=0.2$ and the shear stress relaxation time $\tau_\text{shear}=5 \eta / (sT) $. The general agreement is very good, and for most regions, no disagreement is visible by bare eye. This includes also the regions around the local maxima of the shear stress where other numerical schemes show some deviations. The pseudo-spectral method shows some deviations from the analytic result only in the region of large radii where the density drops and the distance between neighbouring discretization points increases. Let us note here that one could easily move the region where the discretization grid becomes sparse to larger radii. To this end one would have to choose the length parameter $L$ in equation \eqref{eq:rIdiscretizationpoints} somewhat larger. We will in fact do this for applications to realistic heavy ion collision profiles below.
\begin{figure}[t]
	\centering
	\includegraphics[width=0.48\textwidth]{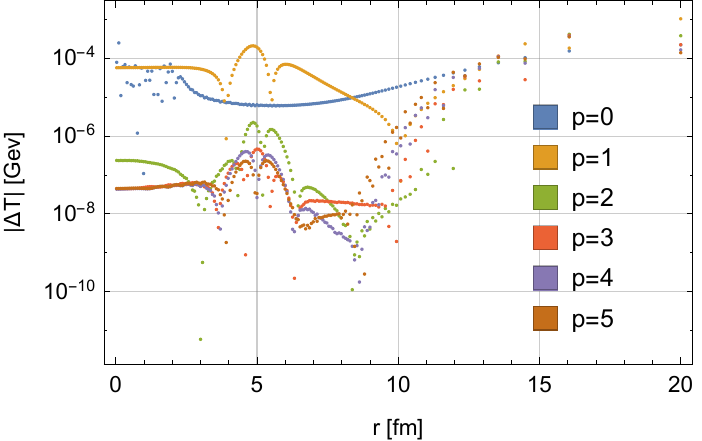} \hfill
	\includegraphics[width=0.48\textwidth]{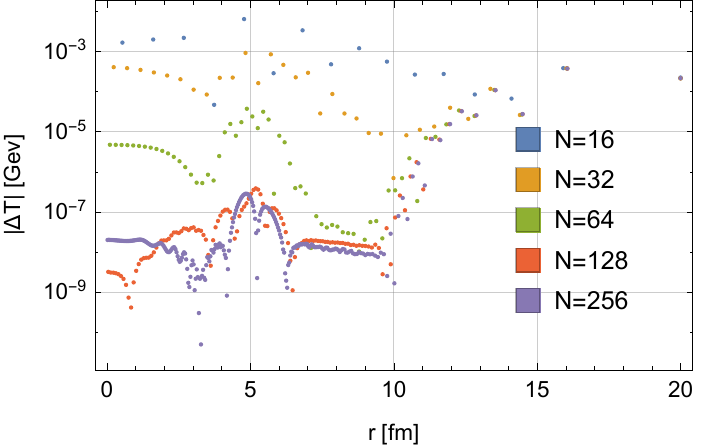}
	\caption{Deviation $|\Delta T|$ between the numerical solution in the pseudo-spectral scheme and the semi-analytic Gubser solution. In the left panel we compare for $N=150$ discretization point different implementations of the numerical spectral viscosity as described in section \ref{sec:NumericalSpectralViscosity} and specifically vary the exponent $p$ while the coefficient is chosen as $\epsilon_N=1/N^{2p}$. The comparison to Gubsers solution is done at Bjorken time $\tau=5$ fm/c. In the right panel we vary the number of discretization points $N$ for fixed implementation of the numerical viscosity scheme with exponent $p=3$.}
\label{fig:comparisonGubserSpectralViscosity}
\end{figure}
\begin{figure}[t]
	\centering
	\includegraphics[width=0.48\textwidth]{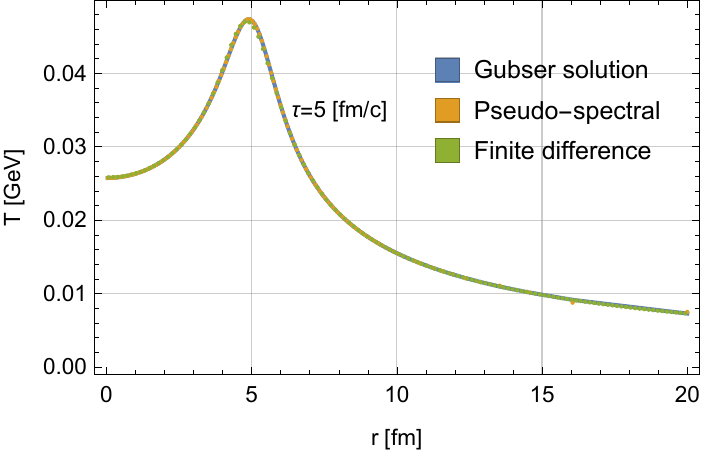}
	\includegraphics[width=0.48\textwidth]{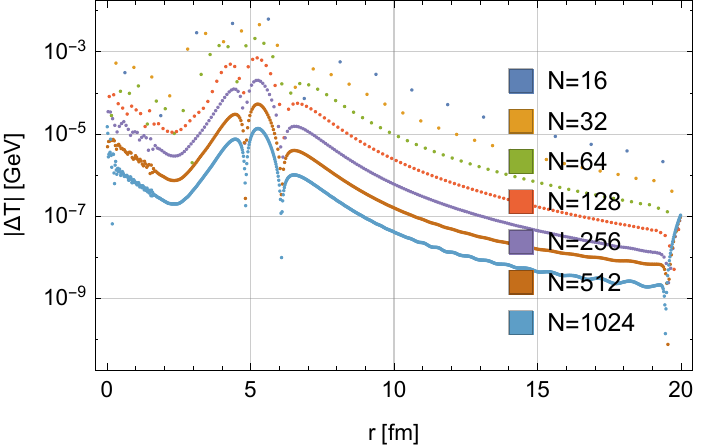}
	\caption{Left panel: Comparison for temperature $T$ as a function or radius at Bjorken time $\tau=5$ fm/c between the semi-analytic Gubser solution (blue line), the numerical pseudo-spectral method (orange points) and a numerical solution using a second order, central finite difference scheme (green points). We have used here $N=128$ discretization points. Although the agreement is in general rather good, the pseudo-spectral method shows some deviation for the points at large radii, while the finite difference scheme shows deviations close to the extremum of temperature. Right panel: Deviation $|\Delta T|$ between the numerical solution using a finite difference scheme and the semi-analytic Gubser solution at Bjorken time $\tau=5$ fm/c for different numbers of discretization points $N$. One observes that the accuracy improves with increasing $N$, albeit not as quickly as for the pseudo-spectral scheme (right panel of fig.\ \ref{fig:comparisonGubserSpectralViscosity}).}
	\label{fig:comaparisonGubserPseudoSpectraFiniteDifference}
\end{figure}

For the plots in figure \ref{eq:GubserValidation} we have used a numerical spectral viscosity according to the description in section \ref{sec:NumericalSpectralViscosity} with $p=3$ and $\epsilon_N = 1/N^{2p}$. While this numerical dissipation term is needed to reach a numerically stable situation, it becomes clear from the comparison to the exact solution that the effect on the physically relevant part of the solution is modest. It is interesting to compare the numerical accuracy for different realizations of the numerical spectral viscosity scheme. This is done in figure \ref{fig:comparisonGubserSpectralViscosity}, where we compare the choices for the damping exponent $p=0, 1, 2, 3, 4$ and $5$. The deviation from the exact Gubser solution $\Delta T$ at Bjorken time $\tau=5$ fm/c decreases with increasing exponent $p$ in the central region until a saturated value of $\Delta T \approx 10^{-7}$ is reached for $p=4$ and a slight increase is observed for $p=5$. We conclude that $p=4$ seems to be an optimal value from this point of view. (This depends in fact on the accuracy goal of the method used to solve the ordinary differential equation in eq \eqref{eq_eombgDiscretized}. If the latter is adapted, one can reach very high numerical accuracies). At larger radii $r$ where the density of discretization points decreases, the numerical accuracy drops and is less sensitive to the implementation of numerical spectral viscosity.
In the right panel of fig.\ \ref{fig:comparisonGubserSpectralViscosity} we also compare the accuracy for different numbers of discretization points $N$. For For $N=128$ one reaches $\Delta T \approx 10^{-7}$ GeV in the central region. 

It is also interesting to compare the performance of the pseudo-spectral method to the one of a more standard (second order, central) finite difference scheme. In the left panel of figure \ref{fig:comaparisonGubserPseudoSpectraFiniteDifference} this is done for the temperature $T$ as a function of radius at Bjorken time $\tau=5$ fm/c. For $N=128$ discretization points, both the pseudo-spectral and the finite difference scheme agree rather well with the semi-analytic Gubser solution. While the former shows some deviations at large radii where the distance between lattice points increases, the latter shows some deviations is the region of the extremum of temperature. In the left panel of fig.\ \ref{fig:comaparisonGubserPseudoSpectraFiniteDifference} we determine the deviation $|\Delta T|$ between the numerical solution obtained with the finite difference scheme and Gubsers solution for different numbers of discretization points $N$. One finds that the accuracy improves with increasing $N$, albeit not as quickly as for the pseudo-spectral method where the corresponding plot is shown in the right panel of fig.\ \ref{fig:comparisonGubserSpectralViscosity}.
%
%
%
%
%
%
\begin{figure}[t]
	\centering
		\includegraphics[height=0.31\textwidth]{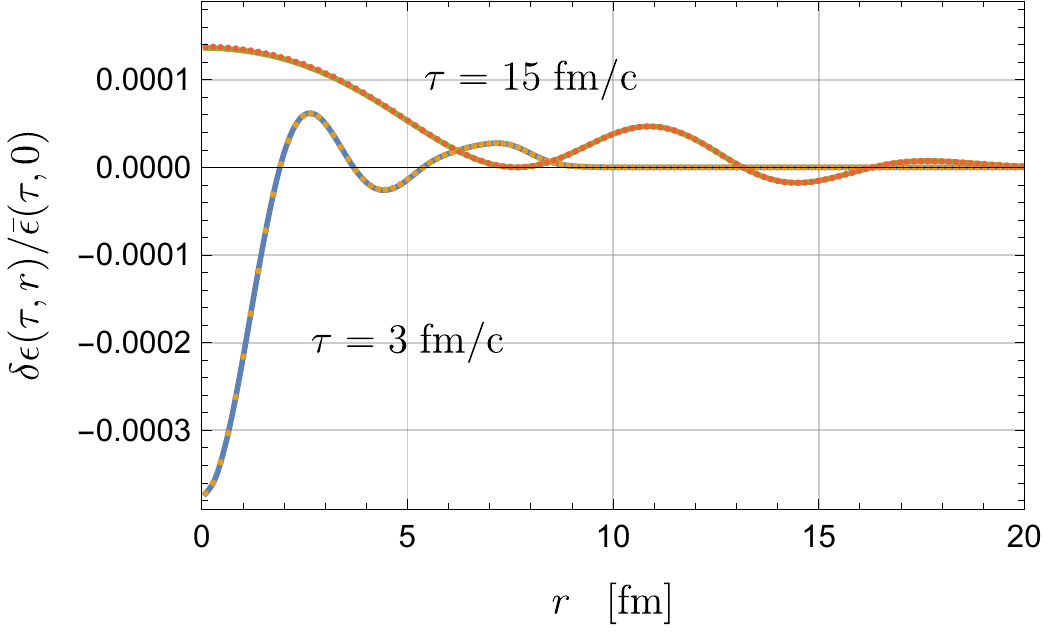} \hfill
		\includegraphics[height=0.31\textwidth]{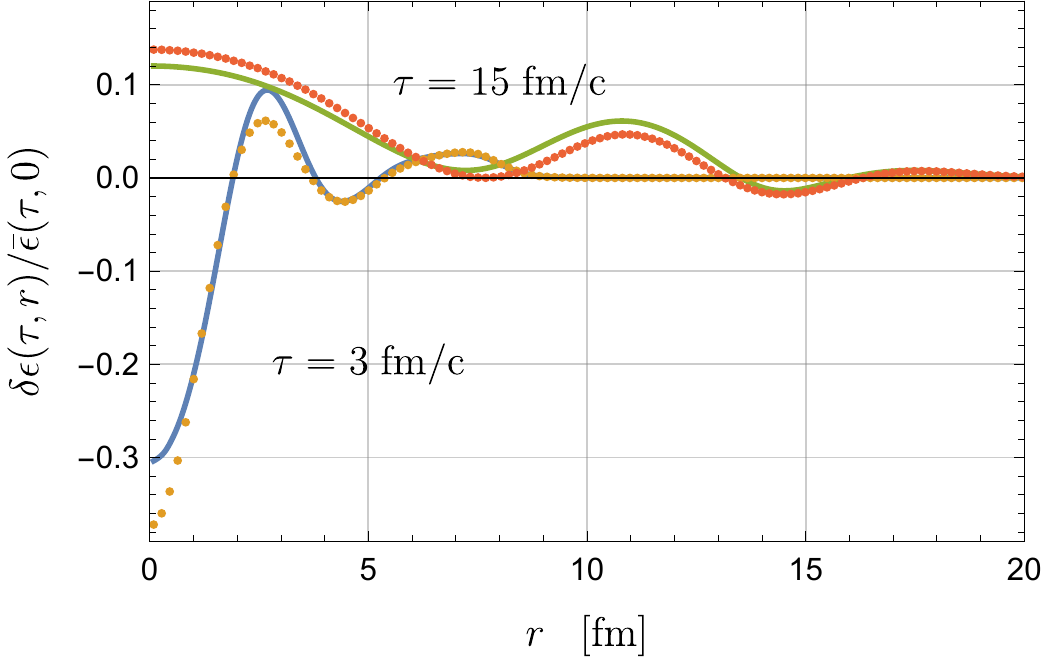}
	\caption{Comparison of solution for perturbations in the $m=0$, $l=3$ mode constructed as a difference between non-linearly evolved solutions (solid lines) and the solution of the linearized equations (dots). We show the perturbation in energy density normalized to the background energy density in the center of the fireball $\delta\epsilon(\tau,r)/\bar\epsilon(\tau,0)$ as a function of radius $r$ for Bjorken times $\tau=3$ fm/c and $\tau=15$ fm/c. For the left plot we have chosen a small perturbation with $\delta\epsilon(\tau_0,0)/\bar\epsilon(\tau_0,0)=10^{-3}$ in the center of the fireball at the initialization time $\tau_0=0.4$ fm/c. Here one observes perfect agreement between the two solutions (within the plot resolution) which shows the consistency of our numerical scheme. For the right plot we have instead chosen a larger initial amplitude $\delta\epsilon(\tau_0,0)/\bar\epsilon(\tau_0,0)=1$. The difference between non-linearly evolved solutions differs now from the solution to the linearized equations, as a consequence of non-linear effects.}
	\label{plot_validationpert}
\end{figure}
\subsection{Perturbations}
The numerical scheme we have discussed in section \ref{sec_adams} and \ref{sec:Pseudospectraldiscretization} can also be used to evolve the linear perturbations using the equations \eqref{eq_eomPerturbationFourier}. One needs to take into account the correct parity and boundary conditions at $r=0$ which depend on the azimuthal wave number $m$, but otherwise the algorithm just works as for the background. While the background solution has been verified against Gubsers solution, we would also like to have a check for the numerical implementation of the linearized equations for perturbations. This is in fact directly possible for the modes with azimuthal wave number $m=0$. These modes can be evolved either by solving the linearized equations for perturbations, or by adding a small perturbation to the initial conditions of the background, evolving it forward in time and subtracting the background solution without modification. The result should agree, at least for small enough perturbations where the linearization is justified. 

We have done this check of our implementation and show the result for the mode with $m=0$ and $l=3$ in fig.\ \ref{plot_validationpert}. Specifically, we show the perturbation in energy density $\delta \epsilon(\tau, r)$ as a function of radius $r$ for Bjorken times $\tau=3$ fm/c  and $\tau=15$ fm/c, normalized by the background energy density in the center of the fireball $\bar\epsilon(\tau,0)$. The solid lines have been obtained by solving the non-linear background equations with a perturbation in the initial state and subtracting the corresponding solution without perturbation. In contrast, the dots give the numerical solution of the linearized fluid dynamic equations of motion. For the left plot we have chosen the amplitude of the perturbation in the initial state to be $\delta\epsilon(\tau_0,0)/\bar\epsilon(\tau_0,0)=10^{-3}$ in the center of the fireball at the initialization time $\tau_0=0.4$ fm/c. The agreement between the solution constructed via the non-linear equations and the solution of the linear equations of motion is very good, which demonstrates the validity and consistency of the numerical scheme, as well as the linearity of the perturbation. For the plot on the right hand side we have instead chosen a larger magnitude of the perturbation, $\delta\epsilon(\tau_0,0)/\bar\epsilon(\tau_0,0)=1$ at $\tau_0=0.4$ fm/c. In that case the agreement between the difference of non-linearly evolved solutions and the solution to the linearized equations is not perfect, which shows how non-linear effects start to set in.

\section{Results}
\begin{figure}[h!]
	\centering
		\hspace*{0.35cm}
		\includegraphics[height=0.3\textwidth]{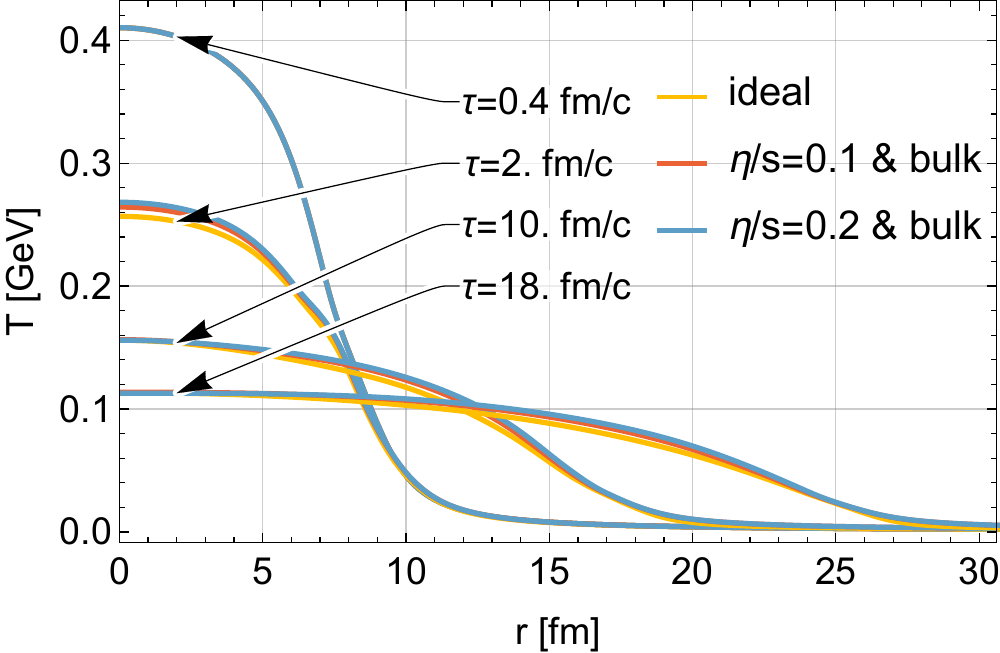} \hfill
		\includegraphics[height=0.3\textwidth]{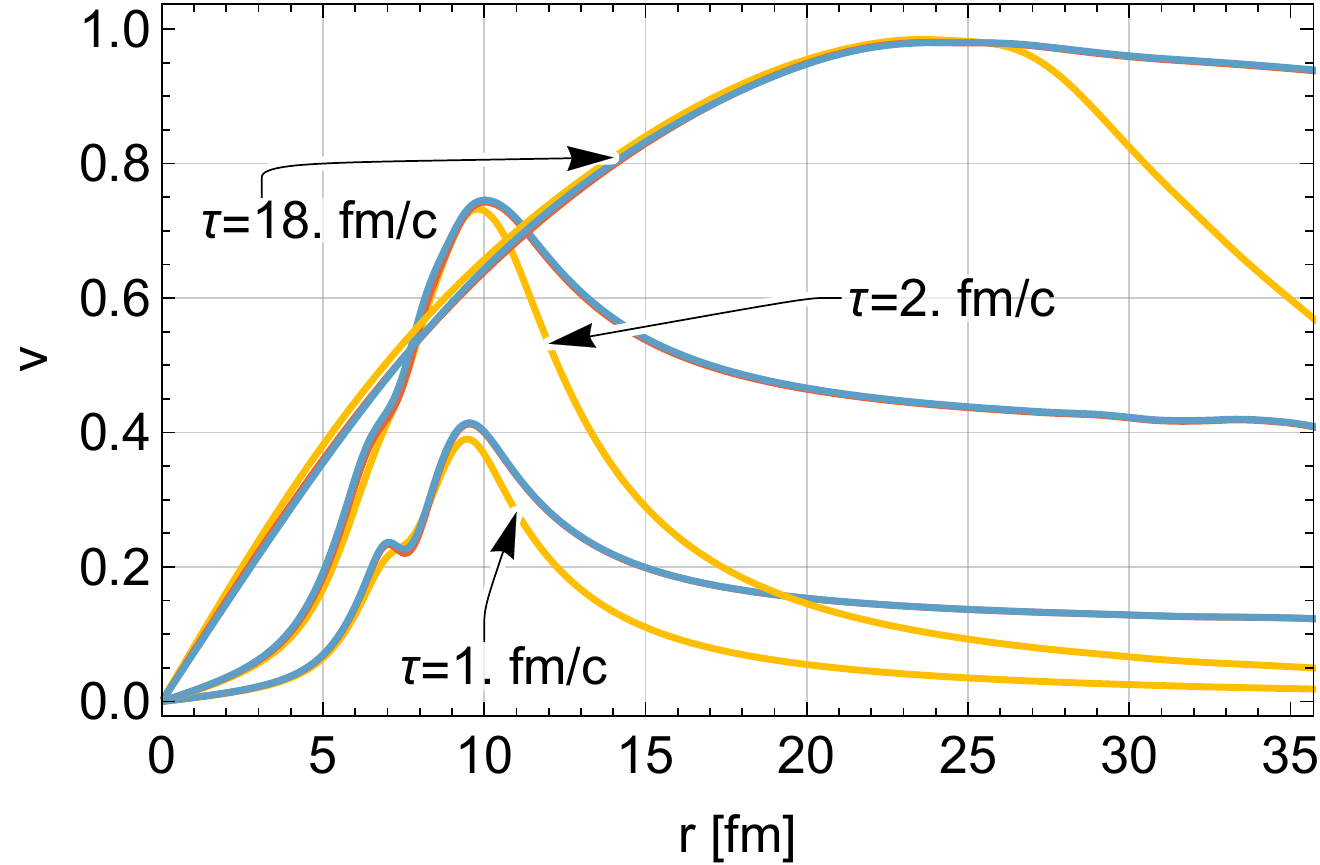}
		\includegraphics[height=0.3\textwidth]{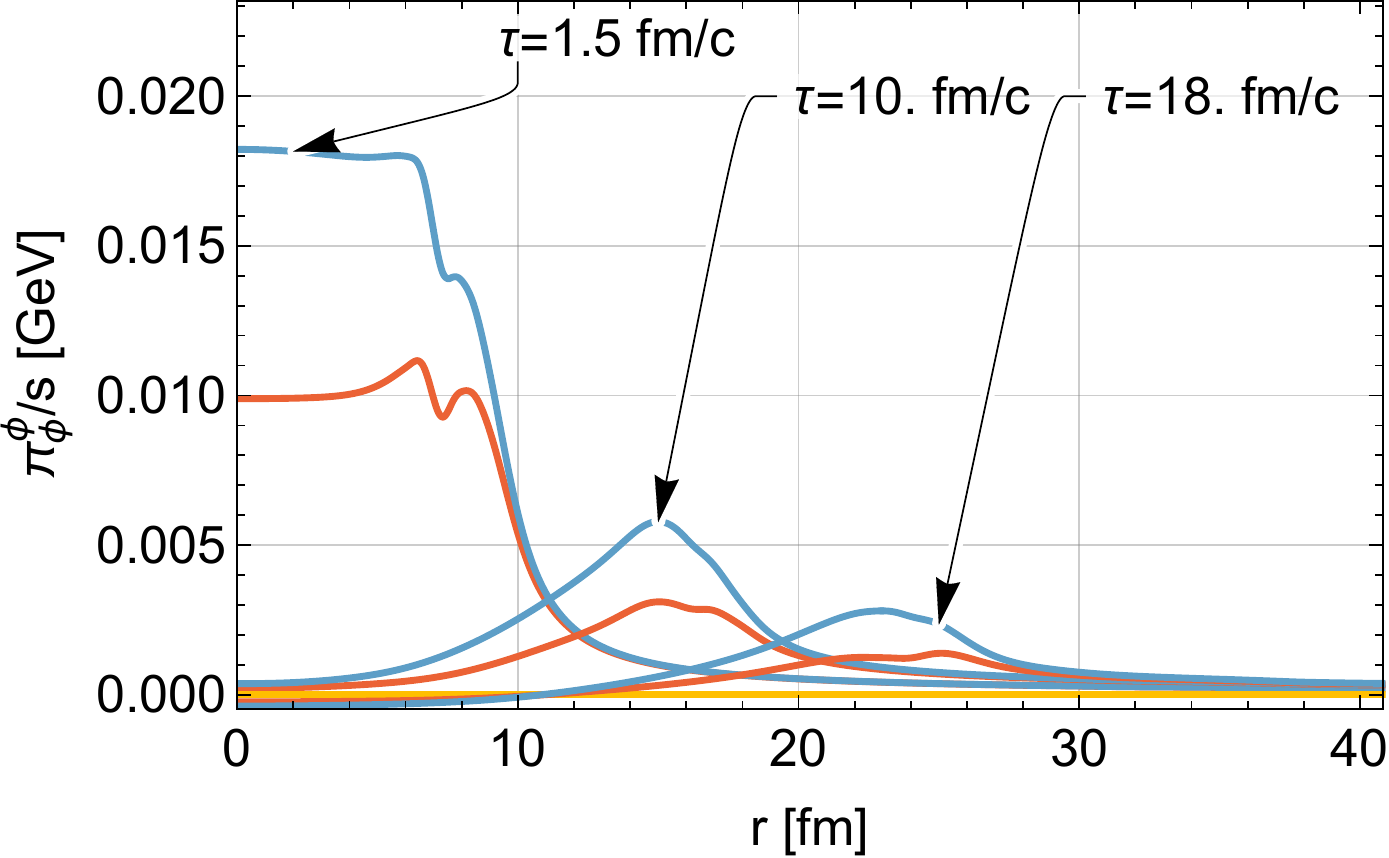} \hfill
		\includegraphics[height=0.3\textwidth]{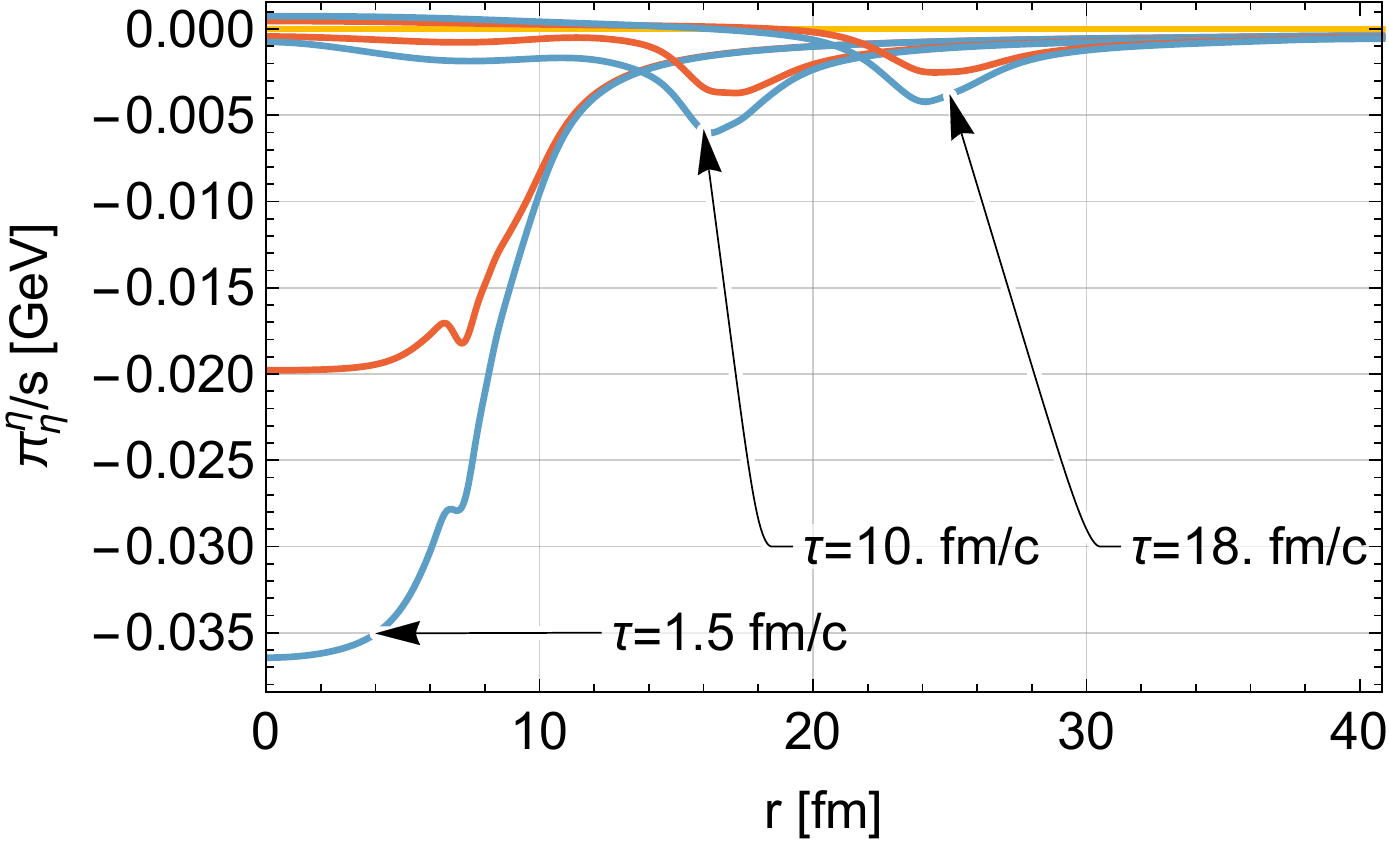}
			\includegraphics[height=0.3\textwidth]{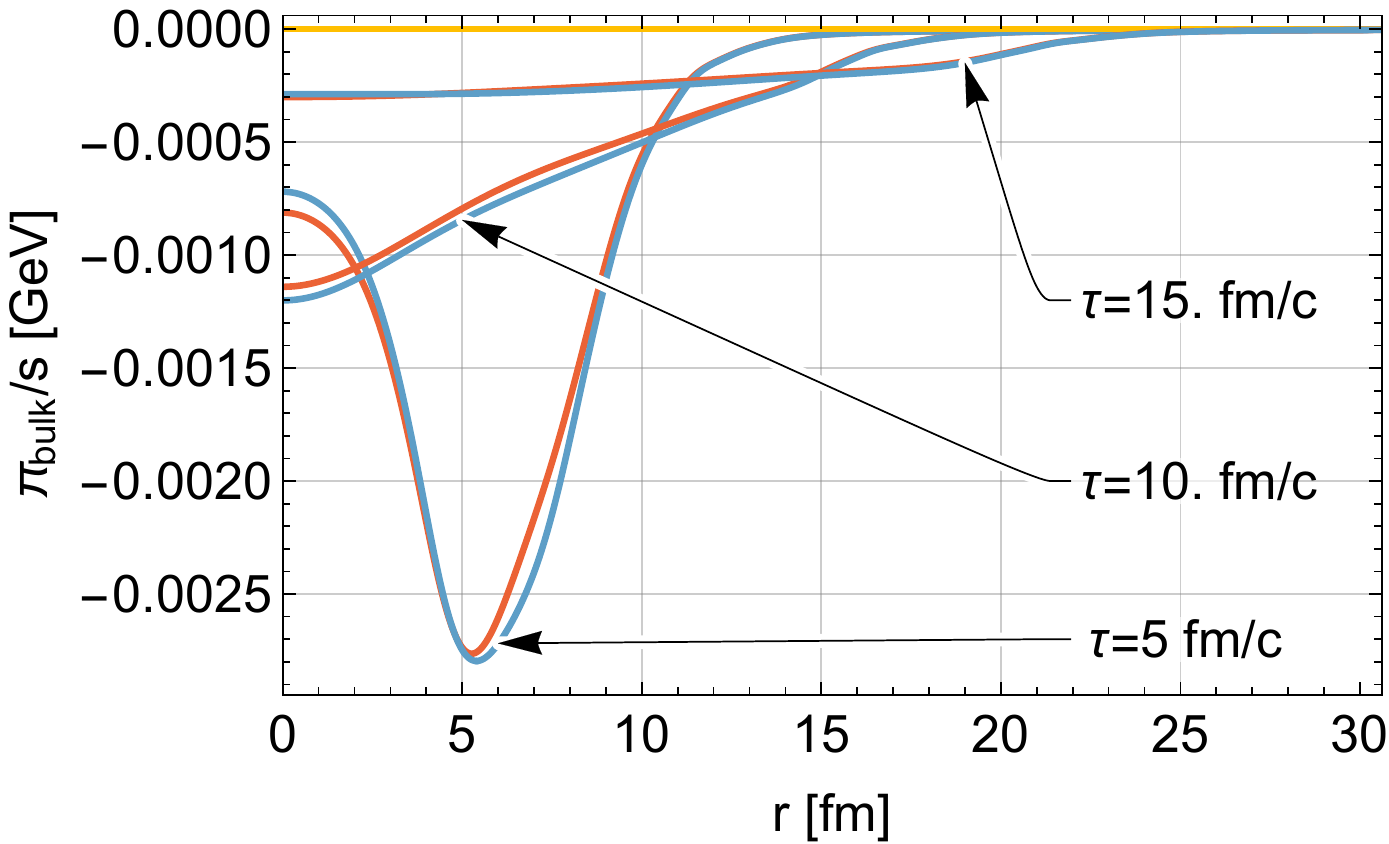}
	\caption{Numerical solution for temperature $T$ (upper left plot), radial fluid velocity $v=u^r/u^\tau$ (upper right plot), shear stress components $\pi^\phi_\phi/s$ (intermediate left plot)  and $\pi^\eta_\eta/s$ (intermediate right plot) and the bulk stress $\pi_{\mathrm{bulk}}/s$ (lowermost plot) as a function of radius at different Bjorken times $\tau$. We compare ideal fluid dynamics (yellow lines) to $\eta/s=0.1$ (red lines) and $\eta/s=0.2$ (blue lines). The latter two cases include also bulk viscosity according to eq.\ \eqref{sec_transportcoeff}. See text for further discussion.}
	\label{backgroundTv}
\end{figure}
\subsection{Background Evolution}
We discuss now the resulting numerical solution to the evolution equations of relativistic fluid dynamics, as discussed in section \ref{sec:RelativisticFluidDynamics}, for a symmetric background following the principles outlined in section \ref{eq:Symmetries}, and using the numerical methods discussed in section \ref{sec_numerics}. 

\begin{table}[t]
	\centering
	\begin{tabular}{|c|c|c|}
		\hline
		Parameter	& Description& ``Default'' value\\
		\hline
		$\epsilon_0$& Initial energy density& 50 GeV/$\mathrm{fm}^3$\\
		$x$& Parameter in optical Glauber model & 0.118\\
		$\eta/s$& Shear viscosity to entropy density & $0.2$\\
		$\tau_\mathrm{shear}$& Shear relaxation time& $5\times \eta/(sT)$\\
		$\zeta/s$& Bulk viscosity to entropy density & Equation\ \eqref{eq_bulkbass}\\
		$\tau_\mathrm{bulk}$& Bulk relaxation time& Equation \eqref{eq:bulkrelaxationTime}\\
		$\tau_0$& Initialization time& $0.4$ fm\\
		\hline
	\end{tabular}
	\caption{A default set of parameters and transport coefficients to solve the fluid evolution equations. The initial state model is the optical Glauber Model at zero impact parameter for collisions of $^{208}$Pb - $^{208}$Pb.}
	\label{tab_parameters}
\end{table}

For concreteness, we choose initial conditions on a hypersurface of constant Bjorken time $\tau_0 = 0.4 $ fm/s and set there the fluid velocity in the radial direction, as well as the independent shear stress components $\pi^\eta_\eta$ and $\pi^\phi_\phi$ and the bulk viscous pressure $\pi_\text{bulk}$ to zero. It remains to choose an initial condition for the energy density (or equivalently temperature) as a function of radius $r$.

The energy density is initialized here according to the optical limit of the Glauber model. Specifically, we assume a linear superposition of $N_\text{coll}$ and $N_\text{part}$ scaling with mixing parameter $\alpha=0.118$, see ref.\ \cite{Qiu:2011hf} for further details. The overall magnitude is left as an open parameter, which should be adjusted to the total particle multiplicity. For the present work, for concreteness we fix the initial energy density to be $\epsilon_0 = 50 \text{ GeV}/\text{fm}^3$ in the center of the fireball. The thermodynamic equation of state and the fluid transport properties are fixed as discussed in section \ref{eq:Thermodynamics}. In table \ref{tab_parameters} we provide a summary of the transport properties and other parameters used in the numerical solution.

In fig.\ \ref{backgroundTv} we show or numerical results of the temperature $T$, the radial fluid velocity $v=u^r/u^\tau$, the shear stress components $\pi^\eta_\eta$ and $\pi^\phi_\phi$ as well as the bulk viscous pressure $\pi_\text{bulk}$ (the latter three divided by entropy density $s$) as a function of radius for different Bjorken times $\tau$. One observes first the expected dilution due to the longitudinal expansion and the build up of a radial expansion as a result of pressure gradients. In the radial velocity $v$ as a function of radius $r$ one observes characteristic features such as local extrema and small oscillations, which can be traced back to a local minimum and maximum in the velocity of sound in the region of the crossover temperature (see figure \ref{epcs}). At late times, the radial fluid velocity smoothens out again and grows for large radii to values approaching the velocity of light.

From the fluid velocity, the local minima and maxima get also inherited to the independent shear stress components divided by entropy density $\pi^\eta_\eta/s$ and $\pi^\phi_\phi/s$. Of course, the development of the shear stress depends also (in an approximately linear way) on the ratio of shear viscosity to entropy density $\eta/s$. The general structure of $\pi^\eta_\eta/s$ and $\pi^\phi_\phi/s$ resembles to an outwards traveling wave. Both components are of similar magnitude but $\pi^\eta_\eta$ has negative sign (and decreases ``longitudinal pressure''), while $\pi^\phi_\phi$ has positive sign.

Finally, we also show the bulk viscous pressure as a function of radius for different Bjorken times $\tau$. As expected, it is negative for an expanding situation. It develops first a maximum in magnitude in the region of radii where the parametrization of bulk viscosity \eqref{eq_bulkbass} has a maximum. At later times, his maximum travels inwards and the ratio $\pi_\text{bulk}/s$ becomes more monotonic, with a maximum in magnitude in the inner region and a decrease towards larger radii. 
\begin{figure}[t]
	\centering
	\begin{subfigure}[t]{0.36\textwidth}
		\includegraphics[width=\linewidth]{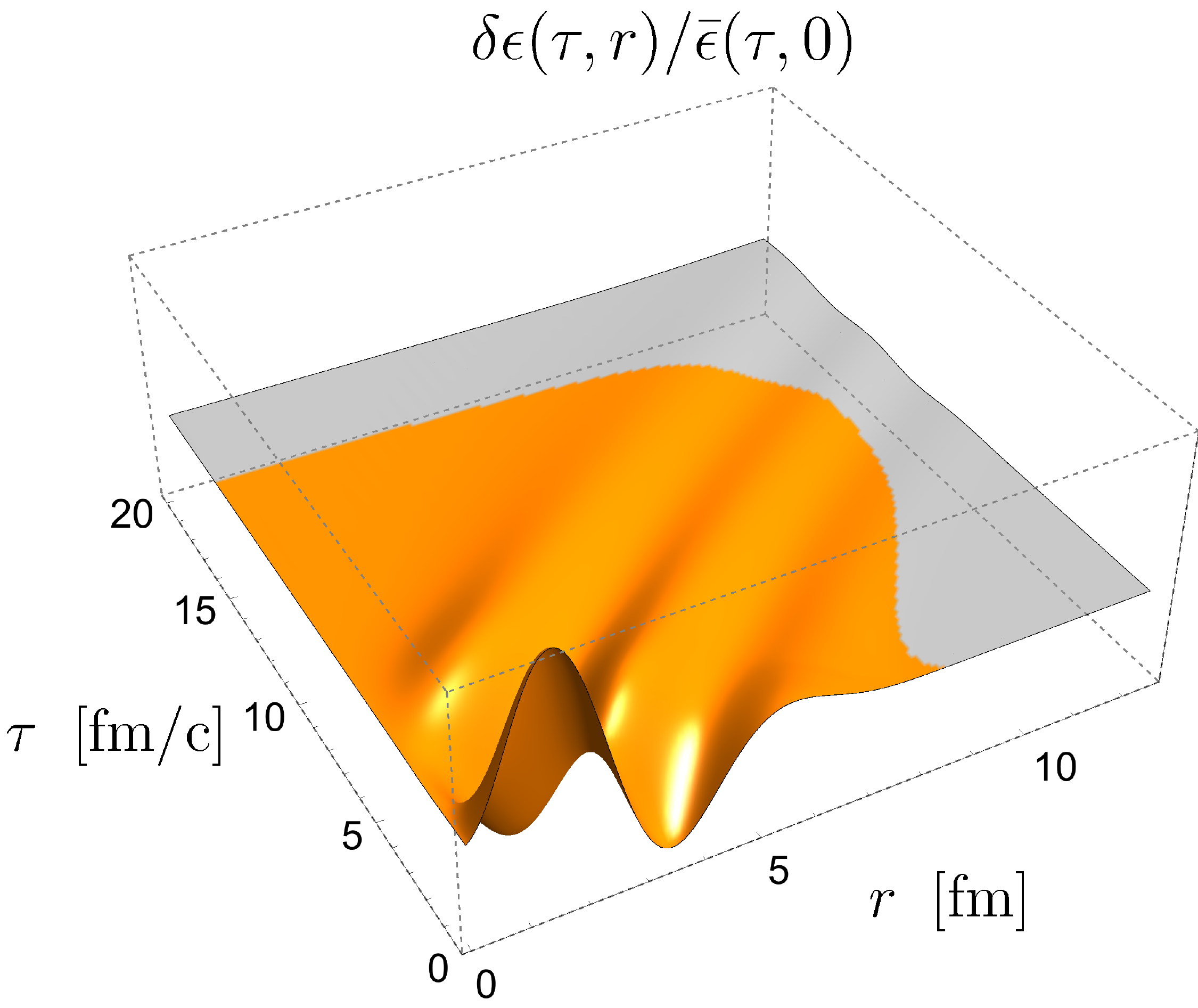}
		\label{fig_p335}
	\end{subfigure} \hspace{0.5cm}
	\begin{subfigure}[t]{0.36\textwidth}
		\includegraphics[width=\linewidth]{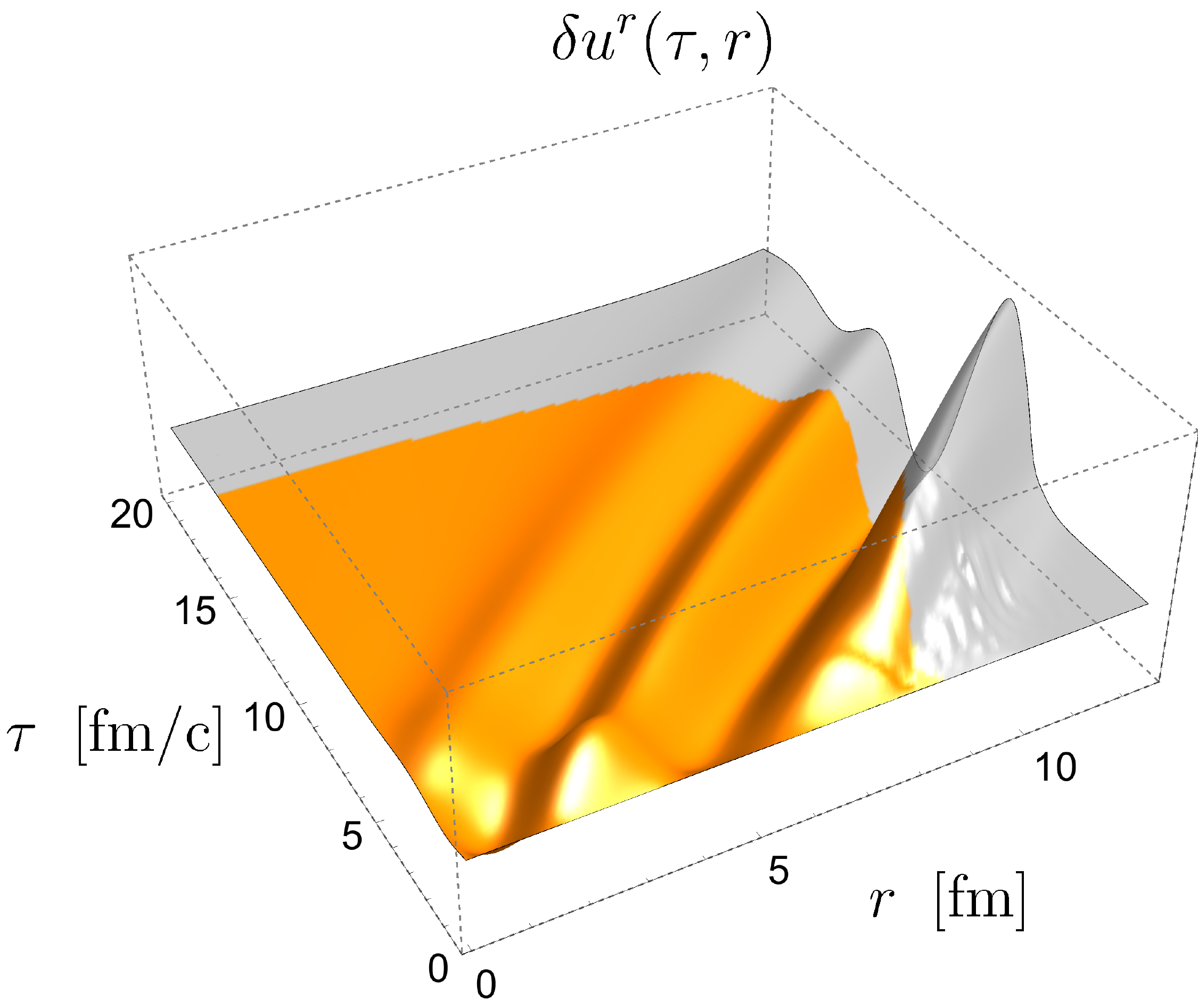}
		\label{fig_p445}
	\end{subfigure}
	\begin{subfigure}[t]{0.36\textwidth}
		\includegraphics[width=\linewidth]{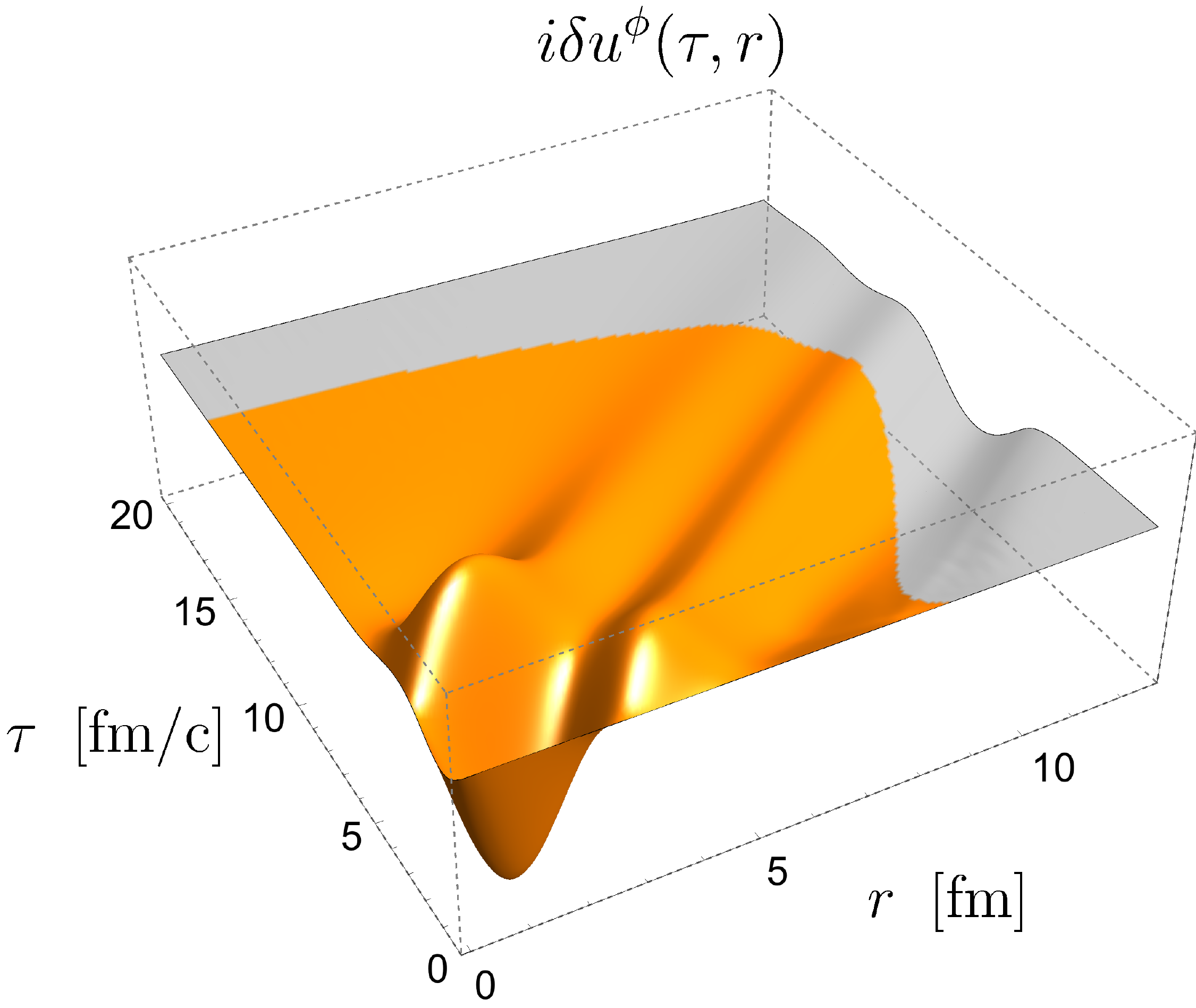}
		\label{fig_pertT5}
	\end{subfigure} \hspace{0.5cm}
	\begin{subfigure}[t]{0.36\textwidth}
		\includegraphics[width=\linewidth]{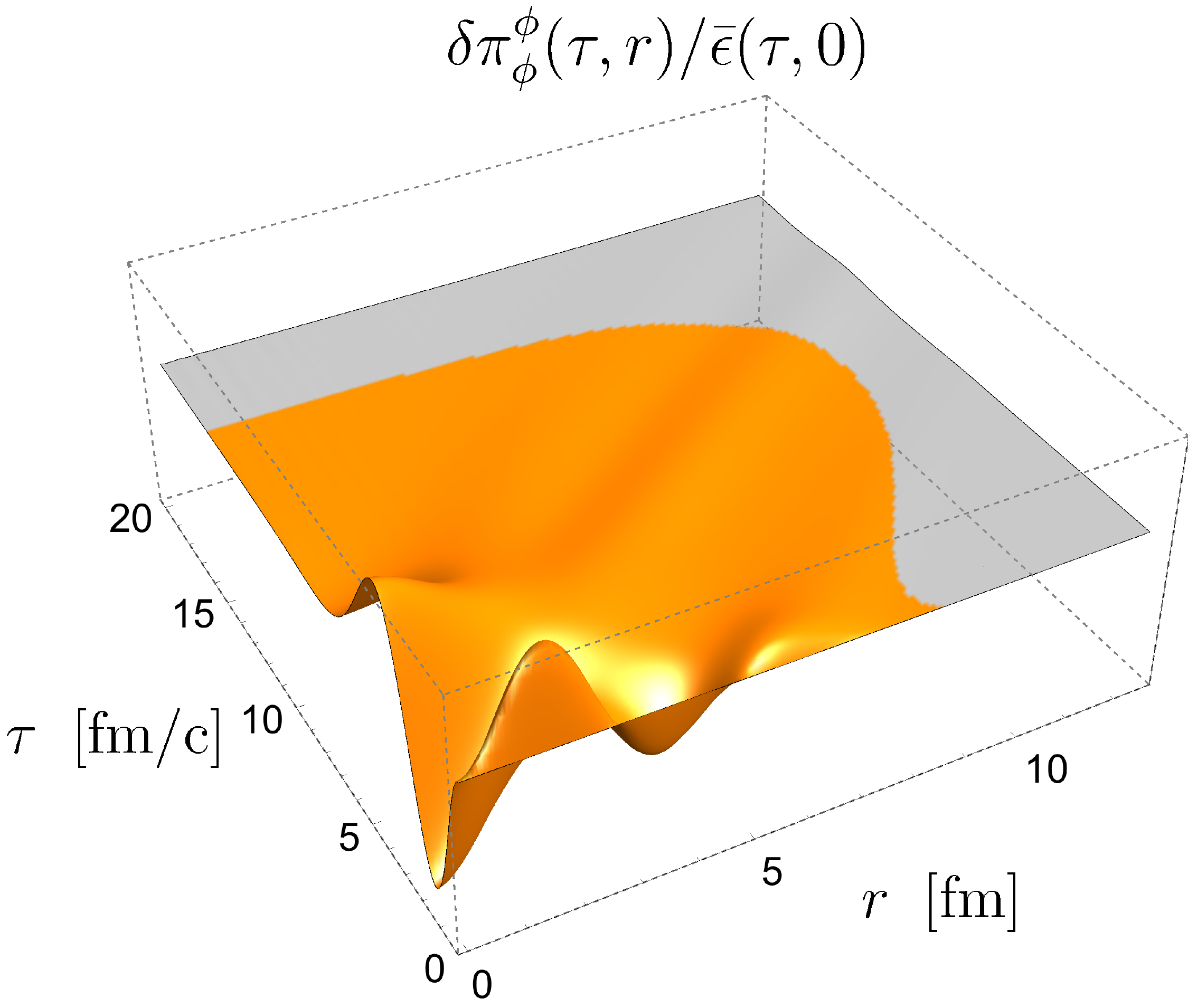}
		\label{fig_pertur5}
	\end{subfigure}
	\begin{subfigure}[t]{0.36\textwidth}
	\includegraphics[width=\linewidth]{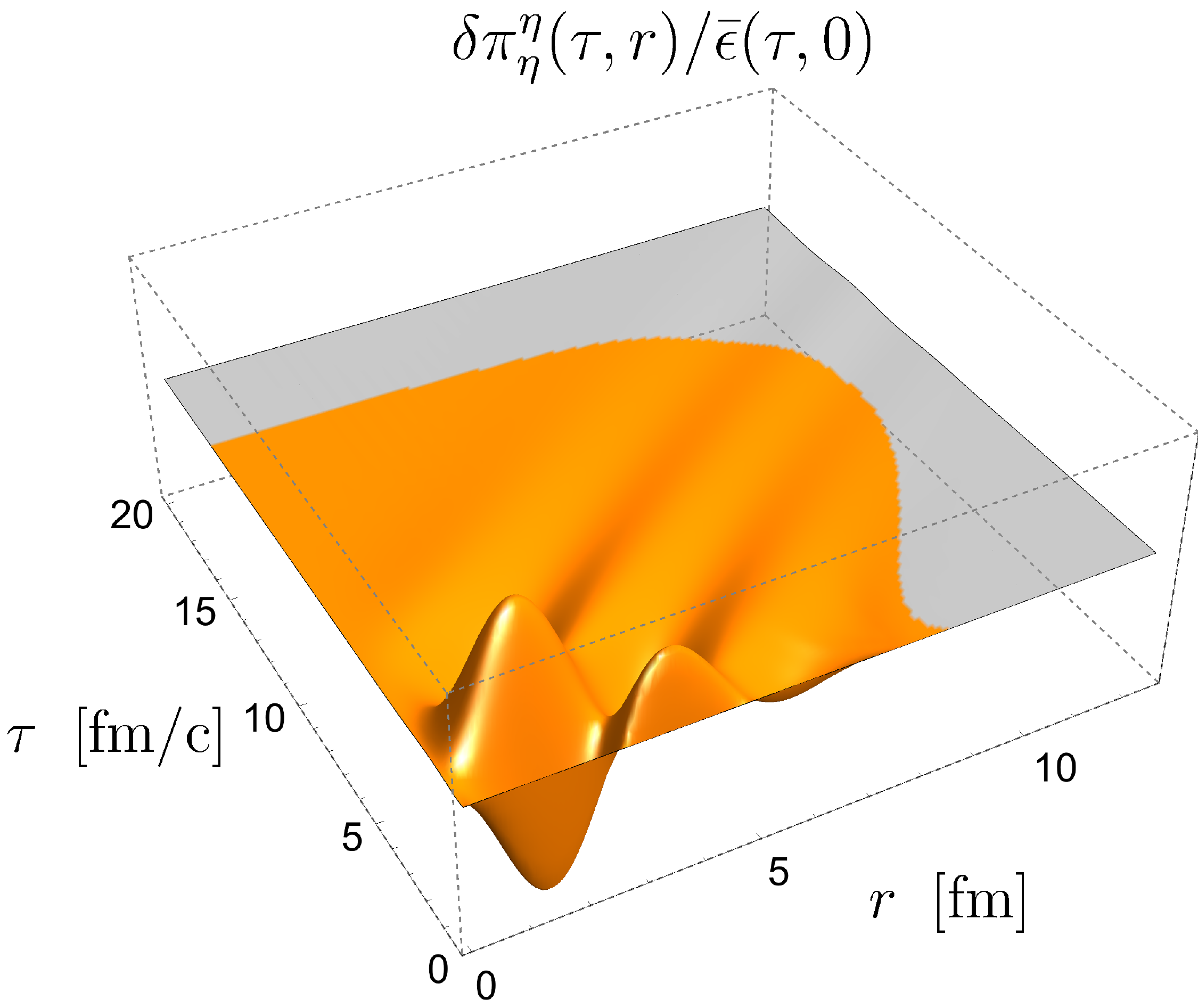}
	\label{fig_p23}
\end{subfigure} \hspace{0.5cm}
	\begin{subfigure}[t]{0.36\textwidth}
	\includegraphics[width=\linewidth]{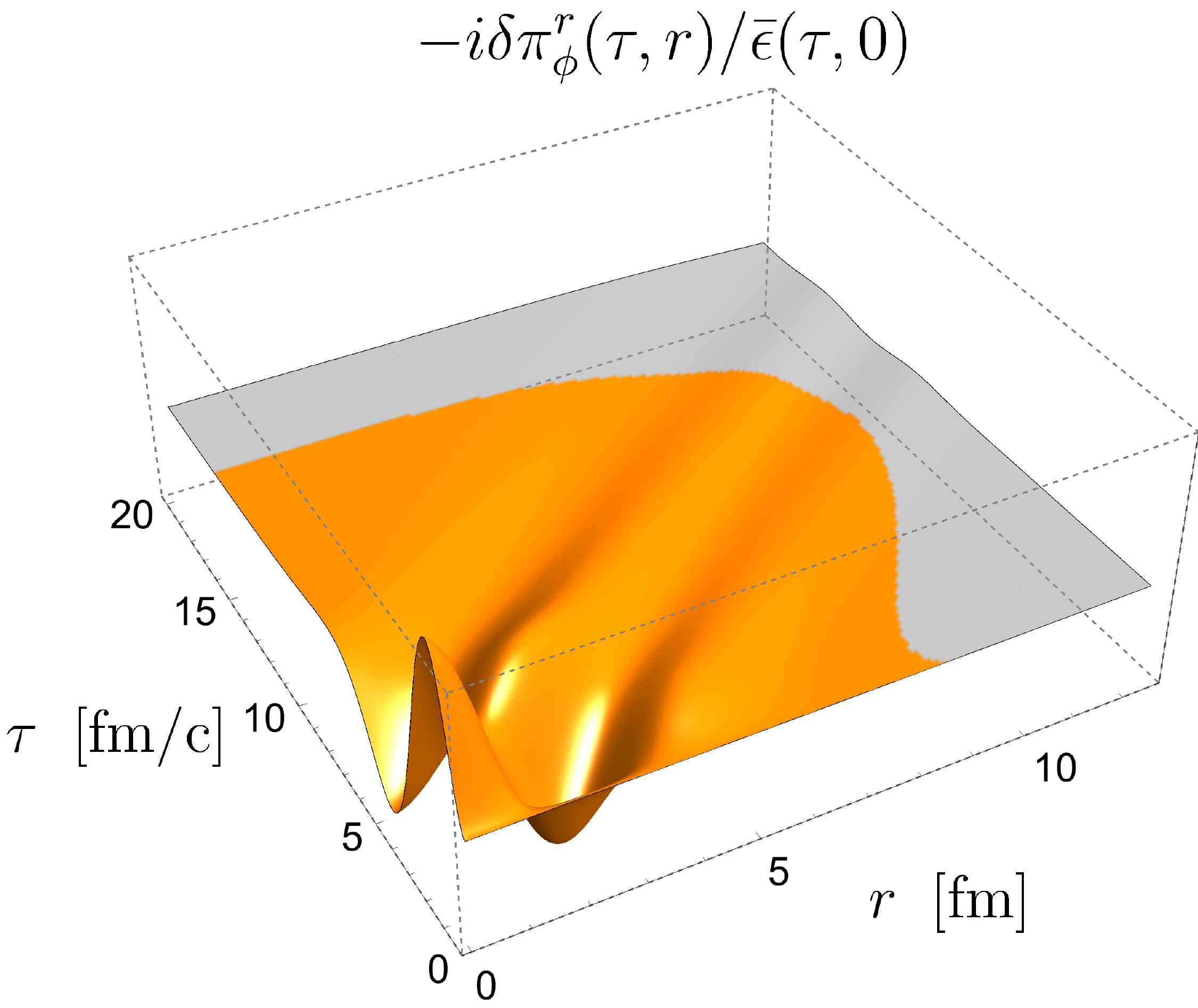}
	\label{fig_p34}
\end{subfigure}
	\caption{Evolution of  perturbations fields initialized in the $m=2$, $l=3$, $k=0$ energy density mode as a function of radius $r$ and Bjorken time $\tau$. We show the perturbation in energy density $\delta \epsilon$, in radial fluid velocity $\delta u^r$, azimuthal fluid velocity $\delta u^\phi$ and different shears stress components. The energy density and shear stress perturbations have been normalized by a time-dependent factor corresponding to the background energy density in the center of the fireball $\bar\epsilon(\tau,0)$ for better visibility. The orange region denotes where the background temperature is above the freeze-out value (taken as $120$ MeV here).}
	\label{plot_evolution}
\end{figure}
\begin{figure}
	\centering
\includegraphics[width=\textwidth]{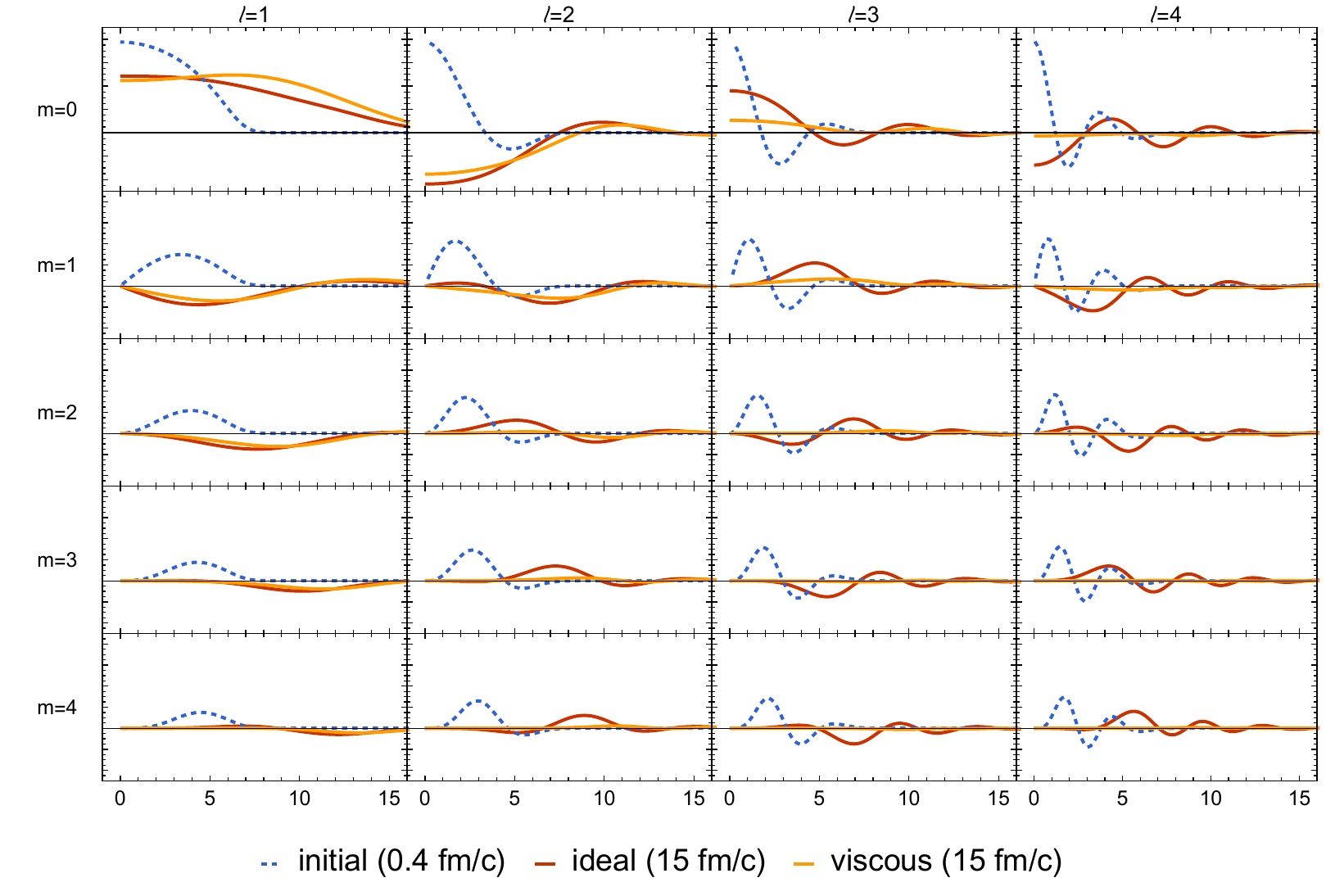}
\caption{Perturbations in the energy density $\delta \epsilon(\tau,r,m,k)$ dived by the background energy density in the center of the fireball $\bar\epsilon(\tau,0)$ as a function of radius $r$ for vanishing longitudinal wave number $k=0$, different values of the azimuthal wave number $m=0,\ldots, 4$ (from top to bottom) and $l=1,\ldots, 4$ (from left to right). We show the form of the perturbation at initialization time $\tau=0.4$ fm/c (dashed lines) and Bjorken time $\tau=15$ fm/c. For the latter curves we compare ideal fluid dynamics (red curves) to viscous fluid dynamics (orange curves). This includes both shear viscosity (with $\eta/s=0.2$) and bulk viscosity as discussed in section \ref{sec_transportcoeff} and summarized in table \ref{tab_parameters}. One observes that in particular the modes with large values of $m$ and $l$ are strongly damped by viscous dissipation.}
\label{plot_perteta}
\end{figure}

\subsection{Perturbations}
The linearized equations of motion \eqref{eq_eomPerturbation} can be solved for the perturbation fields $\mathbf{\tilde{\Phi}}_1$. In principle, this could be done on an event-by-event basis but the idea of mode by mode fluid dynamics is somewhat different. As discussed in section \ref{sec:FluctuatingModes}, one can expand initial conditions for fluctuating fields in a complete set of basis functions. The linearized equations of motion \eqref{eq_eomPerturbation} can be solved for each of these basis functions and an arbitrary solution can then be written as a linear superposition of these solutions. This is very economic in the sense that an (infinitely) large class of initial field configurations can be propagated simultaneously. Of course, formally, the set of basis functions is also infinite but in practice very high wave numbers (in azimuthal, radial and longitudinal direction) should play a less important role because they correspond to finer and finer details in the spatial domain and because they are damped more efficiently by viscosity.

It remains therefore to solve the linearized fluid equations for a set of modes, say $m, l=1,\ldots, 10$. Likely, this is already more than needed for most purposes. For the continuous azimuthal wavenumber $k$ one may use a discretization. However, for the moment we concentrate on strictly boost invariant situations, i.\ e.\ $k=0$. Below, we will also concentrate on initial density perturbations, but initial state perturbations in fluid velocity, shear stress, or bulk viscous pressure can be easily studied in the same framework. 

For scalar transverse density perturbations we can use the basis functions in \eqref{eq:basisfunctionsBessel}. For given wave numbers $m$ and $l$ we initialize the energy density as
\begin{equation}
\delta \epsilon(\tau_0,r,\phi,\eta) = q_{m,l}(r) e^{im\phi} = W(r) J_{m}{\big (}z^{(m)}_l \rho(r) {\big )} e^{im\phi}.
\end{equation}
All other perturbed fluid fields vanish initially for this mode but will get generated by the time evolution. The linearized fluid equations are then solved up to a certain Bjorken time $\tau\sim 20$ fm/c, where the background temperature has dropped well below the freeze-out temperature at $\sim120$ MeV. The numerical solution is obtained by using pseudo-spectral methods as presented in section \ref{sec_numerics}. The obtained solutions can be stored, and the contribution of each mode to final particle spectra and flow coefficients can be evaluated as an integral along the freeze-out surface as described in ref.\ \cite{floerchinger2014kinetic}, see also \cite{Mazeliauskas:2018irt}. 

In fig.\ \ref{plot_evolution} we show the evolution of the mode with $m=2$, $l=2$ and $k=0$ in the plane of radius $r$ and Bjorken time $\tau$. More specifically, we show the perturbations in energy density, radial fluid velocity, azimuthal velocity and different components of the shear stress. For better visibility, we have normalized the density and shear stress perturbations by a time-dependent factor corresponding to the energy density in the center of the fireball $\bar\epsilon(\tau,0)$. One observes characteristic evolution patterns corresponding to propagating sound waves. One also observes a damping at later times, which is primarily an effect of dissipative phenomena (shear and bulk viscosity). 

In a similar way, we calculate and store the solutions to other values of the wave numbers $m$ and $l$. In figure \ref{plot_perteta} we show a comparison for the perturbation in energy density $\delta\epsilon(\tau,r)$ for the values $m=0,\ldots,4$ (from top to bottom) and $l=1,\ldots,4$ (from left to right). We divide by a Bjorken time $\tau$-dependent but radius $r$-independent normalization factor $\bar\epsilon(\tau,0)$ for better visibility. The dashed lines show the perturbation in energy density at the initialization time $\tau=0.4$ fm/c while the solid lines give the time-evolved fields at Bjorken time $\tau=15$ fm/c. We compare there ideal fluid dynamics (red curves) to viscous fluid dynamics (orange curves) including shear and bulk viscous dissipation as discussed in section \ref{sec_transportcoeff} and summarized in table \ref{tab_parameters}. One observes a strong damping effect in the viscous case, as expected, in particular for the modes with larger wave numbers $m$ and $l$.

\section{Conclusions}
In summary, we have developed a theoretical approach to the fluid dynamics of relativistic heavy ion collisions using a background-fluctuation splitting and a mode expansion. The background configuration is taken to be symmetric with respect to azimuthal rotations and Bjorken boosts but has a non-trivial dependence on radius $r$. It corresponds essentially to an event average for an ensemble with random azimuthal orientation. In contrast, no such assumption is made for the perturbations around this configuration and they can depend on azimuthal angle $\phi$ and rapidity $\eta$. Using statistical symmetries with respect to azimuthal rotations and Bjorken boosts, we presented an expansion scheme in terms of orthogonal sets of functions which can also be used favorably for a numerical implementation of time propagation. Modes are classified in terms of an azimuthal wave number~$m$, a rapidity wave number~$k$ and a radial wave number~$l$. For transverse density perturbations, the resulting characterization of the initial state differs from the widely used event eccentricities, but there is an interesting relation as we have discussed.

Both background and perturbation fields are determined by hyperbolic partial differential equations. We have developed a numerical code to solve these equations based on the pseudo-spectral method. It allows to achieve very high numerical accuracy while being rather efficient at the same time. We have demonstrated this explicitly by comparing the numerical solution to the (semi-) analytically known Gubser solution on the level of the background equations. We have also tested the implementation of the algorithm to solve linearized equations for the perturbations by comparing two independent ways to propagate modes with azimuthal wave number $m=0$. We conclude from these exercises that our numerical code works very well and allows to obtain rather accurate results.

To apply the formalism to realistic heavy ions collisions, we have implemented a thermodynamic equation of state that interpolates continuously between known results at high temperature from lattice QCD calculations and a hadron resonance gas approximation at low temperatures. We found that a combination of exponential and rational functions leads to a good approximation, see eq.\ \eqref{eq:eosparamentrization} for the detailed expression and table \ref{tab:fitparamenter} for the best fit parameters. For the transport properties, in particular shear and bulk viscosity, we have made rather simple assumptions for the present work. They are summarized in table \ref{tab_parameters} together with parameters characterizing the initial energy density as obtained from a Glauber model.

On this basis we have solved the evolution equations for the symmetric background fields describing temperature, radial fluid velocity, the two independent components of shear stress and bulk viscous pressure in a realistic heavy ion collision scenario. The result is shown in fig.\ \ref{backgroundTv}. Moreover, we have calculated transfer functions for linear perturbations around this solution corresponding to modes of initial density perturbations with different azimuthal and radial wave numbers. The results are illustrated in figs.\ \ref{plot_evolution} and \ref{plot_perteta}. 

While we have demonstrated explicitly how perturbations with azimuthal and radial dependence resulting from initial density perturbations can be evolved, we stress that the method is more general and allows to calculate also the response to initial velocity, shear stress or bulk viscous pressure perturbations and to include longitudinal dependences. We plan to explore this in more detail in the future. Also, so far we have concentrated on simple scenarios for the shear and bulk viscous transport properties but we plan to investigate other scenarios. 

For a use of our algorithm to compare to experimental data, an important ingredient is the contribution of resonance decays to particle spectra. To this end, we have recently developed a fast algorithm which allows to precompute resonance decays such that particle spectra and flow coefficients can be calculated as simple integrals over the freeze-out surface \cite{Mazeliauskas:2018irt}. This method is particularly suited for use in the context of Fluid{\it u}M, and we plan to do so soon. 

We believe that these developments are useful in order to gain a more quantitative understanding of heavy ion collisions and the principles of relativistic fluid dynamics and quantum field theory that underly them.

\section*{Acknowledgment}
This work is part of and supported by the DFG Collaborative Research Centre ``SFB 1225 (ISOQUANT)''. The authors would like to thank S.~Masciocchi, A.~Mazeliauskas and other members of the ISOQUANT collaboration for useful discussions. S.~F.~thanks U.~A.~Wiedemann for collaboration on related, earlier work.

\appendix

\section{Discrete cosine and sine transform}
\label{app:ConventionFourier}
In this appendix we collect useful formula and conventions for discrete Fourier transformations. We use the following notation for a sequence of $N$ elements 
\begin{equation}
\{u_n\}= (u_1, \cdots u_N).
\end{equation}
A cyclic permutation is denoted by the symbol $\mathcal{C}_l$ such that
\begin{equation}
\mathcal{C}_{+1}\{u_n\}= (u_N,  u_1, \cdots u_{N-1}),\quad\quad\quad \mathcal{C}_{-1}\{u_n\}= (u_2,  u_3, \cdots u_{N},u_{1}).
\end{equation}
Discrete cosine and sine Fourier transforms are usually defined and implemented in terms of the following convention \cite{Rao:1990:DCT:96810,Press:2007:NRE:1403886},
\begin{align}
\text{DCT}_{\text{I}}\left\{ u_n \right\}_J & =\sqrt{\frac{2}{N-1}}\left(
	\frac{u_1}{2}+
	\sum_{n=2}^{N-1} u_n \cos\left[ \left(n-1 \right) \left(J-1\right)  \frac{\pi}{N}\right]
	+(-1)^{J-1}\frac{u_N}{2}
	\right),  \\
\text{DCT}_{\text{II}}\left\{ u_n \right\}_J &=\frac{1}{\sqrt{N}}\sum_{n=1}^N u_J \cos\left[ \left(J-1 \right) \left(n-\frac12 \right)  \frac{\pi}{N}\right],
	\\
\text{DCT}_{\text{III}}\left\{ u_n \right\}_J &=\frac{1}{\sqrt{N}}\left(u_1+2\sum_{n=2}^N u_n \cos\left[ \left(J-\frac{1}{2} \right) \left(n-1 \right)  \frac{\pi}{N}\right]\right), 
	\\
\text{DCT}_{\text{IV}}\left\{ u_n \right\}_J  &=\sqrt{\frac{2}{N}}\sum_{n=1}^N u_n \cos\left[ \left(J-\frac12 \right) \left(J-\frac12 \right)  \frac{\pi}{N}\right], \\
\text{DST}_{\text{I}}\left\{u_n\right\}_J&=\sqrt{\frac{2}{N+1}}\sum_{n=1}^{N} u_n \sin\left[n J \frac{\pi}{N+1} \right], \\
\text{DST}_{\text{II}}\left\{u_n\right\}_J&=\frac{1}{\sqrt{N}}\sum_{n=1}^{N} u_n \sin\left[\left(n -\frac{1}{2}\right) J \frac{\pi}{N} \right], \\
\text{DST}_{\text{III}}\left\{u_n\right\}_J&=\frac{1}{\sqrt{N}}\left(2\sum_{n=1}^{N-1} u_n \sin\left[n\left(J -\frac{1}{2}\right) \frac{\pi}{N} \right]+(-1)^{J-1}u_N\right), \\
\text{DST}_{\text{IV}} \left\{u_n\right\}_J&=\sqrt{\frac{2}{N}}\sum_{n=1}^N u_n \sin\left[ \left(n-\frac12 \right) \left(J-\frac12 \right)  \frac{\pi}{N}\right].
\end{align}
The roman index refers here to a specific periodicity condition and the external index $J$ refers to the resulting sequence of numbers. 

In these conventions, the relations in equations \eqref{eq:coefficientsanbn} and \eqref{eq:backwarddiscreteFourier} between the coefficients $a_n$, $b_n$ and the function values $f^e(r_J)$, $f^o(r_J)$ can be written as
\begin{align}
a_n &=\frac{2}{c_{2n-2}\sqrt{N}}\text{DCT}_{\text{II}} \left\{  f^e(r_J)\right\}_n, \quad\quad\quad f^e(r_J) = \frac{\sqrt{N}}{2} \; \text{DCT}_\text{III} \left\{ a_n \right\}_J , \\
b_n&=\sqrt{\frac{2}{N}} \text{DCT}_{\text{IV}}\left\{ f^o(r_J)\right\}_n, \quad\quad\quad\quad\quad f^o(r_J) = \sqrt{\frac{N}{2}} \text{DCT}_\text{IV} \left\{ b_n \right\}_J,
\end{align}
where we have used the fact that $c_{2n-1}=1$ for all $n$. The formulas for radial derivatives \eqref{eq:derivativeevendiscrete} become in these conventions
\begin{align}
\frac{\partial}{\partial r } f^e (r_J)& = 
-\frac{\partial \theta}{\partial r }\Big|_{r=r_J} \text{DST}_{\text{III}}\left\{\mathcal{C}_{-1}\{\text{DCT}_{\text{II}} \{f^e(r_J)\}_n(2n-2)\}
\right\}_J,\label{fourier1}\\
\frac{\partial}{\partial r } f^o (r_J) &= -\frac{\partial \theta}{\partial r }\Big|_{r=r_J}\text{DST}_{\text{IV}} \left\{\text{DCT}_{\text{IV}}\{f^o(r_J)\}_n(2n-1)\right\}_J\label{fourier2}.
\end{align}
Finally, the spectral viscosity operators in \eqref{eq:spectralviscosityoperators} become
\begin{equation}
\begin{split}
Q^{2p}f^o(r_J)&=-\epsilon_N \text{DCT}_{\text{IV}} \left\{\text{DCT}_{\text{IV}} \{ f^o(r_J)\}_n(2n-1)^{2p} \right\},\\
Q^{2p}f^e(r_J)&=-\epsilon_N \text{DCT}_{\text{III}} \left\{\text{DCT}_{\text{II}} \{ f^o(r_J)\}_n(2n-2)^{2p} \right\}.
\end{split}
\end{equation}
In this form, the relations are directly amendable to a numerical implementation of the pseudo-spectral method.




\begin{thebibliography}{}
\bibitem{ALICE:2011ab}
{\scshape ALICE} collaboration, K.~Aamodt et~al., \emph{{Higher harmonic
  anisotropic flow measurements of charged particles in Pb-Pb collisions at
  $\sqrt{s_{NN}}$=2.76 TeV}},
  \href{https://doi.org/10.1103/PhysRevLett.107.032301}{\emph{Phys. Rev. Lett.}
  {\bfseries 107} (2011) 032301},
  [\href{https://arxiv.org/abs/1105.3865}{{\ttfamily 1105.3865}}].

\bibitem{Chatrchyan:2012ta}
{\scshape CMS} collaboration, S.~Chatrchyan et~al., \emph{{Measurement of the
  elliptic anisotropy of charged particles produced in PbPb collisions at
  $\sqrt{s}_{NN}$=2.76 TeV}},
  \href{https://doi.org/10.1103/PhysRevC.87.014902}{\emph{Phys. Rev.}
  {\bfseries C87} (2013) 014902},
  [\href{https://arxiv.org/abs/1204.1409}{{\ttfamily 1204.1409}}].

\bibitem{ATLAS:2012at}
{\scshape ATLAS} collaboration, G.~Aad et~al., \emph{{Measurement of the
  azimuthal anisotropy for charged particle production in $\sqrt{s_{NN}}=2.76$
  TeV lead-lead collisions with the ATLAS detector}},
  \href{https://doi.org/10.1103/PhysRevC.86.014907}{\emph{Phys. Rev.}
  {\bfseries C86} (2012) 014907},
  [\href{https://arxiv.org/abs/1203.3087}{{\ttfamily 1203.3087}}].

\bibitem{Adare:2011tg}
{\scshape PHENIX} collaboration, A.~Adare et~al., \emph{{Measurements of
  Higher-Order Flow Harmonics in Au+Au Collisions at $\sqrt{s_{NN}} = 200$
  GeV}}, \href{https://doi.org/10.1103/PhysRevLett.107.252301}{\emph{Phys. Rev.
  Lett.} {\bfseries 107} (2011) 252301},
  [\href{https://arxiv.org/abs/1105.3928}{{\ttfamily 1105.3928}}].

\bibitem{Adamczyk:2013gw}
{\scshape STAR} collaboration, L.~Adamczyk et~al., \emph{{Elliptic flow of
  identified hadrons in Au+Au collisions at $\sqrt{s_{NN}}=$ 7.7-62.4 GeV}},
  \href{https://doi.org/10.1103/PhysRevC.88.014902}{\emph{Phys. Rev.}
  {\bfseries C88} (2013) 014902},
  [\href{https://arxiv.org/abs/1301.2348}{{\ttfamily 1301.2348}}].

\bibitem{Gyulassy:2004zy}
M.~Gyulassy and L.~McLerran, \emph{{New forms of QCD matter discovered at
  RHIC}}, \href{https://doi.org/10.1016/j.nuclphysa.2004.10.034}{\emph{Nucl.
  Phys.} {\bfseries A750} (2005) 30--63},
  [\href{https://arxiv.org/abs/nucl-th/0405013}{{\ttfamily nucl-th/0405013}}].

\bibitem{Schafer:2009dj}
T.~Schäfer and D.~Teaney, \emph{{Nearly Perfect Fluidity: From Cold Atomic
  Gases to Hot Quark Gluon Plasmas}},
  \href{https://doi.org/10.1088/0034-4885/72/12/126001}{\emph{Rept. Prog.
  Phys.} {\bfseries 72} (2009) 126001},
  [\href{https://arxiv.org/abs/0904.3107}{{\ttfamily 0904.3107}}].

\bibitem{Huovinen:2013wma}
P.~Huovinen, \emph{{Hydrodynamics at RHIC and LHC: What have we learned?}},
  \href{https://doi.org/10.1142/S0218301313300294}{\emph{Int. J. Mod. Phys.}
  {\bfseries E22} (2013) 1330029},
  [\href{https://arxiv.org/abs/1311.1849}{{\ttfamily 1311.1849}}].

\bibitem{Heinz:2013th}
U.~Heinz and R.~Snellings, \emph{{Collective flow and viscosity in relativistic
  heavy-ion collisions}},
  \href{https://doi.org/10.1146/annurev-nucl-102212-170540}{\emph{Ann. Rev.
  Nucl. Part. Sci.} {\bfseries 63} (2013) 123--151},
  [\href{https://arxiv.org/abs/1301.2826}{{\ttfamily 1301.2826}}].

\bibitem{Gale:2013da}
C.~Gale, S.~Jeon and B.~Schenke, \emph{{Hydrodynamic Modeling of Heavy-Ion
  Collisions}}, \href{https://doi.org/10.1142/S0217751X13400113}{\emph{Int. J.
  Mod. Phys.} {\bfseries A28} (2013) 1340011},
  [\href{https://arxiv.org/abs/1301.5893}{{\ttfamily 1301.5893}}].

\bibitem{shen2014standard}
C.~Shen, \emph{{The standard model for relativistic heavy-ion collisions and
  electromagnetic tomography}}, Ph.D. thesis, Ohio State U., 2014-07-25.

\bibitem{Jaiswal:2016hex}
A.~Jaiswal and V.~Roy, \emph{{Relativistic hydrodynamics in heavy-ion
  collisions: general aspects and recent developments}},
  \href{https://doi.org/10.1155/2016/9623034}{\emph{Adv. High Energy Phys.}
  {\bfseries 2016} (2016) 9623034},
  [\href{https://arxiv.org/abs/1605.08694}{{\ttfamily 1605.08694}}].

\bibitem{Romatschke:2017ejr}
P.~Romatschke and U.~Romatschke, \emph{{Relativistic Fluid Dynamics In and Out
  of Equilibrium -- Ten Years of Progress in Theory and Numerical Simulations
  of Nuclear Collisions}},  \href{https://arxiv.org/abs/1712.05815}{{\ttfamily
  1712.05815}}.

\bibitem{Busza:2018rrf}
W.~Busza, K.~Rajagopal and W.~van~der Schee, \emph{{Heavy Ion Collisions: The
  Big Picture, and the Big Questions}},
  \href{https://arxiv.org/abs/1802.04801}{{\ttfamily 1802.04801}}.

\bibitem{Dubla:2018czx}
A.~Dubla, S.~Masciocchi, J.~M. Pawlowski, B.~Schenke, C.~Shen and J.~Stachel,
  \emph{{Towards QCD-assisted hydrodynamics for heavy-ion collision
  phenomenology}},
  \href{https://doi.org/10.1016/j.nuclphysa.2018.09.046}{\emph{Nucl. Phys.}
  {\bfseries A979} (2018) 251--264},
  [\href{https://arxiv.org/abs/1805.02985}{{\ttfamily 1805.02985}}].

\bibitem{Song:2007ux}
H.~Song and U.~W. Heinz, \emph{{Causal viscous hydrodynamics in 2+1 dimensions
  for relativistic heavy-ion collisions}},
  \href{https://doi.org/10.1103/PhysRevC.77.064901}{\emph{Phys. Rev.}
  {\bfseries C77} (2008) 064901},
  [\href{https://arxiv.org/abs/0712.3715}{{\ttfamily 0712.3715}}].

\bibitem{Bozek:2011ua}
P.~Bozek, \emph{{Flow and interferometry in 3+1 dimensional viscous
  hydrodynamics}},
  \href{https://doi.org/10.1103/PhysRevC.85.034901}{\emph{Phys. Rev.}
  {\bfseries C85} (2012) 034901},
  [\href{https://arxiv.org/abs/1110.6742}{{\ttfamily 1110.6742}}].

\bibitem{Schenke:2010nt}
B.~Schenke, S.~Jeon and C.~Gale, \emph{{(3+1)D hydrodynamic simulation of
  relativistic heavy-ion collisions}},
  \href{https://doi.org/10.1103/PhysRevC.82.014903}{\emph{Phys. Rev.}
  {\bfseries C82} (2010) 014903},
  [\href{https://arxiv.org/abs/1004.1408}{{\ttfamily 1004.1408}}].

\bibitem{Karpenko:2013wva}
I.~Karpenko, P.~Huovinen and M.~Bleicher, \emph{{A 3+1 dimensional viscous
  hydrodynamic code for relativistic heavy ion collisions}},
  \href{https://doi.org/10.1016/j.cpc.2014.07.010}{\emph{Comput. Phys. Commun.}
  {\bfseries 185} (2014) 3016--3027},
  [\href{https://arxiv.org/abs/1312.4160}{{\ttfamily 1312.4160}}].

\bibitem{Shen:2014vra}
C.~Shen, Z.~Qiu, H.~Song, J.~Bernhard, S.~Bass and U.~Heinz, \emph{{The
  iEBE-VISHNU code package for relativistic heavy-ion collisions}},
  \href{https://doi.org/10.1016/j.cpc.2015.08.039}{\emph{Comput. Phys. Commun.}
  {\bfseries 199} (2016) 61--85},
  [\href{https://arxiv.org/abs/1409.8164}{{\ttfamily 1409.8164}}].

\bibitem{del2013relativistic}
L.~Del~Zanna, V.~Chandra, G.~Inghirami, V.~Rolando, A.~Beraudo, A.~De~Pace
  et~al., \emph{{Relativistic viscous hydrodynamics for heavy-ion collisions
  with ECHO-QGP}},
  \href{https://doi.org/10.1140/epjc/s10052-013-2524-5}{\emph{Eur. Phys. J.}
  {\bfseries C73} (2013) 2524},
  [\href{https://arxiv.org/abs/1305.7052}{{\ttfamily 1305.7052}}].

\bibitem{Habich:2014jna}
M.~Habich, J.~L. Nagle and P.~Romatschke, \emph{{Particle spectra and HBT radii
  for simulated central nuclear collisions of C + C, Al + Al, Cu + Cu, Au + Au,
  and Pb + Pb from $\sqrt{s}=62.4$ - $2760$ GeV}},
  \href{https://doi.org/10.1140/epjc/s10052-014-3206-7}{\emph{Eur. Phys. J.}
  {\bfseries C75} (2015) 15},
  [\href{https://arxiv.org/abs/1409.0040}{{\ttfamily 1409.0040}}].

\bibitem{Bernhard:2016tnd}
J.~E. Bernhard, J.~S. Moreland, S.~A. Bass, J.~Liu and U.~Heinz,
  \emph{{Applying Bayesian parameter estimation to relativistic heavy-ion
  collisions: simultaneous characterization of the initial state and
  quark-gluon plasma medium}},
  \href{https://doi.org/10.1103/PhysRevC.94.024907}{\emph{Phys. Rev.}
  {\bfseries C94} (2016) 024907},
  [\href{https://arxiv.org/abs/1605.03954}{{\ttfamily 1605.03954}}].

\bibitem{Auvinen:2017nrm}
J.~Auvinen, I.~A. Karpenko, J.~Bernhard and S.~Bass, \emph{{Parameter
  extractions for RHIC BES using Bayesian statistics}},
  \href{https://doi.org/10.22323/1.311.0019}{\emph{PoS} {\bfseries CPOD2017}
  (2018) 019}.

\bibitem{Moreland:2018gsh}
J.~S. Moreland, J.~E. Bernhard and S.~A. Bass, \emph{{Estimating initial state
  and quark-gluon plasma medium properties using a hybrid model with nucleon
  substructure calibrated to $p$-Pb and Pb-Pb collisions at
  $\sqrt{s_\mathrm{NN}}=5.02$ TeV}},
  \href{https://arxiv.org/abs/1808.02106}{{\ttfamily 1808.02106}}.

\bibitem{Floerchinger2014b}
S.~Floerchinger and U.~A. Wiedemann, \emph{{Mode-by-mode fluid dynamics for
  relativistic heavy ion collisions}},
  \href{https://doi.org/10.1016/j.physletb.2013.12.025}{\emph{Phys. Lett.}
  {\bfseries B728} (2014) 407--411},
  [\href{https://arxiv.org/abs/1307.3453}{{\ttfamily 1307.3453}}].

\bibitem{Floerchinger:2013vua}
S.~Floerchinger and U.~A. Wiedemann, \emph{{Characterization of initial
  fluctuations for the hydrodynamical description of heavy ion collisions}},
  \href{https://doi.org/10.1103/PhysRevC.88.044906}{\emph{Phys. Rev.}
  {\bfseries C88} (2013) 044906},
  [\href{https://arxiv.org/abs/1307.7611}{{\ttfamily 1307.7611}}].

\bibitem{Petersen:2013cta}
H.~Petersen, C.~E. Coleman-Smith and R.~L. Wolpert, \emph{{Quantifying Initial
  State Fluctuations in Heavy Ion Collisions}},
  \href{https://doi.org/10.5506/APhysPolBSupp.6.797}{\emph{Acta Phys. Polon.
  Supp.} {\bfseries 6} (2013) 797--802}.

\bibitem{floerchinger2014kinetic}
S.~Floerchinger and U.~A. Wiedemann, \emph{{Kinetic freeze-out, particle
  spectra and harmonic flow coefficients from mode-by-mode hydrodynamics}},
  \href{https://doi.org/10.1103/PhysRevC.89.034914}{\emph{Phys. Rev.}
  {\bfseries C89} (2014) 034914},
  [\href{https://arxiv.org/abs/1311.7613}{{\ttfamily 1311.7613}}].

\bibitem{Floerchinger2014a}
S.~Floerchinger and U.~A. Wiedemann, \emph{{Statistics of initial density
  perturbations in heavy ion collisions and their fluid dynamic response}},
  \href{https://doi.org/10.1007/JHEP08(2014)005}{\emph{JHEP} {\bfseries 08}
  (2014) 005}, [\href{https://arxiv.org/abs/1405.4393}{{\ttfamily 1405.4393}}].

\bibitem{Floerchinger2014}
S.~Floerchinger, U.~A. Wiedemann, A.~Beraudo, L.~Del~Zanna, G.~Inghirami and
  V.~Rolando, \emph{{How (non-)linear is the hydrodynamics of heavy ion
  collisions?}},
  \href{https://doi.org/10.1016/j.physletb.2014.06.049}{\emph{Phys. Lett.}
  {\bfseries B735} (2014) 305--310},
  [\href{https://arxiv.org/abs/1312.5482}{{\ttfamily 1312.5482}}].

\bibitem{Brouzakis:2014gka}
N.~Brouzakis, S.~Floerchinger, N.~Tetradis and U.~A. Wiedemann,
  \emph{{Nonlinear evolution of density and flow perturbations on a Bjorken
  background}}, \href{https://doi.org/10.1103/PhysRevD.91.065007}{\emph{Phys.
  Rev.} {\bfseries D91} (2015) 065007},
  [\href{https://arxiv.org/abs/1411.2912}{{\ttfamily 1411.2912}}].

\bibitem{Floerchinger:2017cii}
S.~Floerchinger and E.~Grossi, \emph{{Causality of fluid dynamics for
  high-energy nuclear collisions}},
  \href{https://doi.org/10.1007/JHEP08(2018)186}{\emph{JHEP} {\bfseries 08}
  (2018) 186}, [\href{https://arxiv.org/abs/1711.06687}{{\ttfamily
  1711.06687}}].

\bibitem{nla.cat-vn725379}
L.~D. Landau and E.~M. Lifshitz, \emph{Fluid mechanics / by L.D. Landau and
  E.M. Lifshitz ; translated from the Russian by J.B. Sykes and W.H. Reid}.
\newblock Pergamon Press ; Addison-Wesley London : Reading, Mass, 1959.

\bibitem{PhysRev.58.919}
C.~Eckart, \emph{The thermodynamics of irreversible processes. iii.
  relativistic theory of the simple fluid},
  \href{https://doi.org/10.1103/PhysRev.58.919}{\emph{Phys. Rev.} {\bfseries
  58} (1940) 919--924}.

\bibitem{PhysRevD.31.725}
W.~A. Hiscock and L.~Lindblom, \emph{Generic instabilities in first-order
  dissipative relativistic fluid theories},
  \href{https://doi.org/10.1103/PhysRevD.31.725}{\emph{Phys. Rev. D} {\bfseries
  31} (1985) 725--733}.

\bibitem{Hiscock:1983zz}
W.~A. Hiscock and L.~Lindblom, \emph{{Stability and causality in dissipative
  relativistic fluids}},
  \href{https://doi.org/10.1016/0003-4916(83)90288-9}{\emph{Annals Phys.}
  {\bfseries 151} (1983) 466--496}.

\bibitem{Israel:1979wp}
W.~Israel and J.~M. Stewart, \emph{{Transient relativistic thermodynamics and
  kinetic theory}},
  \href{https://doi.org/10.1016/0003-4916(79)90130-1}{\emph{Annals Phys.}
  {\bfseries 118} (1979) 341--372}.

\bibitem{Muller:1967zza}
I.~Müller, \emph{{Zum Paradoxon der Wärmeleitungstheorie}},
  \href{https://doi.org/10.1007/BF01326412}{\emph{Z. Phys.} {\bfseries 198}
  (1967) 329--344}.

\bibitem{Denicol:2012cn}
G.~S. Denicol, H.~Niemi, E.~Molnar and D.~H. Rischke, \emph{{Derivation of
  transient relativistic fluid dynamics from the Boltzmann equation}},
  \href{https://doi.org/10.1103/PhysRevD.85.114047,
  10.1103/PhysRevD.91.039902}{\emph{Phys. Rev.} {\bfseries D85} (2012) 114047},
  [\href{https://arxiv.org/abs/1202.4551}{{\ttfamily 1202.4551}}].

\bibitem{Muronga:2003ta}
A.~Muronga, \emph{{Causal theories of dissipative relativistic fluid dynamics
  for nuclear collisions}},
  \href{https://doi.org/10.1103/PhysRevC.69.034903}{\emph{Phys. Rev.}
  {\bfseries C69} (2004) 034903},
  [\href{https://arxiv.org/abs/nucl-th/0309055}{{\ttfamily nucl-th/0309055}}].

\bibitem{Baier:2007ix}
R.~Baier, P.~Romatschke, D.~T. Son, A.~O. Starinets and M.~A. Stephanov,
  \emph{{Relativistic viscous hydrodynamics, conformal invariance, and
  holography}},
  \href{https://doi.org/10.1088/1126-6708/2008/04/100}{\emph{JHEP} {\bfseries
  04} (2008) 100}, [\href{https://arxiv.org/abs/0712.2451}{{\ttfamily
  0712.2451}}].

\bibitem{Romatschke:2009kr}
P.~Romatschke, \emph{{Relativistic Viscous Fluid Dynamics and Non-Equilibrium
  Entropy}}, \href{https://doi.org/10.1088/0264-9381/27/2/025006}{\emph{Class.
  Quant. Grav.} {\bfseries 27} (2010) 025006},
  [\href{https://arxiv.org/abs/0906.4787}{{\ttfamily 0906.4787}}].

\bibitem{Molnar:2013lta}
E.~Molnár, H.~Niemi, G.~S. Denicol and D.~H. Rischke, \emph{{Relative
  importance of second-order terms in relativistic dissipative fluid
  dynamics}}, \href{https://doi.org/10.1103/PhysRevD.89.074010}{\emph{Phys.
  Rev.} {\bfseries D89} (2014) 074010},
  [\href{https://arxiv.org/abs/1308.0785}{{\ttfamily 1308.0785}}].

\bibitem{Borsanyi:2016ksw}
S.~Borsanyi et~al., \emph{{Calculation of the axion mass based on
  high-temperature lattice quantum chromodynamics}},
  \href{https://doi.org/10.1038/nature20115}{\emph{Nature} {\bfseries 539}
  (2016) 69--71}, [\href{https://arxiv.org/abs/1606.07494}{{\ttfamily
  1606.07494}}].

\bibitem{Bazavov:2014pvz}
{\scshape HotQCD} collaboration, A.~Bazavov et~al., \emph{{Equation of state in
  ( 2+1 )-flavor QCD}},
  \href{https://doi.org/10.1103/PhysRevD.90.094503}{\emph{Phys. Rev.}
  {\bfseries D90} (2014) 094503},
  [\href{https://arxiv.org/abs/1407.6387}{{\ttfamily 1407.6387}}].

\bibitem{huovinen2010qcd}
P.~Huovinen and P.~Petreczky, \emph{{QCD Equation of State and Hadron Resonance
  Gas}}, \href{https://doi.org/10.1016/j.nuclphysa.2010.02.015}{\emph{Nucl.
  Phys.} {\bfseries A837} (2010) 26--53},
  [\href{https://arxiv.org/abs/0912.2541}{{\ttfamily 0912.2541}}].

\bibitem{KARSCH2008217}
F.~Karsch, D.~Kharzeev and K.~Tuchin, \emph{Universal properties of bulk
  viscosity near the qcd phase transition},
  \href{https://doi.org/https://doi.org/10.1016/j.physletb.2008.01.080}{\emph{Physics
  Letters B} {\bfseries 663} (2008) 217 -- 221}.

\bibitem{PhysRevLett.115.112002}
N.~Christiansen, M.~Haas, J.~M. Pawlowski and N.~Strodthoff, \emph{Transport
  coefficients in yang-mills theory and qcd},
  \href{https://doi.org/10.1103/PhysRevLett.115.112002}{\emph{Phys. Rev. Lett.}
  {\bfseries 115} (Sep, 2015) 112002}.

\bibitem{Kovtun:2004de}
P.~Kovtun, D.~T. Son and A.~O. Starinets, \emph{{Viscosity in strongly
  interacting quantum field theories from black hole physics}},
  \href{https://doi.org/10.1103/PhysRevLett.94.111601}{\emph{Phys. Rev. Lett.}
  {\bfseries 94} (2005) 111601},
  [\href{https://arxiv.org/abs/hep-th/0405231}{{\ttfamily hep-th/0405231}}].

\bibitem{policastro2001shear}
G.~Policastro, D.~T. Son and A.~O. Starinets, \emph{Shear viscosity of strongly
  coupled n= 4 supersymmetric yang-mills plasma}, {\emph{Physical Review
  Letters} {\bfseries 87} (2001) 081601}.

\bibitem{Tadmor}
E.~Tadmor, \emph{Convergence of spectral methods for nonlinear conservation
  laws}, \href{https://doi.org/10.1137/0726003}{\emph{SIAM Journal on Numerical
  Analysis} {\bfseries 26} (1989) 30--44},
  [\href{https://arxiv.org/abs/https://doi.org/10.1137/0726003}{{\ttfamily
  https://doi.org/10.1137/0726003}}].

\bibitem{Gottlieb1985}
D.~Gottlieb and E.~Tadmor, \emph{Recovering Pointwise Values of Discontinuous
  Data within Spectral Accuracy}, pp.~357--375.
\newblock Birkh{\"a}user Boston, Boston, MA, 1985.
\newblock 10.1007/978-1-4612-5162-0\_19.

\bibitem{GOTTLIEB200183}
D.~Gottlieb and J.~Hesthaven, \emph{Spectral methods for hyperbolic problems},
  \href{https://doi.org/https://doi.org/10.1016/S0377-0427(00)00510-0}{\emph{Journal
  of Computational and Applied Mathematics} {\bfseries 128} (2001) 83 -- 131}.

\bibitem{Boyd2000}
J.~Boyd, \emph{Chebyshev and Fourier Spectral Methods: Second Revised Edition}.
\newblock Dover Books on Mathematics. Dover Publications, 2001.

\bibitem{hesthaven_gottlieb_gottlieb_2007}
J.~S. Hesthaven, S.~Gottlieb and D.~Gottlieb, \emph{Spectral Methods for
  Time-Dependent Problems}.
\newblock Cambridge Monographs on Applied and Computational Mathematics.
  Cambridge University Press, 2007,
  \href{https://doi.org/10.1017/CBO9780511618352}{10.1017/CBO9780511618352}.

\bibitem{rezzolla2013relativistic}
L.~Rezzolla and O.~Zanotti, \emph{Relativistic hydrodynamics}.
\newblock Oxford University Press, 2013.

\bibitem{gubser2010symmetry}
S.~S. Gubser, \emph{{Symmetry constraints on generalizations of Bjorken flow}},
  \href{https://doi.org/10.1103/PhysRevD.82.085027}{\emph{Phys. Rev.}
  {\bfseries D82} (2010) 085027},
  [\href{https://arxiv.org/abs/1006.0006}{{\ttfamily 1006.0006}}].

\bibitem{cheb}
H.~Ma, \emph{Chebyshev--legendre super spectral viscosity method for nonlinear
  conservation laws},
  \href{https://doi.org/10.1137/S0036142995293912}{\emph{SIAM Journal on
  Numerical Analysis} {\bfseries 35} (1998) 893--908},
  [\href{https://arxiv.org/abs/https://doi.org/10.1137/S0036142995293912}{{\ttfamily
  https://doi.org/10.1137/S0036142995293912}}].

\bibitem{GELB20003}
A.~Gelb and E.~Tadmor, \emph{Enhanced spectral viscosity approximations for
  conservation laws},
  \href{https://doi.org/https://doi.org/10.1016/S0168-9274(99)00067-7}{\emph{Applied
  Numerical Mathematics} {\bfseries 33} (2000) 3 -- 21}.

\bibitem{Tadmor2012}
E.~Tadmor and K.~Waagan, \emph{Adaptive spectral viscosity for hyperbolic
  conservation laws}, \href{https://doi.org/10.1137/110836456}{\emph{SIAM
  Journal on Scientific Computing} {\bfseries 34} (2012) A993--A1009},
  [\href{https://arxiv.org/abs/https://doi.org/10.1137/110836456}{{\ttfamily
  https://doi.org/10.1137/110836456}}].

\bibitem{Gubser:2010ze}
S.~S. Gubser, \emph{{Symmetry constraints on generalizations of Bjorken flow}},
  \href{https://doi.org/10.1103/PhysRevD.82.085027}{\emph{Phys. Rev.}
  {\bfseries D82} (2010) 085027},
  [\href{https://arxiv.org/abs/1006.0006}{{\ttfamily 1006.0006}}].

\bibitem{Gubser:2010ui}
S.~S. Gubser and A.~Yarom, \emph{{Conformal hydrodynamics in Minkowski and de
  Sitter spacetimes}},
  \href{https://doi.org/10.1016/j.nuclphysb.2011.01.012}{\emph{Nucl. Phys.}
  {\bfseries B846} (2011) 469--511},
  [\href{https://arxiv.org/abs/1012.1314}{{\ttfamily 1012.1314}}].

\bibitem{Marrochio:2013wla}
H.~Marrochio, J.~Noronha, G.~S. Denicol, M.~Luzum, S.~Jeon and C.~Gale,
  \emph{{Solutions of Conformal Israel-Stewart Relativistic Viscous Fluid
  Dynamics}}, \href{https://doi.org/10.1103/PhysRevC.91.014903}{\emph{Phys.
  Rev.} {\bfseries C91} (2015) 014903},
  [\href{https://arxiv.org/abs/1307.6130}{{\ttfamily 1307.6130}}].

\bibitem{Qiu:2011hf}
Z.~Qiu, C.~Shen and U.~Heinz, \emph{{Hydrodynamic elliptic and triangular flow
  in Pb-Pb collisions at $\sqrt{s}=2.76$ATeV}},
  \href{https://doi.org/10.1016/j.physletb.2011.12.041}{\emph{Phys. Lett.}
  {\bfseries B707} (2012) 151--155},
  [\href{https://arxiv.org/abs/1110.3033}{{\ttfamily 1110.3033}}].

\bibitem{Mazeliauskas:2018irt}
A.~Mazeliauskas, S.~Floerchinger, E.~Grossi and D.~Teaney, \emph{{Fast
  resonance decays in nuclear collisions}},
  \href{https://arxiv.org/abs/1809.11049}{{\ttfamily 1809.11049}}.

\bibitem{Rao:1990:DCT:96810}
K.~R. Rao and P.~Yip, \emph{Discrete Cosine Transform: Algorithms, Advantages,
  Applications}.
\newblock Academic Press Professional, Inc., San Diego, CA, USA, 1990.

\bibitem{Press:2007:NRE:1403886}
W.~H. Press, S.~A. Teukolsky, W.~T. Vetterling and B.~P. Flannery,
  \emph{Numerical Recipes 3rd Edition: The Art of Scientific Computing}.
\newblock Cambridge University Press, New York, NY, USA, 3~ed., 2007.

\end{thebibliography}

\end{document}